%% file: 2hdms-n2hdm.tex
\documentclass[a4paper,12pt]{article}
\pdfoutput=1
\usepackage{geometry}
\usepackage{amsmath,amssymb,amsfonts}
\usepackage{xcolor,graphicx,soul}
\usepackage{cite}

\usepackage{array,booktabs,longtable}
\usepackage{caption,microtype}
\usepackage{subcaption}
\usepackage{hyperref}

\usepackage{datetime}
\usepackage{appendix}
\usepackage{url}
\usepackage[normalem]{ulem}
\usepackage{siunitx}
\usepackage{placeins}
\usepackage{xspace}

\input{paperdef}
\graphicspath{{figs/}}
\renewcommand{\arraystretch}{1.2}

\allowdisplaybreaks
\usepackage[numbers,sort&compress]{natbib}

\begin{document}
\thispagestyle{empty}

\def\thefootnote{\fnsymbol{footnote}}

\begin{flushright}
\mbox{}
DESY 26-097\\
IFT--UAM/CSIC-26-032\\
\end{flushright}

\mbox{}
\vspace{-0.5cm}

\begin{center}

{\large\sc 
  {\bf  {The 95 GeV Excess in Models with Two Higgs Doublets\\[.5em]
   plus One Singlet: Model Distinction via Four-top Final States\\[.5em] and Triple Higgs Couplings}}}\\

\vspace{0.5cm}

{\sc
 S.~Heinemeyer$^{1}$%
\footnote{emails: Sven.Heinemeyer@cern.ch, lich389@mail.sysu.edu.cn, gudrid.moortgat-pick@desy.de,\\ 
\mbox{}\hspace{17mm} daniel.schieber@ugent.be}%
, C.~Li$^{2}$%
, G.~Moortgat-Pick$^{3,4}$%
~and D.~Schieber$^{3,5}$%
}

\vspace*{.7cm}

{\sl
$^1$Instituto de F\'isica Te\'orica (UAM/CSIC), 
Universidad Aut\'onoma de Madrid, \\ 
Cantoblanco, 28049, Madrid, Spain

\vspace{0.1cm}

$^2$Shenzhen Campus of 
Sun Yat-sen University (SYSU), Gongchang Road 66, Shenzhen, 518107, People's Republic of China 

\vspace{0.1cm}

$^3$Deutsches Elektronen-Synchrotron DESY, Notkestraße 85, 22607 Hamburg, Germany

\vspace{0.1cm}

$^4$II. Institut f\"ur Theoretische Physik, Universit\"at Hamburg,\\
Luruper Chaussee 149, 22761 Hamburg, Germany

\vspace{0.1cm}

$^5$Department of Physics and Astronomy, Ghent University,\\ Krijgslaan 299, 9000 Ghent, Belgium
}

\end{center}

\vspace*{0.1cm}

\begin{abstract}
\noindent
We investigate the prospects for experimentally distinguishing the Next-to-Two-Higgs-Doublet Model (N2HDM) 
and the Two-Higgs-Doublet Model with a Complex Singlet (2HDMS) in the Yukawa type~II realization. 
Both models can successfully accommodate the reported Higgs-boson excesses around $95 \gev$ while satisfying 
current theoretical and experimental constraints, leading to very similar predictions for the observed Higgs spectrum and signal strengths. 
We analyze how they can nevertheless be discriminated through observables sensitive to the scalar potential and the $\mathcal{CP}$-odd sector. 
Analytical expressions for the trilinear Higgs couplings are derived, exhibiting characteristic contributions induced by 
the additional cubic interactions of the 2HDMS. We study the corresponding phenomenology in two complementary collider environments: 
four-top production at the High-Luminosity LHC, probing the different pseudoscalar sectors. In a second step we analyze 
Higgs pair production at a future high-energy $e^+e^-$ collider, providing direct sensitivity to the trilinear Higgs couplings. 
We quantify the potential to distinguish the two models by introducing the minimum deviation from the N2HDM limit required 
for a $2\sigma$ difference between the two models. Although our numerical analysis focuses on benchmark scenarios describing
the $95 \gev$ excesses, the proposed approach provides a general framework for discriminating between singlet-extended Higgs sectors 
at future colliders.

\end{abstract}


\def\thefootnote{\arabic{footnote}}
\setcounter{page}{0}
\setcounter{footnote}{0}


\newpage
\input{section1}

\input{section2.1-7}


\input{section2.8}
\input{section3}
\input{section4}


\input{section5}

\subsection*{Acknowledgements}

The work of S.H.\ has received financial support from the
grant PID2019-110058GB-C21 funded by
MCIN/AEI/10.13039/501100011033 and by ``ERDF A way of making Europe'', and in part by by the grant
IFT Centro de Excelencia Severo Ochoa CEX2020-001007-S funded by MCIN/AEI/10.13039/ 501100011033.
S.H.\ also acknowledges support from Grant PID2022-142545NB-C21 funded by MCIN/ AEI/10.13039/501100011033/ FEDER, UE.
G.M.-P.\ acknowledges support by the Deutsche Forschungsgemeinschaft (DFG, German Research Foundation) under Germany’s Excellence Strategy\textemdash EXC 2121 ``Quantum Universe"\textemdash390833306.
The work of D.S.\ was supported by the European Research Council (ERC) under the European Union’s Horizon Europe research and innovation program grant agreement 101078449 (ERC Starting Grant MultiScaleAmp). Views and opinions expressed are however those of the authors only and do not necessarily reflect those of the European Union or the European Research Council Executive Agency. Neither the European Union nor the granting authority can be held responsible for them.
\textit{}



\begin{appendices}
\section*{Appendix}
\input{appendixA.tex}
\input{appendixB.tex}
\input{appendixC.tex}
\end{appendices}

\newpage

\bibliographystyle{utphys}
\bibliography{reference}

\end{document}

%% file: paperdef.tex
\newcommand{\lsim}
{\;\raisebox{-.3em}{$\stackrel{\displaystyle <}{\sim}$}\;}

\newcommand\Code[1]{\ensuremath{\texttt{#1}}}

\newcommand\al{\alpha}
\newcommand\be{\beta}
\newcommand\tb{\tan\beta}

\newcommand\ReDiag{\mathop{%
  \raise .5pt\hbox{[}%
  \widetilde{\mathrm{Re}}%
  \raise .5pt\hbox{]}}}
\newcommand\ReOffDiag{\mathop{%
  \raise .5pt\hbox{$\llbracket$}%
  \widetilde{\mathrm{Re}}%
  \raise .5pt\hbox{$\rrbracket$}}}

\newcommand\NTHDM{\mathrm{N2HDM}}
\newcommand\THDMS{\mathrm{2HDMS}}

\newcommand\mt{m_t}

\newcommand\hotf{\ensuremath{h_{125}}\xspace}
\newcommand\hnf{\ensuremath{h_{95}}\xspace}

\newcommand\refeq[1]{Eq.~(\ref{#1})}
\newcommand\refeqs[1]{Eqs.~(\ref{#1})}
\newcommand\refta[1]{Tab.~\ref{#1}}

\newcommand\refse[1]{Sec.~\ref{#1}}

\newcommand\citere[1]{Ref.~\cite{#1}}

\newcommand{\CP}{{\cal CP}}
\newcommand{\cp}{{\CP}}

\newcommand{\gev}{\,\, \mathrm{GeV}}

\newcommand\fb{\ensuremath{\mbox{fb}}}

\newcommand{\br}{\text{BR}}
\newcommand{\De}{\Delta}

\newcommand{\sig}{\sigma}

\def\reffi#1{\mbox{Fig.~\ref{#1}}}
\def\reffis#1{\mbox{Figs.~\ref{#1}}}

\def\ga{\gamma}

\def\sgn{\mathrm{sgn}}

\newcommand{\zet}[1]{\ensuremath{\mathbb{Z}_{#1}}}

\definecolor{Orange}{named}{orange}
\definecolor{Purple}{named}{purple}
\definecolor{Lightblue}{cmyk}{0.9,0.1,0.1,0.3}
\definecolor{dgelborange}{cmyk}{0.,0.3,0.5, 0.}
\definecolor{Lila}{rgb}{0.5,0.,1}

\newcommand{\rbr}[1]{\left(#1\right)}
\newcommand{\dv}[3]{\begin{pmatrix} #1 \\ #2 \\ #3\\\end{pmatrix}}
\newcommand{\zv}[2]{\begin{pmatrix} #1 \\ #2 \\ \end{pmatrix}}
\newcommand{\ld}{\mathcal{L}}

\newcommand{\doublenewline}{\newline\newline}
\newcommand{\abs}[1] {\left\lvert #1 \right\rvert}

  \usepackage{tikz}
  \usepackage{feynmf}
  \usetikzlibrary{shapes,arrows,positioning,automata,backgrounds,calc,er,patterns}
  \usepackage{tikz-feynman}
  \tikzfeynmanset{compat=1.0.0} 
  
  \usepackage{dsfont}

%% file: section1.tex
\section {Introduction}
\label{sec:intro}

Since the discovery of the Higgs boson at the LHC, one of the main goals of particle
physics has become the precise determination of the structure of the scalar sector.
While all measurements performed so far are compatible with the predictions of the
Standard Model (SM), many well-motivated extensions predict an enlarged Higgs sector
whose additional states may still have escaped detection. Precision studies of Higgs
properties, together with direct searches for additional scalar particles, therefore
provide one of the most promising avenues for uncovering physics beyond the SM (BSM).

Among the simplest and most extensively studied extensions are models containing two
Higgs doublets together with an additional scalar singlet. In particular, the
Next-to-Two-Higgs-Doublet Model (N2HDM)~{\cite{Muhlleitner:2016mzt}}, featuring a real singlet, and the
Two-Higgs-Doublet Model with a Complex Singlet (2HDMS)~{\cite{Baum:2018zhf}} constitute attractive and
well-motivated scenarios. Although these models differ in the structure of their scalar
potentials and in their $\mathcal{CP}$-odd sectors, they possess the same $\mathcal{CP}$-even Higgs
spectra and can simultaneously accommodate the observed $125 \gev$ Higgs boson, $h_{125}$,
while possessing a different $\mathcal{CP}$-odd spectrum. Consequently, once new Higgs bosons are
discovered, distinguishing experimentally between these models may prove highly
challenging.

A particularly interesting motivation for studying such scenarios is provided by the
long-standing hints for an additional Higgs boson with a mass around $95 \gev$, $h_{95}$.
Several independent experiments have reported mild excesses compatible with such a
state. These observations can be expressed in terms of the signal strength,
\begin{equation}
\mu_X =
\frac{\sigma(pp/e^+e^- \to h)\,
{\rm BR}(h\to X)}
{\sigma_{\rm SM}\,
{\rm BR}_{\rm SM}(H\to X)},
\end{equation}
where the numerator is given by the BSM prediction, and the denominator is evaluated for a SM Higgs-boson with the same
mass as BSM Higgs boson.
The LEP collaboration observed an excess in the Higgsstrahlung process
$e^+e^- \to Z(H\to b\bar b)$ corresponding to~{\cite{Barate:2003sz}}, 
\begin{equation}
\mu_{b\bar b}^{\rm LEP}
=
0.117 \pm 0.057,
\end{equation}
while CMS reported an excess in the di-photon channel~{\cite{Sirunyan:2018aui,CMS:2023yay}},
\begin{equation}
\mu_{\gamma\gamma}^{\rm CMS}
=
0.33^{+0.19}_{-0.12},
\end{equation}
at a compatible Higgs-boson mass. More recently, ATLAS has also reported a smaller
excess in the di-photon channel~{\cite{ATLAS:2023jzc}},
\begin{equation}
\mu_{\gamma\gamma}^{\rm ATLAS}
=
0.18 \pm 0.10
\end{equation}
(see \citere{Biekotter:2023oen} for a combination; however, it should be kept in mind that we use the two individual signal strengths.)
CMS also reported an excess in the  $gg\rightarrow\phi\rightarrow\tau\tau$ channel with $\mu_{\tau\tau}^{\mathrm{exp}}=1.2\pm 0.5$~\cite{CMS:2022goy}
for a Higgs boson at $\sim 95 \gev$. However, such a high $\mu_{\tau\tau}$ value for a $\mathcal{CP}$-even Higgs boson is excluded by experimental
bounds from recent searches performed by CMS for the production of a Higgs boson in association with a top-quark pair or in association with a Z boson,
with subsequent decay into tau pairs~\cite{CMS:2022arx}, and we do not take this into account here.
While none of these observations reaches the statistical significance required for
a discovery, their remarkable consistency in mass has generated considerable interest
in singlet-extended Higgs sectors, where a mostly singlet-like scalar around
$95 \gev$ can naturally coexist with the observed SM-like Higgs boson at $125 \gev$.
Several theoretical interpretations of the $\sim 95 \gev$ excesses have been proposed over the past years. The first comprehensive analyses 
focused on the Next-to-Two-Higgs-Doublet Model (N2HDM), showing that a predominantly singlet-like $\mathcal{CP}$-even Higgs boson can simultaneously account 
for the LEP $b\bar b$ and CMS di-photon excesses while remaining compatible with all theoretical and experimental 
constraints~\cite{Biekotter:2019kde}. Subsequently, the collider prospects of this scenario were investigated for future $e^+e^-$ 
machines~\cite{Biekotter:2020ahz}, and later studies demonstrated that the N2HDM can also accommodate the CMS di-$\tau$ excess together with the 
LEP and di-photon anomalies~\cite{Biekotter:2023jld}. Alternative explanations have been explored in a variety of extensions of the SM Higgs 
sector, including supersymmetric models such as the NMSSM~\cite{Domingo:2018uim,Biekotter:2021qbc,Ellwanger:2023zjc}
or the $\mu\nu$SSM~\cite{Biekotter:2017xmf,Biekotter:2019gtq}, the type~I complex 2HDM~\cite{Azevedo:2023zkg},  
and other non-minimal Higgs sectors. 
In \citere{Heinemeyer:2021msz} we investigated the possibility of explaining these excesses within
the type~II realizations of the N2HDM and the 2HDMS. We demonstrated that both
models can successfully accommodate the observed excesses while satisfying all current
theoretical and experimental constraints. Furthermore, we showed that the two models
can lead to remarkably similar predictions for the experimentally accessible Higgs
spectrum and signal strengths. In \citere{Heinemeyer:2021msz} it was concluded that precision measurements of the $h_{125}$ or the $h_{95}$ 
at the LHC or a future $e^+e^-$ collider cannot distinguish the two model realizations. 
This naturally raises the question of whether the two models can
be distinguished by other future collider measurements.

The purpose of the present work is to take another step to answer this question. Rather than
relying solely on Higgs signal strengths, we investigate observables that are directly
sensitive to the different structures of the scalar potentials and $\mathcal{CP}$-odd sectors. 
We concentrate on the parameter space(s) that yield a description of the $\ga\ga$ and $b\bar b$ excesses, but
leave out the di-tau excesses, as they are most challenged, as discussed above (for a $\mathcal{CP}$-even Higgs boson), by other searches~{\cite{CMS:2022goy}}.
The first class of observables consists of heavy Higgs production in association with
top quarks at the High-Luminosity LHC. Owing to the different $\mathcal{CP}$-odd sectors of the
N2HDM and the 2HDMS, the resulting four-top final state provides direct sensitivity
to the additional pseudoscalar mixing angle of the 2HDMS. The second class comprises
double-Higgs production processes at a future high-energy $e^+e^-$ collider. These
processes probe the trilinear Higgs self-couplings, which receive characteristic
contributions from the different scalar potentials and therefore provide a
complementary handle to distinguish the two models.

Throughout this paper we concentrate on benchmark scenarios that explain the
$95 \gev$ excesses. Nevertheless, the analysis presented here should be viewed in a
much broader context. The strategy developed for distinguishing the N2HDM from the
2HDMS through top-associated Higgs production and precision measurements of Higgs
self-interactions is largely independent of the specific benchmark considered. The
$95 \gev$ excesses merely provide a well-motivated and phenomenologically interesting
framework in which to develop and illustrate this programme. The methods presented
here can readily be applied to a much wider class of extended Higgs sectors exhibiting
similar low-energy phenomenology.

The remainder of this paper is organized as follows.
In \refse{sec:2hdms} we give a brief review of the N2HDM and the 2HDMS, derive compact analytical
expressions for the relevant trilinear Higgs couplings, discuss the theoretical and
experimental constraints, and define the benchmark scenarios. In \refse{sec:hl-lhc} we investigate
the prospects for distinguishing the two models through four-top production at the
HL-LHC. In \refse{sec:epem} present our analysis of Higgs pair production at a future $e^+e^-$ collider
and its sensitivity to the trilinear Higgs couplings. Finally, our conclusions are
presented in \refse{sec:conclusions}. 

%% file: section2.1-7.tex
\section{Models and Constraints}
\label{sec:2hdms}

\subsection{Symmetries and the Higgs Potential}
\label{sec:potential}
In this section, we introduce the two models, N2HDM and 2HDMS, which extend the 2HDM with an additional real or complex singlet, respectively{, and contrast their differences.} After electroweak symmetry breaking, the scalar fields $\Phi_1,\Phi_2$ and $S$ acquire nonzero vacuum expectation values (VEVs). Expanding around the VEVs yields the following expressions:
\begin{gather}
   \Phi_1=\zv{\chi_1^+}{\displaystyle{\frac{v_1+\rho_1+i\eta_1}{\sqrt{2}}}}, \qquad \Phi_2=\zv{\chi_2^+}{\displaystyle{\frac{v_2+\rho_2+i\eta_2}{\sqrt{2}}}}.
\end{gather}
The singlet can be either real, as it is the case in the N2HDM 
\begin{equation}
    S=v_S+\rho_S,
\end{equation}
or explicitly complex in the 2HDMS
\begin{equation}
    S=\displaystyle{\frac{v_S+\rho_S+i\eta_S}{\sqrt{2}}}.
\end{equation}
The doublet  VEVs $v_1$ and $v_2$ have to satisfy the relation 
\begin{equation}
    v_1^2+v_2^2=v^2=(\SI{246.22}{GeV})^2,
    \label{eq:vev_mass}
\end{equation}
in order to retain the gauge boson masses \cite{peskin:1995QFT}. \doublenewline
Additionally, the models have different symmetry structures. Both models respect the discrete $\zet2$ symmetry that was already imposed on the 2HDM in order to prevent flavor changing neutral currents at tree-level. The N2HDM features an additional $\zet2^\prime$ symmetry that acts on the singlet, whereas the 2HDMS obeys an additional $\zet3$ symmetry.  The symmetries are summarised as follows,

\begin{align}
    {\rm N2HDM~and~2HDMS~}\quad \zet2 &~:~
    \Phi_1 \to \Phi_1, \qquad \Phi_2 \to -\Phi_2, \qquad S\to S\,, \\[.3em]
    {\rm N2HDM~}\quad \zet2' &~:~ 
    \Phi_1 \to \Phi_1, \qquad \Phi_2 \to \Phi_2, \qquad S\to -S\,, \\[.3em]
    {\rm 2HDMS~}\quad \zet3 &~:~ 
    	\begin{pmatrix} \Phi_1\\ \Phi_2\\ S \end{pmatrix} \to 
    	\begin{pmatrix} 
    	1&	&	\\	&	e^{i2\pi/3}&	\\	&	&	e^{-i2\pi/3}
	   \end{pmatrix}\,
	   \begin{pmatrix} \Phi_1\\ \Phi_2\\ S
	\end{pmatrix} \,.
\end{align}
The most general potential involving two doublets and one singlet is given by \cite{Baum:2018zhf}
\begin{equation}
	\begin{split}
		V(\Phi_1,\Phi_2,S) &=m_{11}^2\Phi_1^\dagger\Phi_1+m_{22}^2\Phi_2^\dagger\Phi_2-(m_{12}^2\Phi_1^\dagger\Phi_2+\mathrm{h.c.})\\
		&+\frac{\lambda_1}{2}(\Phi_1^\dagger\Phi_1)^2+\frac{\lambda_2}{2}(\Phi_2^\dagger\Phi_2)^2+\lambda_3(\Phi_1^\dagger\Phi_1)(\Phi_2^\dagger\Phi_2)+\lambda_4(\Phi_1^\dagger\Phi_2)(\Phi_2^\dagger\Phi_1)\\
		&+\Big[\frac{\lambda_5}{2}(\Phi_1^\dagger\Phi_2)^2+\lambda_6(\Phi_1^\dagger\Phi_1)(\Phi_1^\dagger\Phi_2)+\lambda_7(\Phi_2^\dagger\Phi_2)(\Phi_1^\dagger\Phi_2)+\mathrm{h.c.}\Big]\\
		&+\rbr{\xi S+\mathrm{h.c.}}+m_S^2 S^\dagger S+\rbr{\frac{{m'_S}^2}{2}S^2+\mathrm{h.c.}}+\frac{\lambda''_3}{4}(S^\dagger S)^2\\
		&+\left(\frac{\mu_{S1}}{6}S^3+\frac{\mu_{S2}}{2}S S^\dagger S+\mathrm{h.c.}\right)+\left(\frac{\lambda''_1}{24}S^4+\frac{\lambda''_2}{6}S^2 S^\dagger S+\mathrm{h.c.}\right)\\
		&+\Big[S(\mu_{11}\Phi_1^\dagger\Phi_1+\mu_{22}\Phi_2^\dagger\Phi_2+\mu_{12}\Phi_1^\dagger\Phi_2 +\mu_{21}\Phi_2^\dagger\Phi_1)+\mathrm{h.c.}\Big]\\
		&+S^\dagger S\Big[\lambda'_1\Phi_1^\dagger\Phi_1+\lambda'_2\Phi_2^\dagger\Phi_2+\lambda'_3\Phi_1^\dagger\Phi_2+\mathrm{h.c.}\Big]\\
		&+\Big[S^2(\lambda'_4\Phi_1^\dagger\Phi_1+\lambda'_5\Phi_2^\dagger\Phi_2+\lambda'_6\Phi_1^\dagger\Phi_2+\lambda'_7\Phi_2^\dagger\Phi_1)+\mathrm{h.c.}\Big]\,.
	\end{split}
\end{equation}
Applying the symmetries and redefining the parameters leads to the potentials of the N2HDM and 2HDMS. In both models, we retain $m_{12}$, while in the 2HDMS, we also keep $\mu_{12}$, 
thereby softly breaking the $\zet2$ symmetry. The potential is then given by,

\begin{equation}
\begin{split}
V_\text{N2HDM}&=m_{11}^2\Phi_1^\dagger\Phi_1+m_{22}^2\Phi_2^\dagger\Phi_2-(m_{12}^2\Phi_1^\dagger\Phi_2+\mathrm{h.c.})+\frac{\lambda_1}{2}(\Phi_1^\dagger\Phi_1)^2+\frac{\lambda_2}{2}(\Phi_2^\dagger\Phi_2)^2\\
&+\lambda_3(\Phi_1^\dagger\Phi_1)(\Phi_2^\dagger\Phi_2)+\lambda_4(\Phi_1^\dagger\Phi_2)(\Phi_2^\dagger\Phi_1)+\frac{\lambda_5}{2}[(\Phi_1^\dagger\Phi_2)^2+\mathrm{h.c.}]\\
&+\frac{1}{2}m_S^2 S^2+\frac{\lambda_6}{8}S^4+\frac{\lambda_7}{2}(\Phi_1^\dagger\Phi_1)S^2+\frac{\lambda_8}{2}(\Phi_2^\dagger\Phi_2)S^2.
\end{split}
\label{eq:n2hdmpot}
\end{equation}

\begin{equation}
	\begin{split}
		V_{\rm 2HDMS}&=m_{11}^2(\Phi_1^\dagger\Phi_1)+m_{22}^2(\Phi_2^\dagger\Phi_2)+\frac{\lambda_1}{2}(\Phi_1^\dagger\Phi_1)^2+\frac{\lambda_2}{2}(\Phi_2^\dagger\Phi_2)^2+\lambda_3(\Phi_1^\dagger\Phi_1)(\Phi_2^\dagger\Phi_2)\\
		&+\lambda_4(\Phi_1^\dagger\Phi_2)(\Phi_2^\dagger\Phi_1)+m_S^2( S^\dagger S)+\lambda'_1( S^\dagger S)(\Phi_1^\dagger\Phi_1)+\lambda'_2( S^\dagger S)(\Phi_2^\dagger\Phi_2)\\
		&+\frac{\lambda''_3}{4}( S^\dagger S)^2+\Big(-m_{12}^2\Phi_1^\dagger\Phi_2+\frac{\mu_{S1}}{6} S^3+\mu_{12} S\Phi_1^\dagger\Phi_2+\text{h.c.}\Big).
	\end{split}
	\label{eq:2hdmspot}
\end{equation}
Compared to the N2HDM, the 2HDMS includes cubic terms ${\propto}\mu_{S1}$ and ${\propto}\mu_{12}$ that arise from the $\zet3$ symmetry. These terms contribute additional corrections to the trilinear Higgs couplings, as will be discussed later in Section \ref{sec:diff}. From eq. (\ref{eq:vev_mass}), we define $\tan\beta=v_2/v_1$. Using the minimization conditions 
\begin{equation}
	\frac{\partial V}{\partial \Phi_1}\bigg|_{\substack{\Phi_1=v_1\\\Phi_2=v_2\\S=v_S}}=\frac{\partial V}{\partial \Phi_2}\bigg|_{\substack{\Phi_1=v_1\\\Phi_2=v_2\\S=v_S}}=\frac{\partial V}{\partial S}\bigg|_{\substack{\Phi_1=v_1\\\Phi_2=v_2\\S=v_S}}=0,
\end{equation}
one can replace $m_{11}^2$, $m_{22}^2$ and $m_{S}^2$ using the expressions from the tadpole equations. This results in the N2HDM having 11 free parameters
\begin{equation}
	\tan\beta,\;\lambda_1,\; \lambda_2,\; \lambda_3,\; \lambda_4,\; \lambda_5,\; \lambda_6,\; \lambda_7,\;\lambda_8\; m^2_{12},\; v_S ,
	\label{eq:lambdasinpn}
\end{equation}
and similarly, the 2HDMS has 12 free parameters 
\begin{equation}
	\tan\beta,\;\lambda_1,\; \lambda_2,\; \lambda_3,\; \lambda_4,\; \lambda'_1,\; \lambda'_2,\; \lambda''_3,\; m^2_{12},\; \mu_{S1},\; \mu_{12},\; v_S .
	\label{eq:lambdasinp2}
\end{equation}
\subsection{Masses}
Compared to the 2HDM, the addition of the singlet gives {in both models} rise to three $\mathcal{CP}$-even Higgs bosons, $h_1, h_2, h_3$. The $\mathcal{CP}$-odd sector of the N2HDM remains unchanged from the 2HDM, resulting in a single doublet-like pseudoscalar Higgs, $a$. In the 2HDMS, where the singlet also has an imaginary component, we obtain two $\mathcal{CP}$-odd Higgs bosons, $a_1$ and $a_2$. The charged sector retains the same structure as in the 2HDM in both models, leading to two charged Higgs bosons, $H^\pm$, with the same mass but opposite charges. Conventionally, we order the masses as $m_{h_1} \leq m_{h_2} \leq m_{h_3}$ and $m_{a_1} \leq m_{a_2}$. The mass eigenstates are obtained by diagonalizing the mass matrices $M_\rho^2$, $M_\eta^2$, and $M_\chi^2$ (see Appendix \ref{appendix:a}).
We introduce the three mixing angles $\alpha_{1,2,3}$, that diagonalize the $\mathcal{CP}$-even mass matrix $M_\rho^2$ and $\alpha_4$ that takes part in the diagonalization of $M_\eta^2$ in the 2HDMS, together with $\beta$ as defined above. 
{Here, the $\cp$-odd component contains the neutral goldstone state, whereas the charged component containts the two charged goldstones.} The diagonalization proceeds as follows,
\begin{align}
    RM_\rho^2R^T&=\mathrm{diag}(m^2_{h_1},m^2_{h_2},m^2_{h_3})\,,\\
    R_{A3}M_\eta^2R_{A3}^T&=\mathrm{diag}(0,m^2_{a_1},m^2_{{a}_2}), \quad {R_{A2}M_\eta^2R_{A2}^T \, = \, \mathrm{diag}(0,m^2_{A})}\,,\\
    R_cM_\chi^2R_c^T&=\mathrm{diag}(0,m^2_{H^\pm})\,,
\end{align}
with the rotation matrices
\begin{equation}
	R=\begin{pmatrix}
		c_{\alpha_1}c_{\alpha_2}& s_{\alpha_1}c_{\alpha_2}& s_{\alpha_2}\\
		-s_{\alpha_1}c_{\alpha_3}-c_{\alpha_1}s_{\alpha_2}s_{\alpha_3}& c_{\alpha_1}c_{\alpha_3}-s_{\alpha_1}s_{\alpha_2}s_{\alpha_3}& c_{\alpha_2}s_{\alpha_3}\\
		s_{\alpha_1}s_{\alpha_3}-c_{\alpha_1}s_{\alpha_2}c_{\alpha_3}& -s_{\alpha_1}s_{\alpha_2}c_{\alpha_3}-c_{\alpha_1}{{s_{a}}}_3& c_{\alpha_2}c_{\alpha_3}
	\end{pmatrix}~,
	\label{eq:rot}
\end{equation}
\begin{equation}
    R_A^{\rm N2HDM} (:= R_{A2}) =
    \begin{pmatrix}
        c_{\beta} & s_{\beta} \\
        -s_{\beta}  & c_{\beta}  
    \end{pmatrix}, 
\qquad
    R_A^{\rm 2HDMS} (:= R_{A3}) = 
    \begin{pmatrix}
        c_{\beta} & s_{\beta} &0 \\
        -s_{\beta} c_{\alpha_4} &  c_{\beta}  c_{\alpha_4} & s_{\alpha_4}\\
        s_{\beta} s_{\alpha_4} &-  c_{\beta} s_{\alpha_4} &   c_{\alpha_4}
    \end{pmatrix}, \label{eq:RA}
\end{equation}
\begin{equation}
        R_C=
    \begin{pmatrix}
        c_{\beta} & s_{\beta}  \\
        -s_{\beta}  & c_{\beta}  
    \end{pmatrix}. 
\end{equation}
Using the masses and the rotation matrices, we can express the free Lagrangian parameters (see eq. (\ref{eq:lambdasinpn}) and (\ref{eq:lambdasinp2})) in terms of the physical parameters. The new set of inputs in the N2HDM is then 
\begin{equation}
	\tan\beta,\quad\alpha_{1,2,3},\quad m_{h_1},\quad m_{h_2},\quad m_{h_3},\quad m_{a},\quad m^2_{12},\quad m_{H^\pm},\quad v_S~,
	\label{eq:inpmassn}
\end{equation}
and in the 2HDMS
\begin{equation}
	\tan\beta,\quad\alpha_{1,2,3,4},\quad m_{h_1},\quad m_{h_2},\quad m_{h_3},\quad m_{a_1},\quad m_{a_2},\quad m_{H^\pm},\quad v_S~.
	\label{eq:inpmassn2}
\end{equation}
This change of basis is described in App.~\ref{appendix:a}.
\subsection{Couplings}
\label{sec:mass-coup}
Next, we focus on the couplings of the mass eigenstates to gauge bosons and fermions. The relation between the mass basis and the interaction basis is given by 
\begin{align}
    \text{Both:\hspace{1cm}}\dv{h_1}{h_2}{h_3}&=R\dv{\rho_1}{\rho_2}{\rho_S},\label{eq:rot_epeven}\\
    \text{2HDMS:\hspace{1cm}}\dv{\xi_a}{a_1}{a_2}&=R_A^{\mathrm{2HDMS}}\dv{\eta_1}{\eta_2}{\eta_S},
    \\
    \text{N2HDM:\hspace{1cm}}\zv{\xi_a}{A}&=R^{\mathrm{N2HDM}}_A\zv{\eta_1}{\eta_2},
\end{align}
where the $\xi_a$ denotes the massless Goldstone boson.
We define the effective coupling as the ratio between the model coupling and the SM couplings, thus
\begin{equation}
    c_{h_ipp}=\frac{g_{h_ipp}}{g^{\rm SM}_{Hpp}}.
\end{equation}
The coupling to gauge boson can be derived as
\begin{equation}
    c_{h_iVV}=c_{h_iZZ}=c_{h_iWW}=\cos\beta R_{i1}+\sin\beta R_{i2}. \label{eq:chivv}
\end{equation}
The couplings to fermions follow from the Yukawa Lagrangian. By extending the $\zet2$ symmetry to the Yukawa sector, we ensure that each type of fermions couples to only one doublet. As in the 2HDM one has four Yukawa types. The couplings to fermions for all Yukawa types are shown in \refta{tab:fermioncoup}.
\begin{table}[htb]
	\centering
\renewcommand{\arraystretch}{1.6}
	\begin{tabular}{l|c|c|c|c}
		\hline
		&Type I& Type II& Lepton specific& Flipped\\
		\hline
		$c_{h_i tt}$& $\frac{R_{i2}}{\sin\beta}$& $\frac{R_{i2}}{\sin\beta}$& $\frac{R_{i2}}{\sin\beta}$& $\frac{R_{i2}}{\sin\beta}$\\
		$c_{h_i bb}$& $\frac{R_{i2}}{\sin\beta}$& $\frac{R_{i1}}{\cos\beta}$& $\frac{R_{i2}}{\sin\beta}$& $\frac{R_{i1}}{\cos\beta}$\\$c_{h_i \tau\tau}$& $\frac{R_{i2}}{\sin\beta}$& $\frac{R_{i1}}{\cos\beta}$& $\frac{R_{i1}}{\cos\beta}$& $\frac{R_{i2}}{\sin\beta}$\\
		\hline
	\end{tabular}
	\caption{Higgs-to-fermion effective couplings for different types of Yukawa interactions.}
	\label{tab:fermioncoup}
\renewcommand{\arraystretch}{1.2}
\end{table}


\subsection{Type II SM-like Higgs Boson}

While the expressions derived above are valid for all four Yukawa types, we now focus on the Yukawa type~II, which will
be main focus of our analysis, see the discussion in \refse{sec:intro}.
In our analysis, we interpret the second-lightest Higgs boson as the SM-like Higgs boson. To ensure that $h_2$ has the properties of the SM-Higgs, we define the alignment limit \cite{Gunion_2003}. 
From \refeq{eq:chivv} and \refta{tab:fermioncoup}, the Type~II couplings of the $h_2$ are given by,
\begin{align}
c_{h_2VV}&=c_{\alpha_3}s_{\beta-\alpha_1}-s_{\alpha_2}s_{\alpha_3}c_{\beta-\alpha_1},\\
c_{h_2tt}&=(c_{\alpha_1}c_{\alpha_3}-s_{\alpha_1}s_{\alpha_2}s_{\alpha_3})/s_\beta,\\	c_{h_2bb}&=(-s_{\alpha_1}c_{\alpha_3}-c_{\alpha_1}s_{\alpha_2}s_{\alpha_3})/c_\beta.	
\end{align}
In the limit of a singlet like lightest $\mathcal{CP}-$even Higgs boson $h_1$ (see \refeq{eq:rot}), thus $\sin^2\alpha_2\rightarrow 1$, the $h_2$ couplings can be simplified further as,
\begin{align}
c_{h_2VV}&\approx \sin(\beta- (\alpha_1+\sgn(\alpha_2)\alpha_3))|\sin{\alpha_2}|,\label{h2vvcouplings}\\
c_{h_2tt}&\approx \frac{\cos(\alpha_1+\sgn(\alpha_2)\alpha_3)}{\sin\beta}|\sin\alpha_2|,\label{h2ttcouplings}\\
c_{h_2bb}&\approx-\frac{\sin(\alpha_1+\sgn(\alpha_2)\alpha_3)}{\cos\beta}|\sin\alpha_2|.\label{h2bbcouplings}
\end{align}
Following \refeq{h2vvcouplings}, the alignment limit is reached by,
\begin{equation}
    \be - (\al_1 + \sgn(\al_2)\al_3) \to \pi/2\,,
\end{equation}
and the coupling modifiers of $h_2$ to fermions and gauge bosons become $1$ simultaneously.


\subsection{Type II 95 GeV Excesses}
\label{sec:Type2excess}

In \citere{Heinemeyer:2021msz} the possibility to describe the 95~GeV excesses in BSM models with two Higgs doublets
plus one Higgs singlet was analyzed (see also the discussion in \refse{sec:intro}). It was shown that in the N2HDM as well as 
in the 2HDMS the excesses can be well described with a singlet dominated $h_1$. Here we follow our earlier analysis to 
eventually extend it to an analysis of THCs.
In order to yield such a description of the 95~GeV excesses, we move away from the exact alignment limit. Further, we use the couplings of $h_1$, that govern the signal strengths and the alignment limit, as input variables and compute the mixing angles from that.
As inputs we choose $c_{h_1VV}\propto \mu^{bb}$, $\frac{c_{h_1bb}}{c_{h_1tt}}\propto (\mu^{\gamma\gamma})^{-1}$ and $\varepsilon$ that defines the offset to the alignment limit, hence
\begin{equation}
    \be - (\al_1 + \sgn(\al_2)\al_3) \, = \, \pi/2-\varepsilon.
\end{equation}
The mixing angles $\alpha_1,\alpha_2,\alpha_3$ can then be evaluated using
\begin{align}
    \alpha_1&=\arctan\rbr{\frac{\tan\beta}{c_{h_1bb}/c_{h_1tt}}} \label{eq:alpha1},\\
    \alpha_2&=\arccos{\rbr{ \frac{c_{h_1VV}}{\cos{\beta}\cos\alpha_1+\sin\beta\sin\alpha_1  }}} \label{eq:alpha2},\\
    \alpha_3&=\sgn(\alpha_2)\rbr{ \beta-\alpha_1-\frac{\pi}{2}+\varepsilon} \label{eq:alpha3}.
\end{align}
A parameter point is considered as allowed when the theory prediction agrees with the experimental signal strengths at the 
$1\,\sigma$ level (see \refse{sec:95ex}).

\subsection{Trilinear Higgs Couplings in the N2HDM and the 2HDMS}
 
\begin{figure}[h]
        \centering
        \begin{tikzpicture}
            \begin{feynman}[small]
                \vertex (l1);
                \vertex[left =of l1] (i3) {\(h_i\)};
                \vertex[above right =of l1] (i4) {\(h_j\)};
                \vertex[below right =of l1] (i5) {\(h_k\)};
                \diagram* { (i3)  --[scalar] (l1)
                          --[scalar] (i4),
                          (l1) --[scalar] (i5)     
                };
            \end{feynman}
            \end{tikzpicture}
        \caption{General trilinear Higgs self-coupling vertex $\propto\lambda_{h_ih_jh_k}^{\rm model}$.}
        \label{fig:THC_general}
 \end{figure}
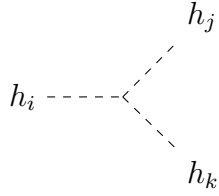

In this subsection we derive the relevant expressions for the involved THCs.
In models with an extended Higgs sector, these expressions can become more complicated, especially since multiple types of Higgs bosons can participate in such interactions. For our analysis, we focus on the $\mathcal{CP}$-even Higgs bosons (see \reffi{fig:THC_general}, here and in the following the superscript
(or subscript) \mbox{}$^{\mathrm{model}}$ refers to either the N2HDM, or the 2HDMS).
 We derive the THC by expanding the Higgs potential of the N2HDM (see eq. \refeq{eq:n2hdmpot}) or 2HDMS (see \refeq{eq:2hdmspot}) using the field rotations in \refeq{eq:rot_epeven} and collecting all terms $\propto h_ih_jh_k$,
 \begin{equation}
    \lambda_{h_ih_jh_k}^{\rm model}=\frac{\partial^3 V^{\rm model}}{\partial h_i\partial h_j\partial h_k}\Bigg|_{\substack{h_i=0\\h_k=0\\h_j=0}}.
\end{equation}
Furthermore, we express the THC in terms of the physical input parameters using the basis change as given in equations (\ref{eq:lambda1_n2hdm})-(\ref{eq:lambda8_n2hdm}) or (\ref{eq:mu12-a4})-(\ref{eq:lambdapp3_2hdms}). Using the orthogonality relation of the mixing matrix
\begin{equation}
    R R^T=\mathds{1}\quad \Rightarrow \quad \sum_{k} R_{ik}R_{kj}^T=\sum_{k} R_{ik}R_{jk}=\delta_{ij},
\end{equation} 
yields 
\begin{align}
    \begin{split}
        \lambda_{h_i h_j h_k}^{\rm N2HDM}&=
       \rbr{ m_i^2+m_j^2+m_k^2}\Bigg [ \frac{R_{i1}R_{j1}R_{k1}}{v\cos \beta}+\frac{R_{i2} R_{j2}R_{k2}}{v\sin\beta}+\frac{R_{i3} R_{j3}R_{k3} } {v_S}\Bigg ]\\
    &+ \frac{\hat \mu^2}{v}  \Bigg[-  \frac{3R_{i1} R_{j1} R_{k1} \sin^2{\beta }}{  \cos{\beta}} - \frac{3 R_{i2} R_{j2}R_{k2} \cos^2{\beta }}{\sin{\beta}}\\
    &+\cos\beta (R_{i1} R_{j2} R_{k2}+R_{i2}R_{j1}R_{k2}+R_{i2} R_{j2} R_{k1})\\
    &+\sin\beta (R_{i1} R_{j1} R_{k2}+R_{i1} R_{j2} R_{k1}+ R_{i2} R_{j1} R_{k1} ) \Bigg],
    \end{split}\label{eq:thc_n2hdm}
\end{align}
and
\begin{align}
    \begin{split}
        \lambda_{h_i h_j h_k}^{\rm 2HDMS}&= 
        \lambda_{h_i h_j h_k}^{\rm N2HDM}(\hat\mu \to \tilde\mu) + \Delta\lambda_{h_i h_j h_k}\\[.3em]
        &= \lambda_{h_i h_j h_k}^{\rm N2HDM}(\hat\mu \to \tilde\mu) + \\
        &+\rbr{\frac{ m_{a_2}^2-m_{a_1}^2}{v}  \cos \alpha_4\sin\alpha_4} \Bigg[ \frac{ 5 R_{i3} R_{j3} R_{k3}   v^{2} \sin{ 2\beta } }{3v_{S}^2} \\
    & - \tan\beta(R_{i1} R_{j1} R_{k3}+R_{i1}R_{j3}R_{k1}+R_{i3} R_{j1} R_{k1})\\
    &-\cot\beta(R_{i2} R_{j2} R_{k3}+R_{i2}R_{j3}R_{k2}+R_{i3} R_{j2} R_{k2})\\
    &-\frac{v\sin\beta}{v_S} (R_{i1} R_{j3} R_{k3}+R_{i3}R_{j1}R_{k3}+R_{i3} R_{j3} R_{k1}) \\
    &-\frac{v\cos\beta}{v_S}(R_{i2} R_{j3} R_{k3}+R_{i3}R_{j2}R_{k3}+R_{i3} R_{j3} R_{k2})\\
    &+ R_{i3}R_{j2}R_{k1}+R_{i2}R_{j3}R_{k1}+R_{i3}R_{j1}R_{k2}\\&+R_{i1}R_{j2}R_{k3}+R_{i1}R_{j3}R_{k2}+R_{i2}R_{j1}R_{k3}\Bigg]\\
    &+\frac{R_{i3} R_{j3}R_{k3} }{3 v_S}  \rbr{m^2_{a_1} \sin^2\alpha_4+m^2_{a_2} \cos^2\alpha_4}. \label{eq:thc_2hdms}
\end{split} 
\end{align}


\subsection{Differences Between the N2HDM and the 2HDMS}
\label{sec:diff}

As discussed above, N2HDM and 2HDMS have similar particle content. However, the imaginary component to the singlet leads to an additional $\mathcal{CP}$-odd Higgs boson in the case of the 2HDMS. The contributions of the doublet and singlet components to the pseudoscalar Higgses are controlled by the mixing angle $\alpha_4$. Taking into account the rotation matrix (see eq. (\ref{eq:RA})) we find that for $\alpha_4<\frac{\pi}{4}$ the $a_2$ is dominated by the singlet, whereas for $\alpha_4>\frac{\pi}{4}$ the $a_1$ is dominated by the singlet contribution. In the limit of $\alpha_4=0$ or $\frac{\pi}{2}$, $a_2$ or $a_1$, is fully singlet like thus decouples from all fermions, with a zero coupling to gauge bosons, becoming unobservable in experiments. This limit is refered to as the N2HDM limit of the 2HDMS, since close to this limit the observable particle content of both models appears to be the same. Furthermore, the various Higgs-boson couplings to SM particles are also identical in this limit. However, the 2HDMS potential features additional cubic terms in the potential $\mu_{12}$ and $\mu_{S1}$. These terms contribute to the trilinear Higgs couplings. As shown in \refeqs{eq:thc_2hdms} and (\ref{eq:thc_n2hdm}) the THCs in the 2HDMS can be written as, 
\begin{equation}
    \lambda_{h_1h_jh_k}^{\rm 2HDMS}=\lambda_{h_1h_jh_k}^{\rm N2HDM}+\Delta\lambda_{h_1h_jh_k}.
\end{equation}
However, we must keep in mind that $\hat\mu\neq\tilde\mu$. In order to gain an understanding of the size of the involved terms, we study $\Delta\lambda_{h_1h_jh_k}$ in a frame near the alignment limit while embedding the excesses. To do so, we take the inputs as discussed in section \ref{sec:Type2excess} and fixing $c_{h_1bb}/c_{h_1tt}=1$. Further we assume that $c_{h_3VV}$ is negligibly small so $c_{h_2VV}^2=1-c_{h_1VV}^2$. Lastly we drop all terms of order $\mathcal{O}(\varepsilon^2)$. This gives
\begin{align}
    \cos\alpha_1=\cos\beta, &\qquad \sin\alpha_1=\sin\beta,\\
    \cos\alpha_2=c_{h_1VV}, &\qquad \sin\alpha_2=c_{h_2VV},\\
    \cos\alpha_3=\varepsilon+\mathcal{O}(\varepsilon^2), &\qquad \sin\alpha_3=-1+\mathcal{O}(\varepsilon^2).
\end{align}
The $\mathcal{CP}$-even mixing matrix then becomes
\begin{equation}
    R_{ij}\approx\begin{pmatrix}
        c_{h_1VV} \cos\beta  &  c_{h_1VV} \sin\beta &c_{h_2VV}\\
         c_{h_2VV} \cos\beta -\varepsilon \sin\beta & c_{h_2VV} \sin\beta +\varepsilon \cos\beta & -c_{h_1VV}\\
       - c_{h_2VV}\varepsilon  \cos\beta -\sin\beta & -c_{h_2VV}\varepsilon  \sin\beta +\cos\beta& c_{h_1VV}\varepsilon
     \end{pmatrix}+\mathcal{O}(\varepsilon^2) \label{eq:R_simpls}.
\end{equation}
In this frame, the terms $\propto \hat\mu^2$ or $\propto \tilde\mu^2$  as well as many other mixing terms vanish, yielding
\begin{align}
    \begin{split}
        \Delta\lambda_{h_i h_j h_k}&=\frac{1}{3\,v_S}\Bigg\{\rbr{\frac{ m_{a_2}^2-m_{a_1}^2}{v}  \cos \alpha_4\sin\alpha_4} \Bigg[ \frac{ 5 R_{i3} R_{j3} R_{k3}   v^{2} \sin{ 2\beta } }{v_{S}} \\
    &-3\,v\,\sin\beta (R_{i1} R_{j3} R_{k3}+R_{i3}R_{j1}R_{k3}+R_{i3} R_{j3} R_{k1}) \\
    &-3\,v\,\cos\beta (R_{i2} R_{j3} R_{k3}+R_{i3}R_{j2}R_{k3}+R_{i3} R_{j3} R_{k2})\Bigg]\\
    &+(R_{i3} R_{j3}R_{k3}) \rbr{m^2_{a_1} \sin^2\alpha_4+m^2_{a_2} \cos^2\alpha_4} \Bigg\}, \label{eq:thc_diff}
\end{split} 
\end{align}
 for the additional terms in the 2HDMS. First, we find that all terms are inversely proportional to the singlet VEV $v_S$, so we expect large differences for low $v_S$. Next there are two different types of terms in $\Delta\lambda_{h_i h_j h_k}$. Terms $\propto \cos\alpha_4\sin\alpha_4$ vanish in the exact N2HDM limit of the 2HDMS. Terms $\propto m_{a_1}^2\sin\alpha_4^2+m_{a_2}^2\cos\alpha_4^2$ do not vanish in this limit but are proportional to the mass of the singlet-like pseudoscalar Higgs boson. 
 Analyzing the analytical structure of the di-Higgs production amplitude at $e^+e^-$ colliders,
 this contribution cancels out in the process of Higgs strahlung.
 The matrix element takes a form of 
 \begin{equation}
\begin{tikzpicture}
        \draw (0,0) node[anchor=center]{}; 
        \draw (0,0.8) node[anchor=center]{\(\mathcal{M}=\dots+\)};
\end{tikzpicture}
    \scalebox{0.6}{
    \begin{tikzpicture}   
    \begin{feynman}[small]
               \vertex (l1) {\(Z\)};
        \vertex[right = 3.5em of l1] (i3);
        \vertex[above right =of i3] (i4);
        \vertex[above right = of  i4] (i8) {\(h_i\)};
        \vertex[below right =of i3] (i5) {\(Z\)};
        \vertex[above right =of i5] (i6) {\(h_j\)};
        \diagram* {(l1)   --[boson,] (i3)
                  --[scalar] (i4) --[scalar] (i8),
                  (i3) --[boson] (i5), 
                  (i4) --[scalar,edge label'=\(h_1\)] (i6)
        
        };
    \end{feynman}
    \end{tikzpicture}}
    \begin{tikzpicture}
        \draw (0,0) node[anchor=center]{}; 
        \draw (0,0.8) node[anchor=center]{\(+\)};
\end{tikzpicture}
  \scalebox{0.6}{
  \begin{tikzpicture}
    \begin{feynman}[small]
               \vertex (l1) {\(Z\)};
        \vertex[right = 3.5em of l1] (i3);
        \vertex[above right =of i3] (i4);
        \vertex[above right = of  i4] (i8) {\(h_i\)};
        \vertex[below right =of i3] (i5) {\(Z\)};
        \vertex[above right =of i5] (i6) {\(h_j\)};
        \diagram* {(l1)   --[boson,] (i3)
                  --[scalar] (i4) --[scalar] (i8),
                  (i3) --[boson] (i5), 
                  (i4) --[scalar,edge label'=\(h_2\)] (i6)
        
        };
    \end{feynman}
    \end{tikzpicture}}
    \begin{tikzpicture}
        \draw (0,0) node[anchor=center]{}; 
        \draw (0,0.8) node[anchor=center]{\(+\dots\)};
\end{tikzpicture}
\end{equation}
The mixing part is given by $R_{i3} R_{j3}R_{k3}$.
Without loss of generality, the index $k$ indicates the Higgs boson in the propagator. We can then write the matrix element as
\begin{equation}
    \mathcal{M}_{Zh_ih_j}\propto\dots+ c_{h_1VV}\lambda_{h_1h_ih_j}\mathcal{D}_{95}+c_{h_2VV}\lambda_{h_2h_ih_j}\mathcal{D}_{125},
\end{equation}
where $\mathcal{D}$ is the scalar propagator with the mass of the respective Higgs. Assuming that $\mathcal{D}_{95}\approx\mathcal{D}_{125}$ and inserting $R_i$ from matrix (\ref{eq:R_simpls}), we obtain 
\begin{equation}
    \frac{m_{a_1}^2\sin^2\alpha_4+m_{a_2}^2\cos^2\alpha_4}{3v_S}\bigg[c_{h_1VV}c_{h_2VV}R_{j3}R_{k3}-c_{h_2VV}c_{h_1VV}R_{j3}R_{k3}
    \bigg]=0.
    \label{eq:DtermDiff}
\end{equation}
As a consequence, the tree-level contribution of this term cancels and differences vanish in the exact N2HDM limit. A similar argument can be made for the $c_{h_iff}$ coupling, yielding approximately equal rates for $gg\rightarrow h_2h_2$ in both models.
However, the differences from approximations made and the remaining terms can become sizeable moving slightly away from this limit. 
These differences scale with the mass difference of the pseudoscalar Higgses and inversely with $v_S$. 
Furthermore, the mixing part of \refeq{eq:thc_diff} is governed by the third column in mixing matrix (see \refeq{eq:R_simpls}). Consequently $h_1$ in the vertex introduces factors of $c_{h_2VV}$ to the $R_{i3}$ terms, whereas the $h_2$ in the vertex leads to factors of $-c_{h_1VV}$. In summary, differences will appear larger the more $h_1$ contributes to the interaction.

%% file: section2.8.tex
\subsection{The Constraints}
\label{sec:constraints}

The parameters in the N2HDM and 2HDMS must satisfy theoretical and experimental constraints. 
In this subsection we briefly summarize the applied constraints.


\subsubsection{Theoretical Constraints}
\label{sec:theo}

\begin{itemize}

\item \textbf{Tree-level perturbative unitarity}\newline
Tree-level perturbative unitarity ensures perturbativity of the model up to very high energy scales, thus controlling the growth of the scattering amplitude. 
$\mathcal{M}$ is the four-point scattering amplitude of all quartic interactions in the interaction basis. A parameter point is considered as unitary if all eigenvalues of $\mathcal{M}$ are smaller than $8\pi$. The expressions for the constraints are derived in \citere{Muhlleitner:2016mzt} for the N2HDM and in \cite{Heinemeyer:2021msz} for the 2HDMS.

\item \textbf{Boundedness from below}\newline
The Higgs potential must have a global minimum to yield a stable solution. 
Boundedness from below (BFB) gives us conditions that guarantee the existence of a global minimum.
In rather simple models with a small number of parameters, such conditions can be trivial, however in the more advanced models like the 2HDM, N2HDM and 2HDMS, the potentials can have much more complicated structures. To obtain the BFB conditions, one only considers the quartic terms, since they dominate the potential for large $\phi_{1,2},S$.
The expressions for the BFB constraints are derived in \citere{Muhlleitner:2016mzt} for the N2HDM and in \cite{Li:2023hsr}for the 2HDMS.

\item \textbf{Vacuum stability}
The SM requires a non-zero VEV, i.e. vacuum, that is stable at the electroweak scale. In BSM models with more complicated potential structures, vacuum stability gives rise to additional constraints on the allowed parameter space. If the vacuum is a global minimum (true vacuum) of the potential, it is called absolutely stable. If the vacuum is only a local minimum (false vacuum) it could tunnel to a lower minimum. 
A parameter point is considered as sufficiently long-lived, so-called metastable, if the lifetime of the electroweak vacuum surpasses the current age of the universe. Anything else is considered unstable. For our study, the public code \verb|EVADE| \cite{Hollik:2019,Ferreira:2019,evade:online} is used. For a given parameter point, it finds the tree-level minima and computes the bounce action $B$ in the case of a false vacuum. We consider a point sufficiently long-lived, hence stable in the sense of its lifetime, if $B>440$~\cite{Hollik:2019}.

\end{itemize}


\subsubsection{Experimental Constraints}
\label{sec:expconstrain}


\begin{itemize}

\item \textbf{Higgs boson rate measurements}

Since the discovery of the Higgs Boson in 2012 \cite{ATLAS:2012tfa,Chatrchyan:2012xdj}, the properties of this new particle have been studied extensively and compared to the SM predictions. So far, no significant deviations of the Higgs boson from the SM have been observed. This narrows the available parameter space for BSM models to provide a scalar particle with a mass of \SI{125}{GeV}, that agrees with the properties of the observed Higgs. In general, this can be achieved in both N2HDM and 2HDMS by going into the alignment limit. However, to retain the BSM properties of the models, one cannot be in the exact alignment limit. Hence, parameters must be chosen carefully in order not to exceed the experimental bounds of the SM Higgs measurements. To test a given parameter on its agreement with the SM, we use the public code \verb|HiggsSignals|\footnote{
\texttt{HiggsSignals\,v2.6.2}, \texttt{HiggsSignals} dataset \texttt{v1.1}.} 
~\cite{Bechtle:2013xfa,stal:2013HS,bechtle:2021HS,Bechtle:2014ewa} which is part of \verb|HiggsTools| \cite{Bahl_2023,HT_2019}. It calculates a $\chi^2$ value based on the agreement of the model prediction with the experimental data including Higgs boson mass measurements, and signal rates from ATLAS and CMS at the LHC run I \cite{PhysRevLett.114.191803,Khachatryan:2016vau} as well as newer results from LHC run II
\cite{CERN_YELLOW_2017}. \verb|HiggsSignals| is calculating a $\chi^2_{\rm HS}$ based on $n_{\rm obs}=159$ observables. To judge if a parameter point agrees with SM observables, we compare the calculated $\chi^2_{\rm HS}$ to the SM prediction $\chi^2_{\rm HS,SM}$
\begin{equation}
    \Delta \chi^2=\chi^2_{\rm HS}-\chi^2_{\rm HS,SM}.
\end{equation}
A parameter point is considered as allowed within a \SI{95}{\%} confidence level of the SM prediction when 
\begin{equation}
    \Delta \chi^2<5.99.
\end{equation}
$\chi^2_{\rm HS}$ can be obtained employing the implemented reference model with SM-like couplings which gives
\begin{equation}
    \chi^2_{\rm HS,SM}=152.54.
\end{equation}
{Our study uses the effective couplings input for \verb|HiggsTools|, which calculates the production cross sections and branching ratios of 125~GeV Higgs. These branching ratios are then used to obtain the model predicted $\chi^2_{\rm HS}$ in \verb|HiggsSignals|.}



\item \textbf{BSM Higgs boson searches}

Until now, the searches for BSM Higgs bosons did not yield a discovery of
any additional Higgs bosons, besides the one at $125 \gev$. Nonetheless, searches for the BSM particles, provide exclusion limits on the BSM parameter space. For a given parameter point, it is necessary to check if this point is not already excluded by any of the available searches. To simplify this task we use the code \verb|HiggsBounds|\footnote{
\texttt{HiggsBounds\,v5.10.2}, \texttt{HiggsBounds} dataset \texttt{v1.6}.}
~\cite{Bechtle_2010,Bechtle_2011,bechtle2013recent,Bechtle_2014_3,Bechtle_2015,Bechtle_2020,Bahl_2022} which is also part of \verb|HiggsTools| \cite{Bahl_2023,HT_2019}. The tool contains more than 250 limits at a \SI{95}{\%} confidence level from LEP and LHC. The code calculates the expected $r_{\rm exp}$ and the observed $r_{\rm obs}$ ratios using the narrows width approximation from supplied production cross-sections and branching ratios. The ratios are defined as:
\begin{align}
    r_{\rm exp}&=\frac{(\sigma\cdot \mathcal{BR})_{\rm pred.}}{({\sigma}\cdot {\mathcal{BR}})_{\rm exp}},\\
    r_{\rm obs}&=\frac{(\sigma\cdot \mathcal{BR})_{\rm pred.}}{(\sigma\cdot \mathcal{BR})_{\rm \SI{95}{\%}\:C.L. limit}},
\end{align}
where $(\sigma\cdot \mathcal{BR})_{\rm pred.}$ are the cross-sections, which are implied by a given parameter point for the implemented limit channel. 
The expected ratio divides the model prediction by the expected limit for such a process in absence of BSM physics. 
The observed ratio is obtained by dividing this value by the \SI{95}{\%} C.L. limit on the rate measurement. 
To maintain a \SI{95}{\%} C.L.\ on the exclusion,  one cannot test a point for all available limits. \verb|HiggsBounds| selects the channel with the highest sensitivity based on the largest $r_{\rm exp}$. If the corresponding  $r_{\rm obs}>1$, the point is considered as already excluded by the BSM search. For the \verb|HiggsBounds| calculation, we feed the Higgs production cross-sections and explicit branching ratios generated by \verb|SPheno| to \verb|HiggsTools| for all the Higgs bosons. The BSM contributions of the decay widths from the decay to additional Higgs bosons
are taken into account in this setup.




\item \textbf{Flavor physics observables}

In Two Higgs Doublet Models, measurements of flavor physics observables give rise in particular to constraints on the charged Higgs boson. Since the charged sector in the N2HDM and 2HDMS remains basically unchanged from the 2HDM, we can refer to previously performed studies in flavor physics in the 2HDM \cite{Haller_2018}. There, the mass of the charged Higgs $m_{H^\pm}$ is constrained for different $\tan\beta$ regions. In the 2HDM type II for ${\tan\beta \in(1,15)}$, we consider a point as allowed within a \SI{95}{\%} C.L. when $m_{H^\pm}>\SI{600}{GeV}$, which is just a loose constraint.



\item \textbf{Electroweak precision observables}

BSM physics can enter into EW precision observables, like the $W$ mass, via contributions at the one-loop level. For the study of such corrections the oblique parameters $S$, $T$ and $U$ were introduced in \citere{PhysRevD.46.381}.
The experimental reference values for these parameters can be extracted with an electroweak fit on the data \cite{Haller_2018}. This yields
\begin{equation}
    \begin{matrix}
        S=0.04\pm0.11, & T=0.09\pm 0.14,& U=-0.02\pm0.11,\\
        \text{corr}(S,T)=+0.92, & \text{corr}(S,U)=-0.68, &\text{corr}(T,U)=-0.87. \label{eq:STU_res}
    \end{matrix}
\end{equation}
To evaluate, if the BSM contributions of a parameter point exceed the fit uncertainties, one computes the model predictions $S^\prime,T^\prime,U^\prime$ for this point. The calculation follows the results for a general multi Higgs doublet models with an arbitrary number of singlets (mHDSM) \cite{Grimus_2008}. The details can be found in App.~\ref{sec:A_STU}.
We determine a $\chi^2$ value by
\begin{equation}
    \chi^2=\begin{pmatrix} S^\prime-S & T^\prime-T & U^\prime-U \end{pmatrix}   \cdot  \textbf{cov}^{-1} \cdot \begin{pmatrix} S^\prime-S \\ T^\prime-T \\ U^\prime-U \end{pmatrix},
\end{equation}
with 
\begin{equation}
    \textbf{cov}=\begin{pmatrix}
        \sigma_{S}^2 & \text{corr}(S,T) \sigma_S\sigma_T &\text{corr}(S,U) \sigma_S\sigma_U\\
        \text{corr}(S,T) \sigma_S\sigma_T  & \sigma_{T}^2 &\text{corr}(T,U) \sigma_T\sigma_U\\
        \text{corr}(S,U) \sigma_S\sigma_U  & \text{corr}(T,U)\sigma_{T}\sigma_{U}  &\sigma_U^2
    \end{pmatrix},
\end{equation}
where $\sigma_{S,T,U}$ denotes the uncertainty of $S,T,U$.
We consider a parameter point as allowed, thus within a \SI{95}{\%} C.L interval from the observed values, when  $\chi^2<7.81$, corresponding to $2\,\sigma$ for 3 dof.

\subsubsection{Embedding the 95\,GeV excesses}
\label{sec:95ex}

In the N2HDM and 2HDMS, we demand the lightest $\mathcal{CP}$-even Higgs $h_1$ to fit the \SI{95}{GeV} excesses from CMS, ATLAS and LEP. To compute the signal strengths, we are using the narrow width approximation, 
\begin{align}
    \begin{split}
        \mu_{\rm pred}^{\gamma\gamma}&=\frac{\sigma_{\rm model}(gg\rightarrow h_1)}{\sigma_{\rm SM}(gg\rightarrow H^0_{\rm SM})}\frac{\mathcal{BR}(h_1 \rightarrow \gamma\gamma)}{\mathcal{BR}(H_{\rm SM}^0 \rightarrow \gamma\gamma)}\\&=\abs{c_{h_1tt}}^2 \frac{\mathcal{BR}(h_1 \rightarrow \gamma\gamma)}{\mathcal{BR}(H_{\rm SM}^0 \rightarrow \gamma\gamma)},\\
    \end{split}
\end{align}
\begin{align}
    \begin{split}
    \mu_{\rm pred}^{b\bar b}&=\frac{\sigma_{\rm model}(e^+e^-\rightarrow Z h_1)}{\sigma_{\rm SM}(e^+e^-\rightarrow Z H^0_{\rm SM})}\frac{\mathcal{BR}(h_1 \rightarrow b \bar b)}{\mathcal{BR}(H_{\rm SM}^0 \rightarrow b \bar b)}\\&=\abs{c_{h_1VV}}^2 \frac{\mathcal{BR}(h_1 \rightarrow b \bar b)}{\mathcal{BR}(H_{\rm SM}^0 \rightarrow b \bar b)},
\end{split}
\end{align}
where $c_{h_1VV}$ and $c_{h_1tt}$ are the effective couplings and $\mathcal{BR}(h_1\rightarrow XX)$   the parameter dependent branching ratios, that are computed by \verb|SPheno-4.0.5| \cite{Porod_2003,Porod_2012}. The mass of $H_{\rm SM}^0$ has to
be set to the same value as $m_{h_1}$. The rescaled branching ratios can be taken from \cite{CERN_YELLOW_2017}. With $m_{H^0_{\rm SM}}=\SI{95}{GeV}$, we obtain
\begin{align}
    \mathcal{BR}(H_{\rm SM}^0 \rightarrow \gamma\gamma)&=1.39\cdot10^{-3},\\
    \mathcal{BR}(H_{\rm SM}^0 \rightarrow b\bar b)&=8.01\cdot10^{-1}.
\end{align}
Neglecting correlations, we define a $\chi_{95}^2$ as 
\begin{equation}
    \chi_{95}^2=\rbr{\frac{\mu_{\rm pred}^{\gamma\gamma}- \mu_{\rm CMS}^{\gamma\gamma}}{{\Delta\mu}_{\rm CMS}^{\gamma\gamma}}}^2+\rbr{\frac{\mu_{\rm pred}^{\gamma\gamma}- \mu_{\rm ATLAS}^{\gamma\gamma}}{{\Delta\mu}_{\rm ATLAS}^{\gamma\gamma}}}^2+\rbr{\frac{\mu_{\rm pred}^{b\bar b}- \mu_{\rm LEP}^{b\bar b}}{{\Delta\mu}_{\rm LEP}^{b\bar b}}}^2.
\end{equation}
Note, that the unbalanced uncertainty  of CMS measurement needs to be taken into account, so that ${\Delta\mu}_{\rm CMS}^{\gamma\gamma}=-0.12$ if $\mu_{\rm pred}^{\gamma\gamma}\leq\mu_{\rm CMS}^{\gamma\gamma}$ and  ${\Delta\mu}_{\rm CMS}^{\gamma\gamma}=0.19$ else. We demand a point to agree with the observed excesses within a $1 \sigma$ interval, hence
\begin{equation}
    \chi_{95}^2\leq3.523.
\end{equation} 
\end{itemize}


\subsection{Benchmark Scenario and ``\boldmath{$\al_4$} limits''}
\label{sec:benchmarkscenario}

In this section, we define our benchmark scenario for the phenomenological study. \refta{tab:BP3} lists the parameters of the N2HDM. 
\newcolumntype{x}[1]{>{\centering\arraybackslash\hspace{0pt}}p{#1}}
\begin{table}[h]
    \centering
    \renewcommand*{\arraystretch}{1.5}
    \begin{tabular}{x{2.cm}|x{2.cm}|x{2.cm}|x{2.1cm}|x{2.cm}|x{2.cm}|x{2.cm}}
     $m_{h_1}$ [GeV]& $m_{h_2}$  [GeV]&$m_{h_3}$ [GeV]&$m_{H^\pm}$ [GeV]&$m_{A}$ [GeV]&$\hat{\mu}$ [GeV]& $v_S$ [GeV]\\\hline
     95.4& 125.09& 650&650&600&600&120\\
    \end{tabular}\\\vspace{0.5cm}
    \begin{tabular}{x{2.cm}|x{2.cm}|x{2.cm}|x{2.05cm}||x{1.97cm}|x{2.cm}|x{2.cm}}
     $\tan\beta$ &  $c_{h_1VV}$&$\frac{c_{h_1bb}}{c_{h_1tt}}$\vspace{0.05cm}&$\varepsilon$&$\alpha_1$&$\alpha_2$&$\alpha_3$\\\hline
    1.5&0.35&0.8&0.05&1.0808&1.2114&-1.6188\\
    \end{tabular}
    \caption{Benchmark point parameters. The values for the $\mathcal{CP}$-even mixing angles $\alpha_1,\alpha_2,\alpha_3$ can be computed using \refeqs{eq:alpha1} - (\ref{eq:alpha3}).
    } 
    \label{tab:BP3}
\end{table}
We identify the lightest $\mathcal{CP}$-even Higgs ${h_1}$ with the state at $\SI{95}{GeV}$ and choose the coupling related inputs to 
describe the excesses according to section \ref{sec:95ex}.
Consequently, the $\SI{125}{GeV}$ Higgs boson is the second lightest Higgs, $h_2$,
with couplings close to the SM Higgs-boson couplings, as our benchmark scenario is close to the alignment limit ($\varepsilon = 0.05$).
The $\cp$-even sector of the 2HDMS is identical to the N2HDM, while the $\cp$-odd sector of 2HDMS has two pseudoscalars $a_1$ and $a_2$, 
resulting from the diagonalization with the angle $\alpha_4$. 
The masses of the 2HDMS $\cp$-odd sector $m_{a_{1,2}}$  are chosen such that the doublet-like Higgs has the same mass as the $\cp$-odd Higgs in the N2HDM $m_A$, hence for $\alpha_4<\frac{\pi}{4}$ ($a_1$ has dominant doublet component) we set
\begin{align}
\label{mAlow}
    m_{a_1} = m_A {= m_{a_D}} \quad  \text{and}\quad   m_{a_2}=\text{free}=m_{a_S}\,,
\end{align}
or for $\alpha_4>\frac{\pi}{4}$ ($a_2$ has dominant doublet component) we set
\begin{align}
\label{mAhigh}
    m_{a_1}=\text{free}=m_{a_S}\quad \text{and}\quad   m_{a_2}=m_A = m_{a_D}\,.
\end{align}
We will often refer to the mass of the singlet-like state in the 2HDMS as $m_{a_S}$.
All further masses are chosen to be heavy and in agreement with the experimental constraints, see \refse{sec:expconstrain}.
The VEV of the singlet, $v_S$, 
is chosen to be small enough to yield differences in the THC observable (a dedicated discussion of the $v_S$ dependence is presented in section \ref{sec:vs_limits} below).
Furthermore, in the N2HDM limit of the 2HDMS we require that $\hat{\mu}=\tilde{\mu}$. 
Comparing with \refeq{eq:mutilde2HDMS} (see App.~\ref{sec:2hdms-app}), this can be achieved by choosing, 
\begin{equation}
    \hat\mu=m_{A}\,,
\end{equation}
which is fulfilled in our benchmark scenario. 

As discussed above, the 2HDMS becomes indistinguishable from the N2HDM by choosing all common parameters equal and setting $\al_4 = 0$,
or $\al_4 = \pi/2$. The crucial question we analyze in this work is which deviation from $\al_4 = 0, \pi/2$ allows an experimental 
distinction of the two models. For this we define the 
``$\al_4$ limit'' as the value where 2HDMS and N2HDM can be distinguished by (di-Higgs, or four-top) cross-section measurements.
Concretely, we obtain the value of $\alpha_4$,
where the Higgs-boson production cross-section predictions of both models differ by 
at least $2\,\sigma$, where
all free parameters of the models are chosen equal.
We define the distinction significance $\Delta s$ as,
\begin{equation}
  \Delta s= \sqrt{ \sum_{\substack{\text{final}\\\text{states}}}\Bigg(\frac{\abs{\sigma^{\rm2HDMS}(\al_4)-\sigma^{\rm N2HDM}}}{\Delta \sigma^{\rm N2HDM}}\Bigg)^2} \,,
  \label{Deltas}
\end{equation}
where $\Delta \sigma^{\rm N2HDM}$ is the anticipated experimental uncertainty estimate for the N2HDM.
The ``$\al_4$ limit'' is found for $\Delta s \ge 2$.
At the HL-LHC we will analyze the four-top channel $pp \to \bar tt\bar tt$ (see \refse{sec:hl-lhc}).
At a future $e^+e^-$ collider we will analyze the combination of the various di-Higgs production channels, $e^+e^-\to Zh_ih_j$ 
(see \refse{sec:epem}).

\begin{figure}[ht!]
\centering
\begin{subfigure}{0.45\textwidth}
        \centering
        \includegraphics[width=\textwidth]{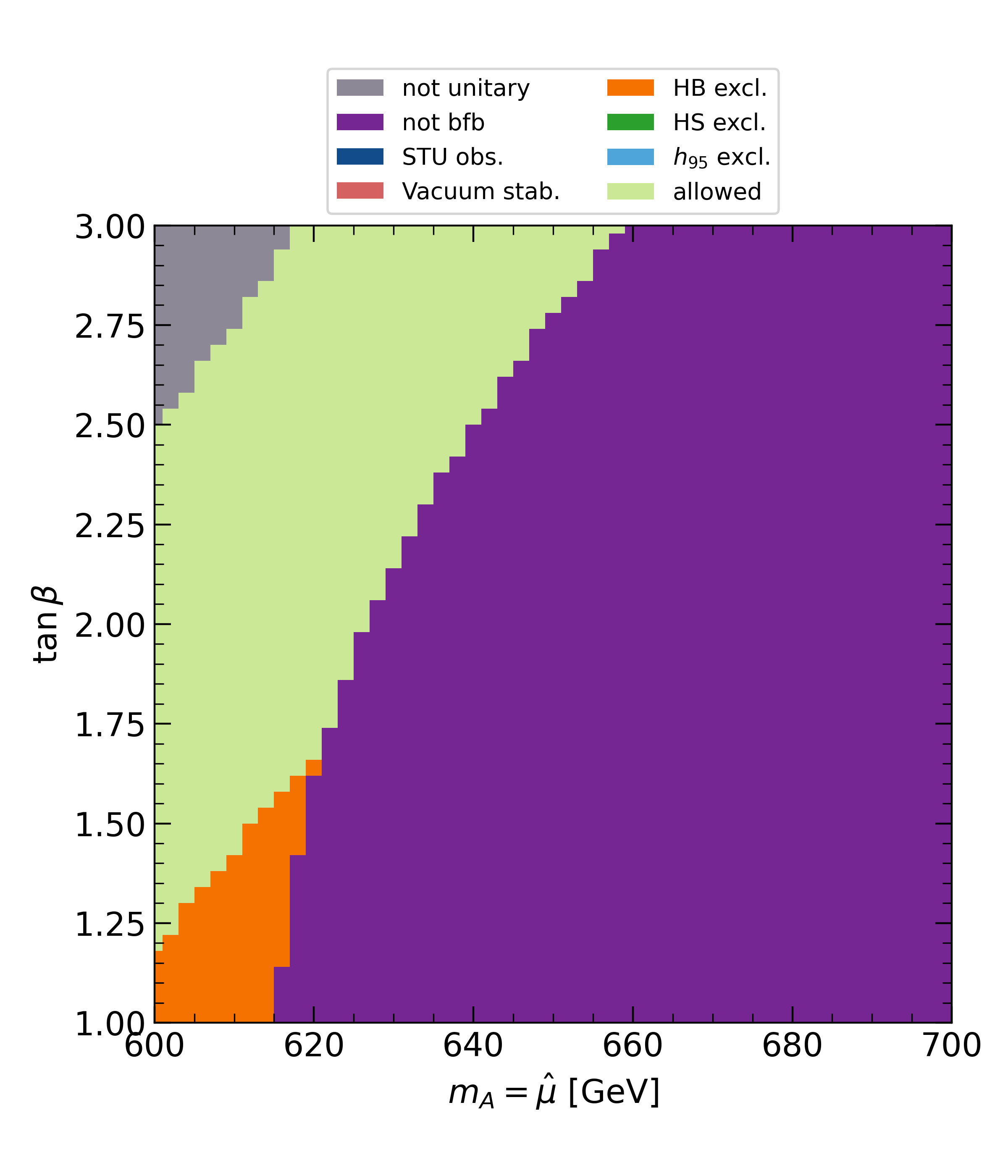}
    \caption{Benchmark plane 1}
     \label{fig:n2hdm_limit_region_1}
\end{subfigure}
\begin{subfigure}{0.45\textwidth}
        \centering
        \includegraphics[width=\textwidth]{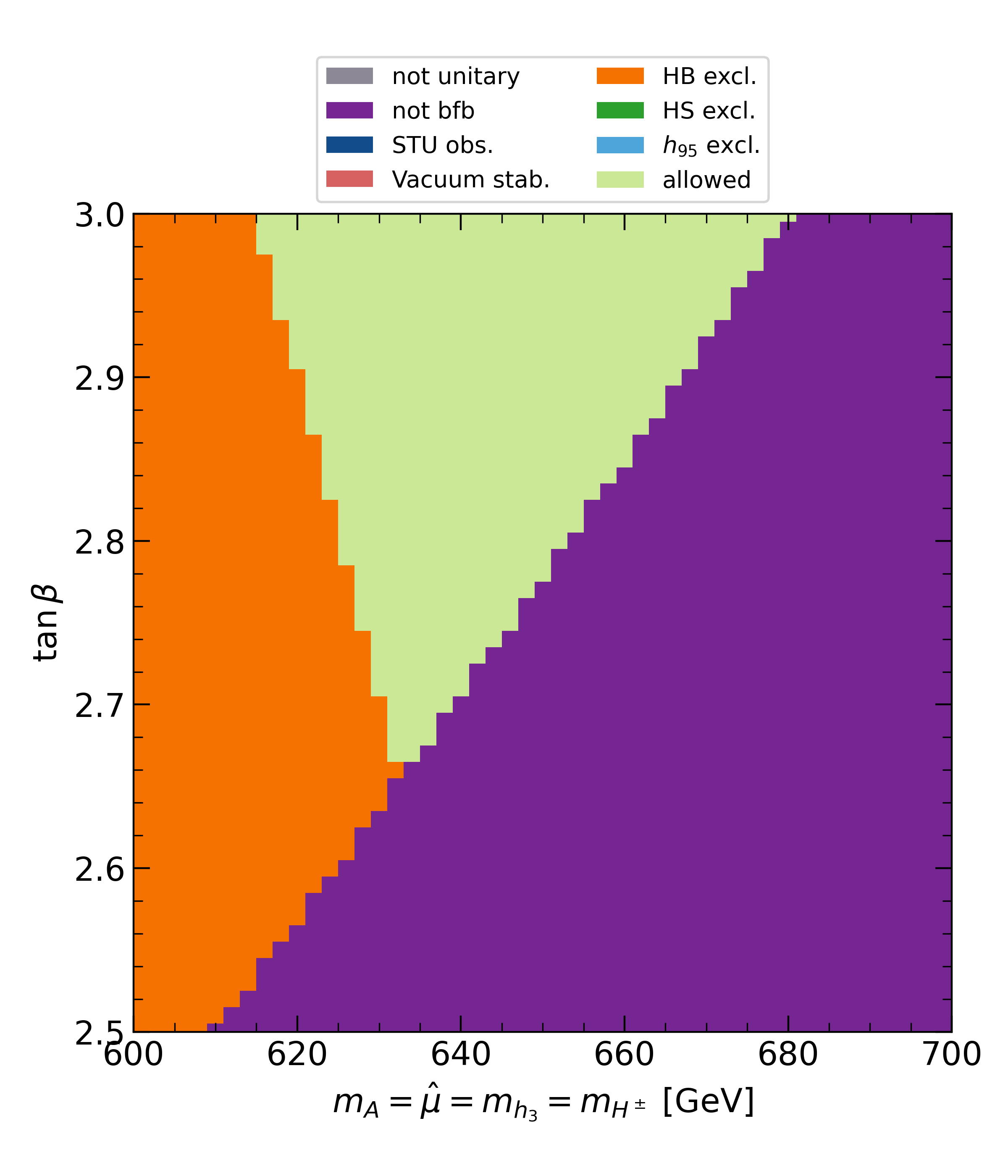}
    \caption{Benchmark plane 2}
    \label{fig:n2hdm_limit_region_2}
\end{subfigure}
    \caption{N2HDM benchmark plane 1 (left) and 2 (right). The green regions indicate the parameter space allowed by the theoretical and experimental constraints. For the other color coding: see legend. The colors show all possible exclusions from the imposed constraints however only unitarity, boundedness from below and Higgs bounds excluded parameter points in this scan.}
    \label{fig:n2hdm_limit_region}
\end{figure}

Starting from our benchmark scenario in \refta{tab:BP3}, we study these limits in two different benchmark planes.
We vary $\tan\beta$ and $m_A$ in the N2HDM and evaluate the 
$\al_4$ limits in two planes spanned by these two parameters as follows, see \reffi{fig:n2hdm_limit_region}:
\begin{description}
    \item[plane 1]: $\tan\beta\in [1,3]$,  $m_A=\hat{\mu}\in[600,700] \gev$ and $m_{h_3}=m_{H^\pm}=650 \gev$ (see \reffi{fig:n2hdm_limit_region_1}),
    \item[plane 2]: $\tan\beta\in [2.5,3]$ and $m_A=\hat{\mu}=m_{h_3}=m_{H^\pm}\in[600,700] \gev$ (see \reffi{fig:n2hdm_limit_region_2}).
\end{description}
Concerning the plots, it should be noted that the constraints checks are stacked: 
if a point fails unitarity the remaining constraints are not tested, etc. The allowed points pass all constraints indicated in the legend.

Concretely, the $\al_4$ limit for a specific N2HDM point is found as follows.
Given an N2HDM point, one can define a 2HDMS plane spanned by the two additional free parameters, $m_{a_S}$ 
and $\al_4$.   
\begin{figure}[ht!]
    \centering
    \includegraphics[width=0.45\textwidth]{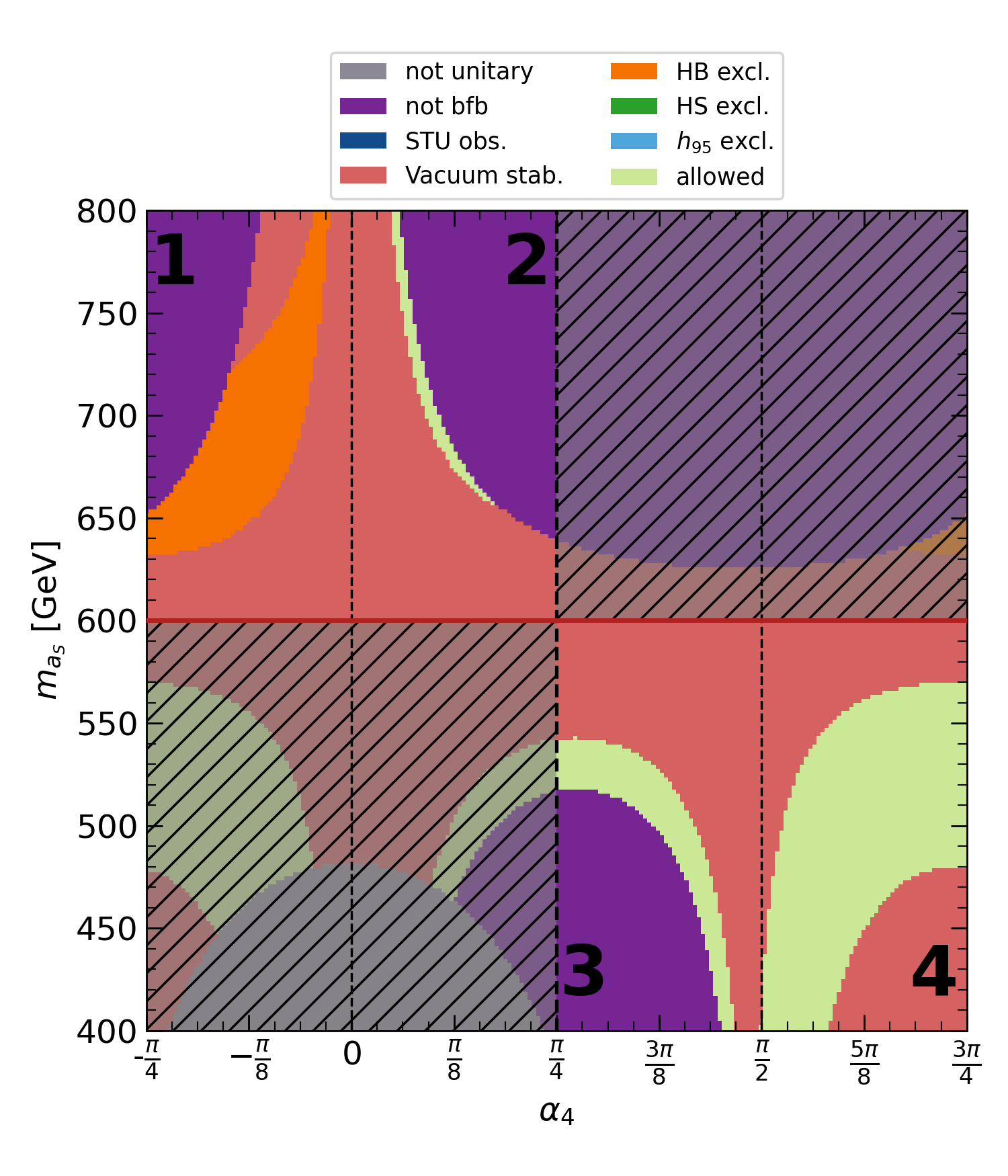}
    \caption{Constraints of the extended region to the 2HDMS for the benchmark scenario. The mass of the doublet like $\mathcal{CP}$-odd Higgs is shown in red. The greyed out regions violate the mass hierarchy of the $\mathcal{CP}$-odd Higgs bosons.}
    \label{fig:2hdms_region}
\end{figure}
Here it is important to note that not all points in the $\al_4$--$m_{a_S}$ plane do automatically satisfy all the experimental
and theoretical constraints. As an example, we show in \reffi{fig:2hdms_region} the $\al_4$--$m_{a_S}$ plane for our benchmark
point of \refta{tab:BP3}.
Of this plane, only the upper left and the lower right quadrant are allowed by the mass hierarchy of the $\cp$-odd Higgs boson.
The two remaining quadrants are seperated into two parts each (labelled 1\ldots4 in the plot). The upper left one has 
$m_{a_S} > m_A$ and is defined around the N2HDM limit of the 2HDMS at $\al_4 = 0$, the lower right one has
$m_{a_S} < m_A$ and is located around the other N2HDM limit at $\al_4 = \frac{\pi}{2}$.
Throughout our benchmark scenarios, sectors one and two the are excluded by the constraints almost entirely, except for a sliver as 
visible in sector two in \reffi{fig:2hdms_region}. Consequently, we focus our study on sectors three and four, 
i.e.\ $\al_4>\frac{\pi}{4}$ and $m_{a_S} < m_A$ and define $\Delta m_a := m_A - m_{a_S}$.%
\footnote{Here it should be kept in mind that in the case of $m_{a_1} \equiv m_{a_S} < m_A \equiv m_{a_2}$, 
the decay $a_2 \to a_1 h_i$ decay can be kinematically allowed,
depending on $\Delta m_a$ and $m_{h_i}$. Also the other branching ratios of $a_2$ are sensitive to $\De m_a$.}

\begin{figure}[htb!]
    \centering
    \includegraphics[width=1\linewidth]{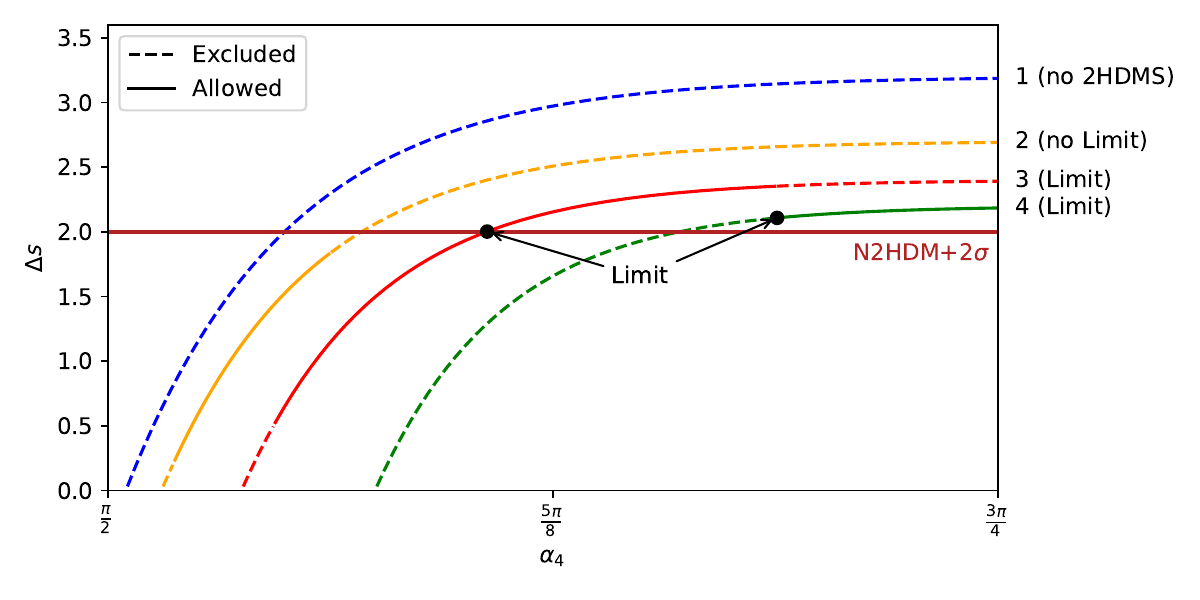}
    \caption{Schematic plot of possible scenarious encountered during computation of the $\alpha_4$ limits.}
    \label{fig:limitscenarios}
\end{figure}

For our analysis of the $\al_4$ limits in a benchmark plane, for practical purposes, we study horizontal lines in this plane 
by choosing different fixed values of $\De m_a =\{25,50,75,100,125\} \gev$.
To determine the $\al_4$ limit for each fixed $\De m_a$, 
we compute the cross-section for the processes under consideration 
(as stated above, $pp \to \bar tt\bar tt$ at the HL-LHC and $e^+e^- \to Z h_ih_j$ at the ILC/LCF) 
in the N2HDM, as well as 2HDMS, varying $\al_4\in [\frac{\pi}{4},\frac{3\pi}{4}]$ 
over values not excluded by the constraints. This yields for each allowed point the significance $ \Delta s$ as defined in \refeq{Deltas}.
We call the points where $\Delta s\, \simeq \,2$, i.e.\ where $\Delta s$ ``crosses'' 
the $2\,\sigma$ value the $\al_4$ limit. 
For 2HDMS parameter points further away from $\pi/2$ the 2HDMS can be distinguished experimentally from the N2HDM. 
It is important to note that
the $\al_4$~limits can be determined separately for the areas~3 and~4 in \reffi{fig:2hdms_region}. 
This will yield an interval as $[\al_4\,\mbox{limit}(\al_4 < \frac{\pi}{2}), \al_4\,\mbox{limit}(\al_4 > \frac{\pi}{2})]$,
in which the two models are indistinguishable experimentally.
We compute the $\al_4$ limits for each fixed $\De m_a$ individually.
During this process we encounter four different possible scenarios (\reffi{fig:limitscenarios}) (possibly for all $\De m_a$ values, but individually for 
area~3 and~4.): 
\begin{enumerate}
    \item  The 2HDMS is excluded by the constraints for every $\al_4$ (2HDMS excluded).%
    \footnote{The subtle differences in the potential can lead to different behaviours of vacuum stability, hence an allowed in the N2HDM is not automatically allowed in the 2HDMS even in the N2HDM limit.}

    \item  The by the constraints allowed area has $\Delta s< 2$ for all allowed $\al_4$: no limit, the point cannot be distinguished.

    \item  The by the constraints allowed area has a point where $\Delta s\, \simeq \,2$, yielding the $\al_4$ limit.
    We find that this happens only once for each fixed $\De m_a$ (but possibly once for area~3 and once for area~4).

    \item  The by the constraints allowed area has only points where $\Delta s>2$: for the $\De m_a$ the two models 
    can be distinguished experimentally for all $\al_4$ values.
     In this case, the $\al_4$ limit is given by the $\al_4$ value closest to $\pi/2$ that is allowed by the constraints.
\end{enumerate}

For illustration purposes we show examples for the four possibilities in \reffi{fig:limitscenarios}. Potential lines for $\De s$ as a function of $\al_4$ 
(for $\al_4 > \pi/2$) are depicted, where the solid (dashed) parts of the curves represent allowed (excluded) parameter regions. The blue curve corresponds
to ``2HDMS excluded'' as no valid points are found. For the orange curve the solid line does not reach $\De s = 2$, corresponding to ``models cannot be
distinguised''. The red line has a point with the solid part crossing $\De s = 2$, and the corresponding angle constitutes the $\al_4$ limit. Finally, 
the green curve has allowed regions only above $\De s = 2$, and thus the smallest allowed $\al_4$ yields the desired limit.

%% file: section3.tex
\newpage

\section{HL-LHC Prospects for Model Distinction}
\label{sec:hl-lhc}
\subsection{Four-top final state at the hadron collider
}

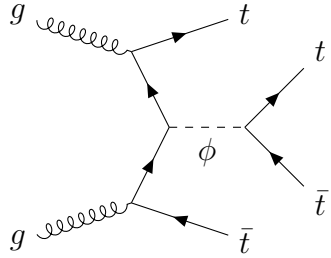
\begin{figure}[h]
    \begin{center}
\begin{tikzpicture}
            \begin{feynman}[small]

                \vertex (i1) at (0,0);
                \vertex (i2) at (1,0);
                \vertex (t1) at (2,1) {\(t\)};
                \vertex (tb1) at (2,-1) {\(\bar t\)};

                \vertex (i3) at (-0.5,1) ;
                \vertex (i4) at (-0.5,-1) ;

                \vertex (t2) at (1,1.5) {\(t\)};
                \vertex (tb2) at (1,-1.5) {\(\bar t\)};

                \vertex (g1) at (-2,1.5) {\(g\)};
                \vertex (g2) at (-2,-1.5) {\(g\)};

                \diagram* {(i1) --[scalar,edge label'=\(\phi\)] (i2),
                (tb1) --[fermion] (i2) --[fermion] (t1),
                (tb2) --[fermion] (i4) --[fermion] (i1) --[fermion] (i3) --[fermion] (t2),
                (g1) --[gluon] (i3),
                (g2) --[gluon] (i4),
                };
            \end{feynman}
            \end{tikzpicture}
    \end{center}
    \caption{Example diagram for the $ttH$ associated production process.} 
    \label{fig:tth-FD}
\end{figure}

At the (HL-)LHC the heavy BSM Higgs-bosons, $h_3$ and $A$, can be produced in association with top quarks. This includes
the channels of $tt\phi$, $tW\phi$ and $tq\phi$, with $\phi$ denoting collectively the heavy neutral Higgs bosons, with the mass $m_\phi$. 
Here the $tt\phi$ channel is particularly sensitive to the Higgs top-Yukawa coupling.
This process proceeds through diagrams as shown in \reffi{fig:tth-FD}.

For $m_\phi \ge 2 \mt$ the BSM Higgs can (potentially dominantly) decay to a $\bar{t}t$ final state, which then
contributes to the four-top final state at the (HL-)LHC.
Within the 2HDMS and N2HDM, the BSM contribution to the four-top signal rate would be a combination of the
$\mathcal{CP}$-even state $h_3$ and the $\mathcal{CP}$-odd state doublet dominated $a_D$, which add incoherently. In particular, we consider the case that the masses of  $h_3$ and $a_D$ are close,
below the experimental resolution, 
\begin{align}
    \sigma(pp\to \bar{t}\bar{t}tt) =& \sigma(pp\to t\bar{t} a_D\to  \bar{t}\bar{t}tt)+\sigma(pp\to t\bar{t} h_3 \to  \bar{t}\bar{t}tt)\,.
\end{align}
The top-associated production cross-section can be calculated by rescaling the effective cross-sections for $\mathcal{CP}$-even and -odd 
Higgs bosons with the couplings of $c_{h_3tt}$ (and $c_{h_3VV}$ entering the $tWh_3$ channel) and $c_{a_Dtt}$, respectively~\cite{Paasch:2023gdz}, where the calculation is implemented in \texttt{HiggsTools}~\cite{Bahl:2022igd}. 
The fermionic couplings of $\mathcal{CP}$-even sector in the 2HDMS and N2HDM are identical, while the couplings of the $\mathcal{CP}$-odd sector between the two models are different. In the N2HDM, the cross-section of $ttA$ scales as,
\begin{equation}
    \sigma(pp\to t\bar{t} A)^\mathrm{N2HDM} \propto |c_{Att}^2| \propto \frac{1}{\tan^2\beta},
\end{equation}
while in the 2HDMS has the following dependences for the $tta_i$ cross sections,
\begin{equation}
    \sigma(pp\to t\bar{t} a_S)^\mathrm{2HDMS} \propto |c_{a_S tt}^2| \propto \frac{\cos^2\alpha_4}{\tan^2\beta},\qquad \sigma(pp\to t\bar{t} a_D)^\mathrm{2HDMS} \propto |c_{a_D tt}^2| \propto \frac{\sin^2\alpha_4}{\tan^2\beta}.
    \label{eq:si4t-scaling}
\end{equation}
Hence, the couplings of the $\mathcal{CP}$-odd Higgs bosons to top quarks differ by occurrence of the mixing angle $\alpha_4$.

\begin{figure}[ht!]
    \centering
    \includegraphics[width=0.5\linewidth]{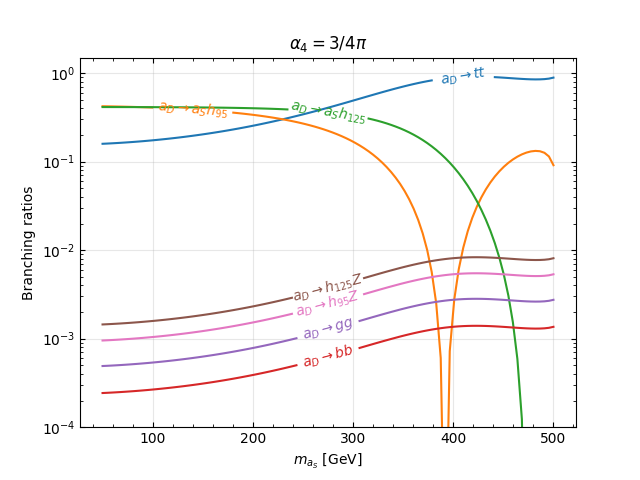}
    \caption{The BRs of $a_D$ as a function of $m_{a_S}$, with $\alpha_4 = \frac{3}{4}\pi$, 
    and the other parameters are chosen as for the benchmark scenario in \protect\refta{tab:BP3}.}
    \label{fig:bra2ma1}
\end{figure}

In addition, the decay width of the BSM Higgs decay to top quarks, $a_D/h_3\to t\bar{t}$, plays an important role 
in the four-top cross-section calculations as well. 
Usually, in the 2HDMS the $a_D\to t\bar{t}$ decay is the most dominant decay and the BR is largely insensitive to the 
top Yukawa couplings, particularly when the masses of $a_S$ and $a_D$ are close and $a_D\to a_S h_i$ decay is kinematically forbidden.
However, the model difference enters into the THCs and thus has an impact on the decay widths of $a_D\to a_S h_i$ 
and $h_3 \to h_i h_j$. These decay widths are also sensitive to the lighter Higgs-boson masses. 
This is illustrated in \reffi{fig:bra2ma1}, where we show various BRs of the $a_D$ as a function of $m_{a_S}$ 
for the parameters in \refta{tab:BP3}. It can be seen that $\br(a_D\to a_S h_i)$ becomes more relevant when the mass of $a_S$ 
goes down. Thus, for lower $m_{a_S}$ one finds a larger $\br(a_D\to a_S h_{125}$) and $\br(a_D\to a_S h_{95}$), 
where $a_D\to a_S h_{125}$ and $a_D\to a_S h_{95}$ become the most dominant decays for $m_{a_S} \lesssim 250 \gev$. 
In the case that $a_D \to t\bar{t}$ is not the dominant decay, the $\br(a_D \to t\bar{t})$ is more sensitive to $\alpha_4$, 
and the distinction between 2HDMS and N2HDM could be more significant. It should be noted that $\br(a_D\to a_S h_{95})$ 
has a gap at around $m_{a_S} \sim 400 \gev$, since the $h_{95} a_S a_D$ coupling has a sign change and crosses the zero value in this
parameter region.

\begin{figure}[ht!]
    \centering
    \includegraphics[width=0.6\linewidth]{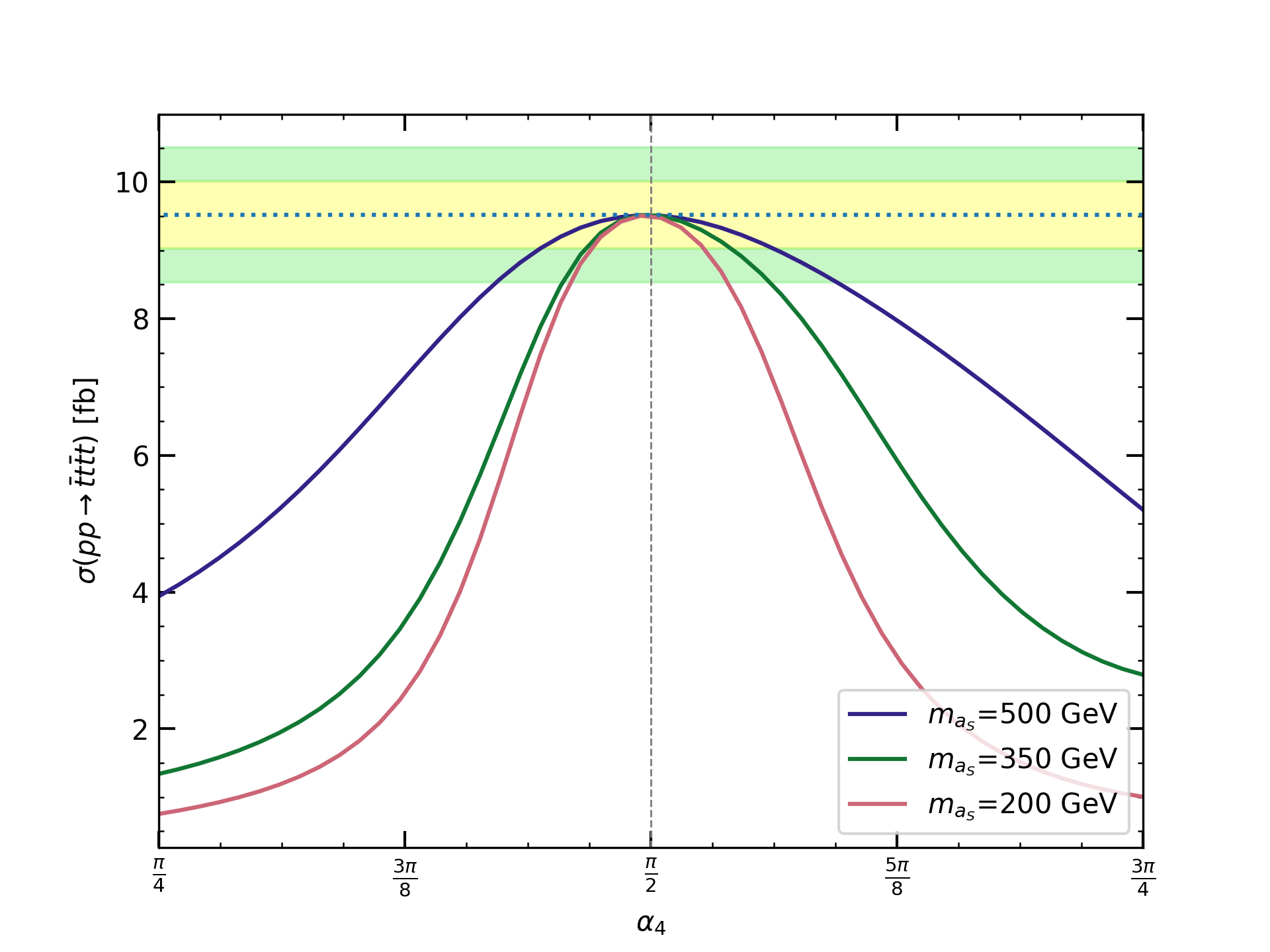}
    \caption{The four-top ($pp\to tt\bar{t}\bar{t}$) cross-section as a function of $\alpha_4$ for the 2HDMS.
    As vertical lines we show the N2HDM cross-section (dotted line) and the corresponding $1\,(2)\,\sigma$ 
    uncertainty band in yellow (green). The other parameters are chosen according to our benchmark scenario, see \refta{tab:BP3}. }
    \label{fig:4ta4}
\end{figure}

This above discussion demonstrates that the four-top signal rate for the 2HDMS and the N2HDM can be different 
while most of the parameters, in particular the $\mathcal{CP}$-even Higgs-boson masses and mixing angles are identical.
Consequently, one can potentially distinguish the 2HDMS from the N2HDM by comparing the four-top signal rate. 
Based on our benchmark point (see \refta{tab:BP3}), in \reffi{fig:4ta4} we present the 2HDMS four-top cross-section 
with varying $\alpha_4$ and different singlet mass $m_{a_S}$, as indicated by the legend. The 2HDMS cross-section can be 
compared to the N2HDM cross-section as well as the corresponding $2\sigma$ uncertainty band 
(see \refeqs{eq:detdsig}, (\ref{eq:detdxs}) and the corresponding discussion below).
As one can observe in \reffi{fig:4ta4}, the four-top cross-section is maximal in the N2HDM limits, $\al_4 = \pi/2$ 
and identical to the N2HDM value.  It decreases slightly asymmetricly 
for increasing $|\pi/2 - \al_4|$.
In the case that the 2HDMS four-top rate is outside the N2HDM $2\,\sig$ uncertainty band, 
the 2HDMS can be considered to be experimentally distinguishable from the N2HDM, where the ``$\alpha_4$ limit''
(see \refse{sec:benchmarkscenario}) is given by the smallest value of $|\pi/2 - \al_4|$ that yields a distinction
between the two models. Corresponding to the discussion in \refse{sec:benchmarkscenario}, we use three different values for
$m_{a_S} = 250, 350, 500 \gev$, where $a_D\to a_S h_i$ decay is kinematically forbidden for $m_{a_S}=500 \gev$. 
The 2HDMS four-top cross-section with lower $m_{a_S}$ decreases faster for increasing $|\pi/2 - \al_4|$ than for 
$m_{a_S}=500 \gev$, as the lower $m_{a_S}$ would open more decay channels for $a_D$, and the $a_D\to a_S h_i$ processes 
would be affected more strongly by $\alpha_4$. 
Consequently, the properties of $a_D$ with lower $m_{a_S}$ in 2HDMS can be more easily distinguished from the N2HDM case.
Details about the uncertainty bands in \reffi{fig:4ta4} are discussed in the next subsection.

\subsection{\boldmath{$\alpha_4$} limits at the HL-LHC}

At the HL-LHC, the four-top process has been analyzed in \citere{Craig:2016ygr} with an integrated luminosity of 3000~fb$^{-1}$ 
and a center-of-mass energy of 14~TeV. 
In order to estimate the four-top signal uncertainty at the HL-LHC and include the background, one can use the Log-likelihood function to determine the signal significance, which is given by,
\begin{equation}
   \mathcal{S} {\equiv \mathcal{S}(x|\mu) :=}
   \sqrt{-2\ln \frac{L(x|\mu)}{L(\mu|\mu)}},\qquad L(x|\mu) = \frac{x^\mu e^{-x} }{\mu!}\,, 
\end{equation}
where $x$ and $\mu$ denote the number of events expected in the two models to be distinguished. In our case $x$ corresponds to the
2HDMS, whereas $\mu$ to the N2HDM. Both predictions contain the SM contribution (background events, $b$) and their respective BSM 
contribution (signal events, $s$).
Concretely, the number of events is obtained as, 
\begin{align}
    x &= s^{{\mathrm{2HDMS}}}+b = (\sigma_s^{{\mathrm{2HDMS}}} \cdot \epsilon_s + \sigma_b\cdot \epsilon_b)
               \times \mathcal{L}\,,\\
    {\mu} &= s^{{\mathrm{N2HDM}}}+b = (\sigma_s^{{\mathrm{N2HDM}}} \cdot \epsilon_s + \sigma_b\cdot \epsilon_b)
                \times \mathcal{L}\,.
\end{align}
Here $\sigma_s^{\mathrm{2HDMS,N2HDM}}$ and $\sigma_b$ are the signal and background cross-sections. 
In addition, $\epsilon_s$ and $\epsilon_b$ are the detection efficiency of signal and background, 
which we have taken over from the HL-LHC study~\cite{Craig:2016ygr}. 
In order to define the number of events, $\Delta\mu$, that yield a $2\,\sig$ distinction of 2HDMS from the 
N2HDM hypothesis, we set
$x = \mu + \Delta \mu$, and solve
\begin{equation}
    \mathcal{S}=\sqrt{-2 \left( \mu\ln (1 {+} \frac{\Delta\mu}{\mu}) -\Delta \mu \right)} \; {\stackrel{!}{=}} \; 2\,.
    \label{eq:detdsig}
\end{equation}
This allows to define the difference in the signal cross section between 2HDMS and N2HDM yielding a $2\,\sig$ distinction,
\begin{equation}
    \Delta\sigma_{s} = \frac{\Delta \mu}{\epsilon_s \cdot \mathcal{L}}\,.
    \label{eq:detdxs}
\end{equation}
Correspondingly, the 1$\sigma$ region can be determined with the same method, by setting the significance to~1.

The four-top analysis at HL-LHC evaluates the signal and background efficiencies based on the assumption, that the $\mathcal{CP}$-even Higgs $H$ and the $\mathcal{CP}$-
odd $A$ have the same masses, while the benchmark point has different $m_{h_3}$ and $m_{a_D}$. However, we can assume that the HL-LHC analysis 
is still valid within a mass window of 50~GeV, i.e. $|m_{h_3}-m_{a_D}| \lesssim 50 \gev$. Therefore, we determine the $1\sigma$ area (yellow 
band in \reffi{fig:4ta4}) and the $2\sigma$ area (green band in \reffi{fig:4ta4}) for the benchmark point (see
\refta{tab:BP3})
by using the method from \refeqs{eq:detdsig} and (\ref{eq:detdxs}) and the numbers for
$\epsilon _s$, $\epsilon_b$ and $\sigma_b$ are taken over from \citere{Craig:2016ygr}. 
The $\al_4$ limits, corresponding to a distinction of the model at the $2\,\sig$ level, see \refse{sec:benchmarkscenario},
in the case shown in \reffi{fig:4ta4}, one finds for $m_{a_S}=500 \gev$ the values of $\alpha_4 \lesssim  1.32 $ and $\alpha_4 \gtrsim 1.87$. 
For lower $m_{a_S}$, the four-top rate is more sensitive and stronger $\al_4$ limits can be found:
$\alpha_4 \lesssim  1.43\, (1.44)$ and $\alpha_4 \gtrsim 1.76\, (1.69)$ for $m_{a_S}=350\, (200) \gev$.

\begin{figure}[ht!]
    \centering
    \begin{subfigure}{.485\linewidth}
    \centering
        \includegraphics[width=\linewidth]{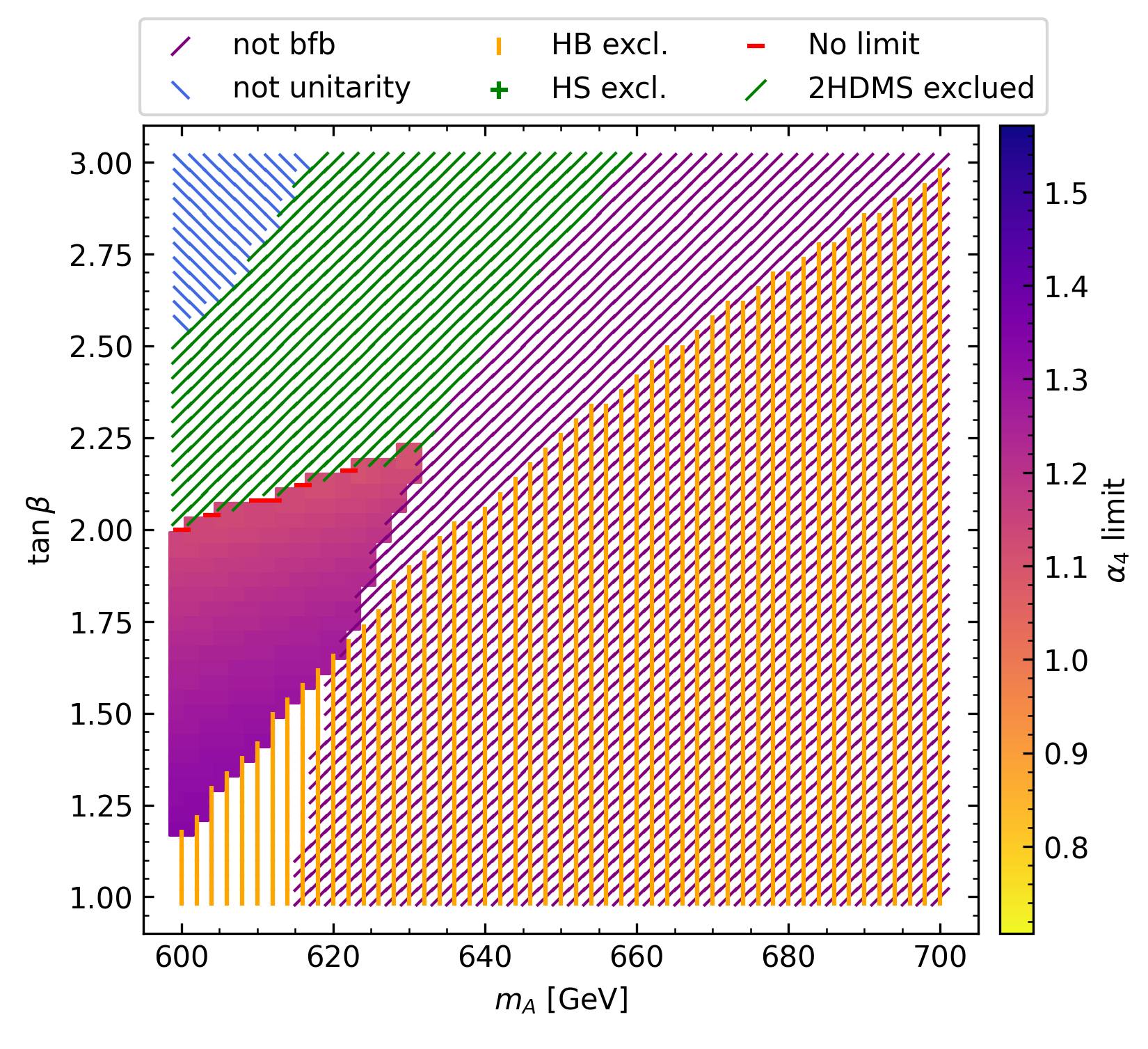}\caption{}\label{a4det11}
    \end{subfigure}\hfill\begin{subfigure}{.485\linewidth}
    \includegraphics[width=\linewidth]{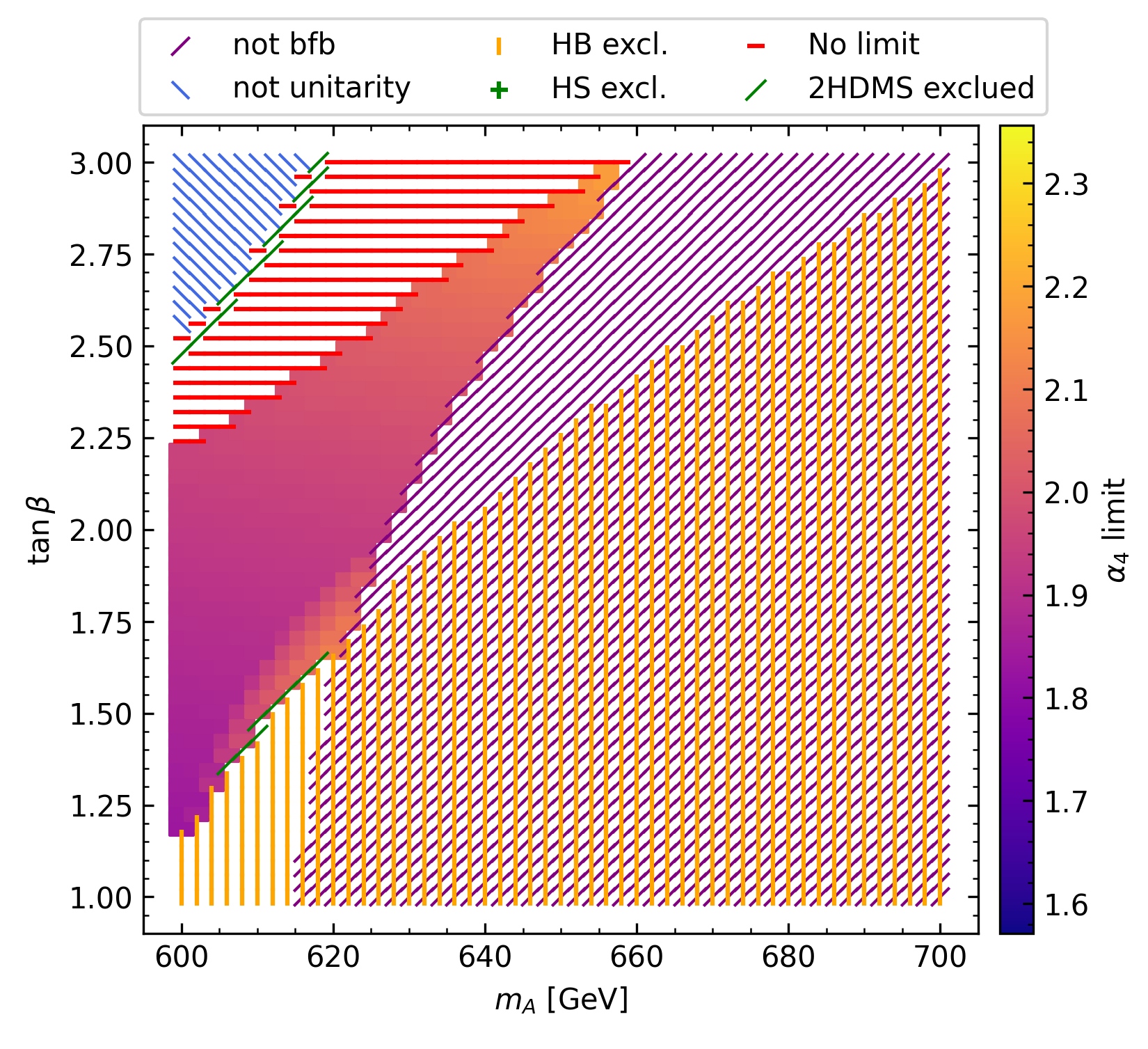}\subcaption{}\label{a4det12}
    \end{subfigure}
    \caption{$\al_4$~limits in the $m_A$-$\tb$ parameter space around our benchmark scenario as given in \protect\refta{tab:BP3} 
    according to the plane shown in \protect\reffi{fig:n2hdm_limit_region_1} with $\Delta m_a = m_{A_D} - m_{A_S} = 100 \gev$ for 
    $\al_4 < \, (>)\, \pi/2$ in the left (right) plot.}
    \label{fig:a4lim2dm100}
\end{figure}

\begin{figure}[ht!]
    \centering
    \begin{subfigure}{.485\linewidth}
    \centering
        \includegraphics[width=\linewidth]{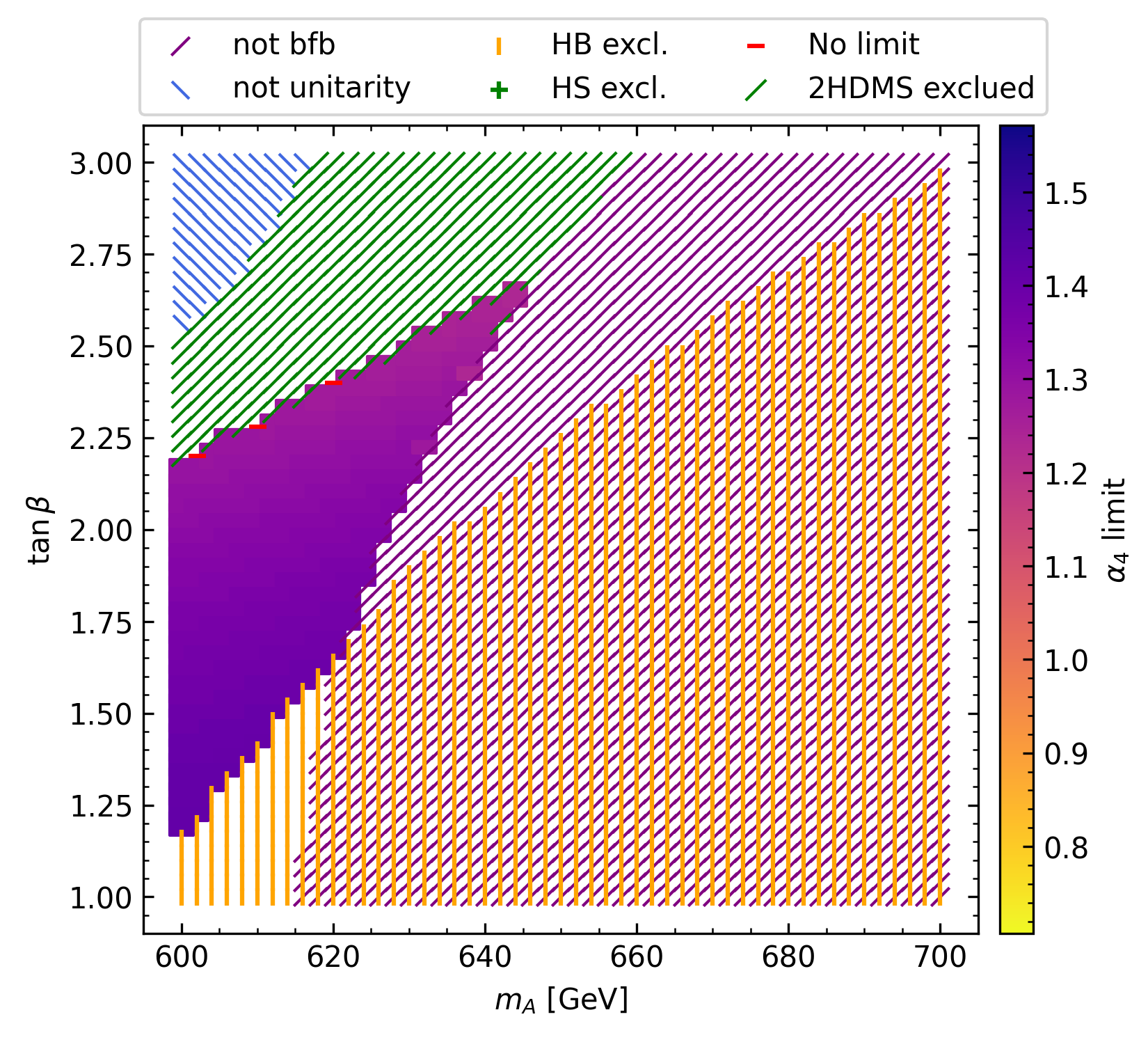}\caption{}\label{a4det21}
    \end{subfigure}\hfill\begin{subfigure}{.485\linewidth}
    \includegraphics[width=\linewidth]{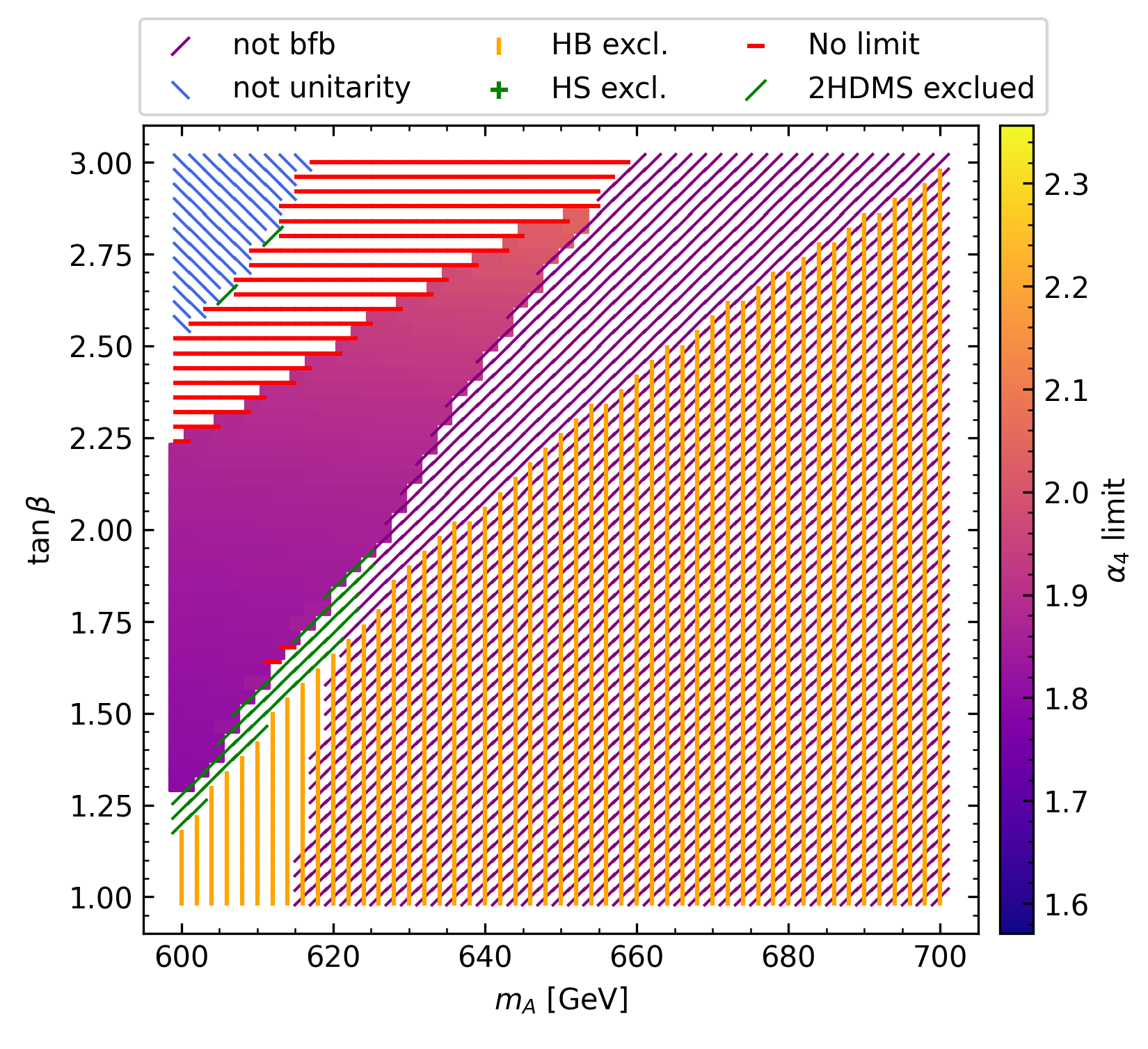}\caption{}\label{a4det22}
    \end{subfigure}
    \caption{$\al_4$~limits in the $m_A$-$\tb$ parameter space as in \protect\reffi{fig:a4lim2dm150}
    with $\Delta m_a = m_{A_D} - m_{A_S} = 150 \gev$.}
    \label{fig:a4lim2dm150}
\end{figure}

\smallskip
The main question of this work concerns the distinction of the N2HDM and the 2HDMS, as defined by the $\al_4$~limits. To this end
we demonstrate the $\al_4$~limits in the $m_A$-$\tb$ parameter space around our benchmark scenario as given in \refta{tab:BP3} according
to the plane shown in \reffi{fig:n2hdm_limit_region_1}. In our benchmark scenario we have $m_{h_3} = 650 \gev$, such that
$m_A (= m_{a_D}) \in [600 \gev, 700 \gev]$ assures that $|m_{a_D} - m_{h_3}| \le 50 \gev$, fulfilling this analysis requirement. 
The results for $m_{a_D} - m_{a_S} = 100\,, 150 \gev$ are presented in \reffis{fig:a4lim2dm100} and \ref{fig:a4lim2dm150}, respectively.
The left (right) plots show the $\al_4$~limits for $\al_4 < \, (>)\, \pi/2$, where we have considered parameter regions of
$\al_4 \in [\pi/4, \pi/2]$ ($[\pi/2, 3\pi/4]$).
Following \refse{sec:benchmarkscenario}, we first test every point against the experimental and theoretical constraints, see 
\refse{sec:constraints}, however, now within both models. Correspondingly the allowed parameter space is a subset of the 
parameter space allowed in \reffi{fig:n2hdm_limit_region_1}. Furthermore, each allowed parameter point in the left (right) plots has
an allowed interval of, 
\begin{align}
    \al_4 &\in [\pi/4 < \al_{4,\mathrm{min}}, \pi/2]\; \hspace{2mm}(\mathrm{left:}\, \al_4 < \pi/2)\,, \nonumber\\
    \al_4 &\in [\pi/2, \al_{4,\mathrm{max}} < 3\pi/4]\; (\mathrm{right:}\, \al_4 > \pi/2)\,,
\label{al4minmax}
\end{align} 
for which the 2HDMS passes all constraints. 
In the left plots of both figures the additionally excluded parameter space is indicated by green hatched regions. 

For the allowed parameter spaces the $\al_4$~limits are shown as color codings in the two figures. As one can observe, 
the $\al_4$ limits are better, i.e.\ closer to $\pi/2$ for smaller $\tb$. This can be understood from the scaling given 
in \refeq{eq:si4t-scaling}. Comparing \reffis{fig:a4lim2dm100} and \ref{fig:a4lim2dm150}, one finds that the larger mass difference,
$m_{a_D} - m_{a_S} = 150 \gev$ leads to tighter $\al_4$~limits (i.e.\ closer to $\pi/2$) than for the smaller mass difference.
Furthermore, one can see that for the same $m_{a_D} - m_{a_S}$, $m_A$ and $\tb$, comparing the left and the right plot, 
the $\al_4$ limits are better for $\al_4 < \pi/2$ than for $\al_4 > \pi/2$. 
Finally, the ``no limit'' regions in the two right plots can be understood from the allowed intervals of $\al_4$ in each point
as indicated in \refeq{al4minmax}. At the borders of the ``no limit'' the $\al_4$~limit reaches exactly these $\al_{4,\mathrm{max}}$
values. 

\begin{figure}[ht!]
    \centering
    \begin{subfigure}{.485\linewidth}
    \centering
        \includegraphics[width=\linewidth]{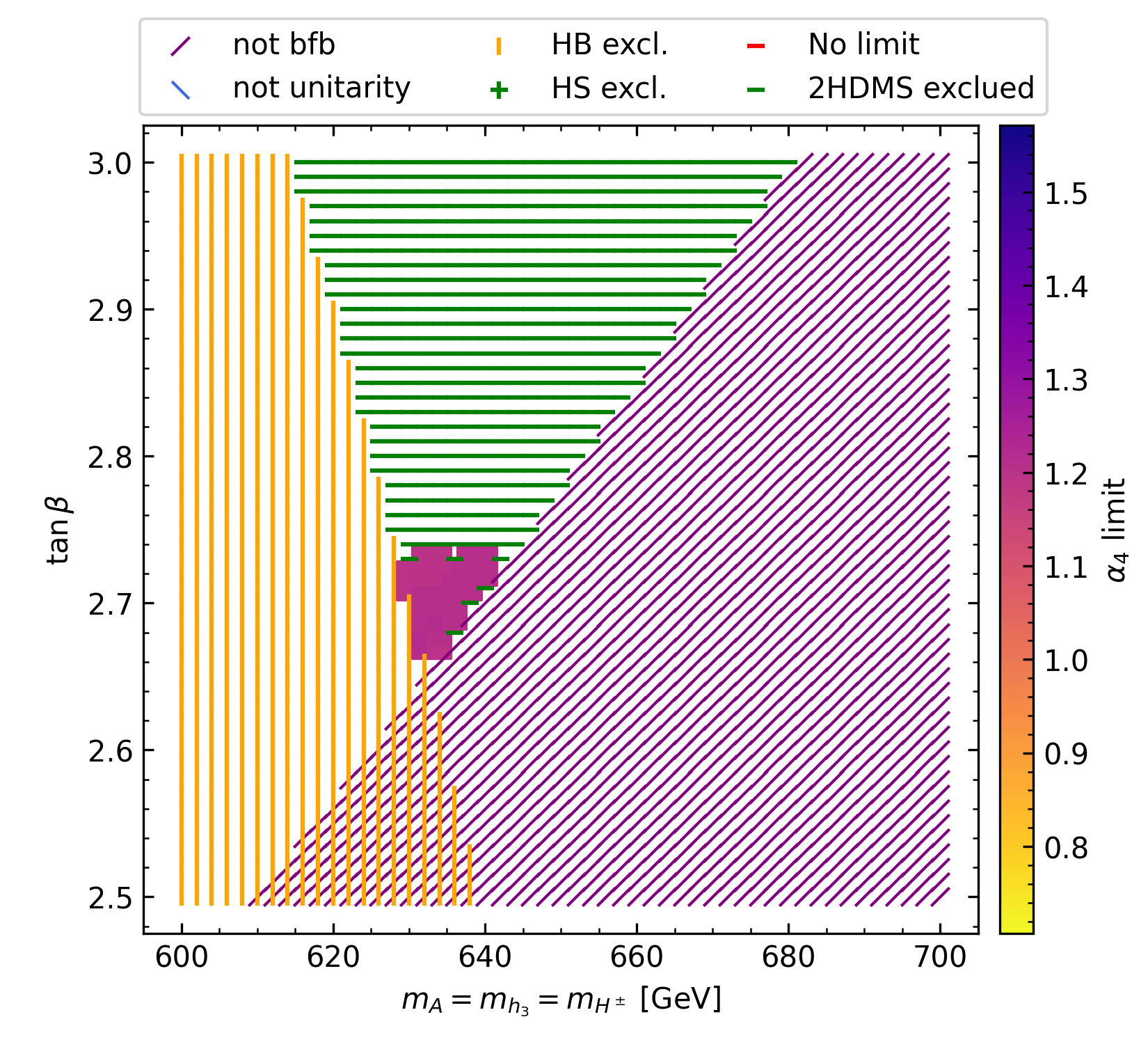}
        \caption{}
        \label{a4det31}
    \end{subfigure}\hfill\begin{subfigure}{.485\linewidth}
    \includegraphics[width=\linewidth]{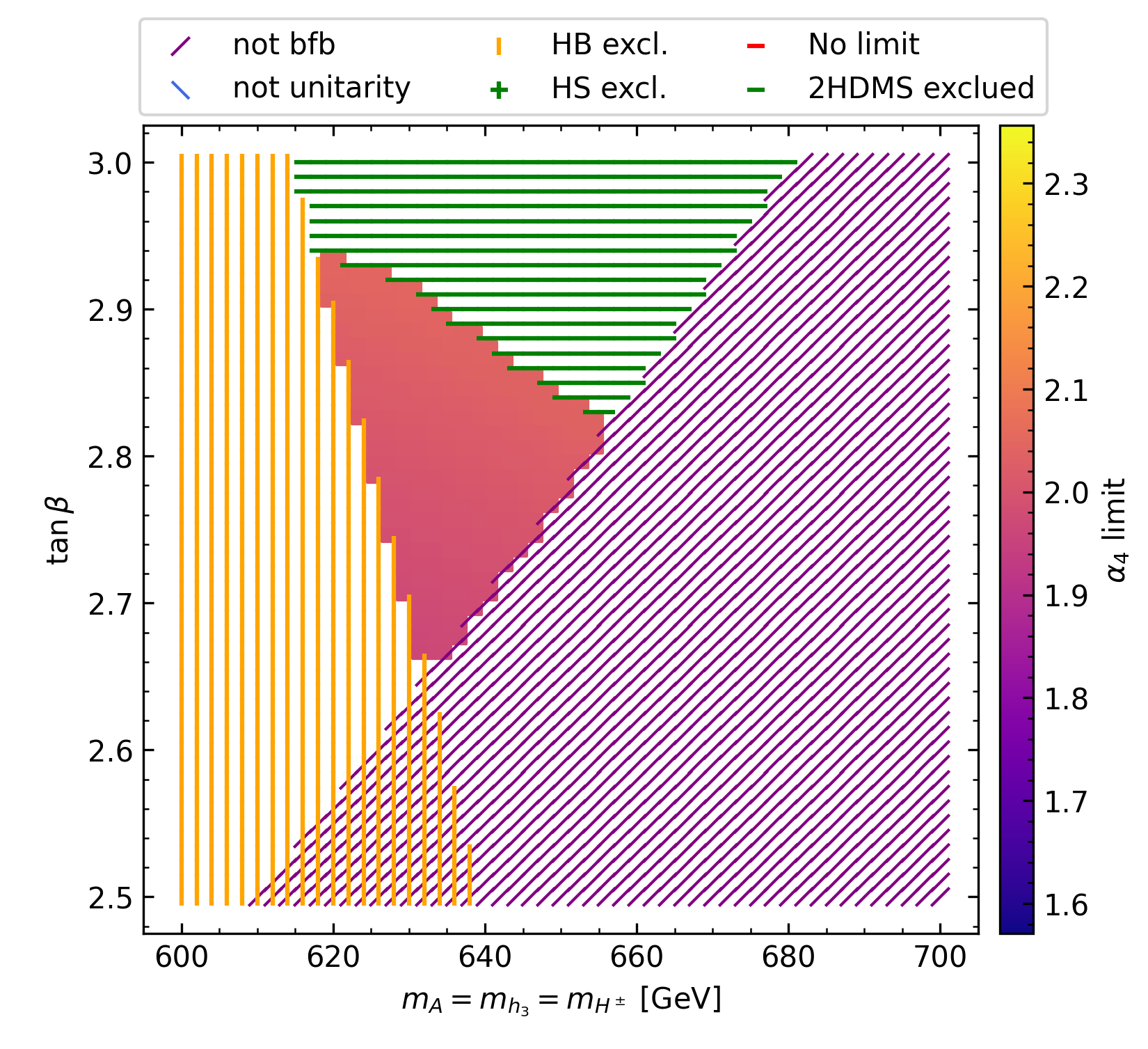}
    \caption{}
    \label{a4det32}
    \end{subfigure}
    \caption{$\al_4$~limits in the $m_A$-$\tb$ parameter space around our benchmark scenario as given in \protect\refta{tab:BP3} 
    according to the plane shown in \protect\reffi{fig:n2hdm_limit_region_2} with $ m_{a_S} = 500 \gev$ for 
    $\al_4 < \, (>)\, \pi/2$ in the left (right) plot.}
    \label{fig:a4lim2dm150m}
\end{figure}

To complement our analysis of $\al_4$~limits at the HL-LHC, we present in \reffi{fig:a4lim2dm150m}
the limits in the $m_A$-$\tb$ parameter space around our benchmark scenario as given in \refta{tab:BP3}, 
but now according to the plane shown in \reffi{fig:n2hdm_limit_region_2}, and for a fixed value $m_{a_S} = 500 \gev$, i.e.\ varying
$\Delta m_a$ over the horizontal axis (see \refeqs{mAlow}, (\ref{mAhigh}).)
As before, the left (right) plots show the $\al_4$~limits for $\al_4 < \, (>)\, \pi/2$. In both plots the green hatched area
indicates the parameter space excluded additionally from the 2HDMS. 
For $\al_4 < \pi/2$ one can observe that only a relatively small part of the $m_A$-$\tb$ parameter space is still allowed, yielding 
$\al_4$ limits around $\al_4 \sim 1.2$. For $\al_4 > \pi/2$ a somewhat larger part is allowed, but also showing an only slowly varying
$\al_4$~limit, around $\al_4 \sim 2.0$. 

\smallskip
Overall, the four-top analysis at the HL-LHC with an integrated luminosity of 3~ab$^{-1}$ per detector
will allow to distinguish the 2HDMS from the N2HDM in some parts of the parameter space.
In general lower $\tb$ values lead to tighter $\al_4$ limits, i.e.\ closer to $\pi/2$, following the dependences of the cross
sections on the mixing angles, see \refeq{eq:si4t-scaling}. Also lower $m_{a_S}$ leads in general to tighter bounds. 
On the other hand, larger $\tb$, or opening up the decay channels such as $a_D \to a_S h_i$ lead to a reduction of the discrimination
power of the four-top channel. In this case either direct searches for the singlet-like $\mathcal{CP}$-odd Higgs boson should be refined,
or di-Higgs measurements at future $e^+e^-$ colliders could possibly lead to a distinction. The latter possibility will be 
discussed in the next section.

%% file: section4.tex
\newpage

\FloatBarrier
\section{Prospects at future \boldmath{$e^+e^-$} colliders}
\label{sec:epem}

While in the previous section we analyzed the prospects to disentangle the N2HDM from the 2HDMS, it is clear that a
lepton colliders offer a much cleaner experimental environment, optimal for Higgs precision measurements. 
A proposal of a next generation $e^+e^-$ collider is the International Linear Collider (ILC) \cite{aryshev2023international}. 
The design provides large enough center-of-mass energies and a high integrated luminosity well suited for precision Higgs physics. 
Other proposed lepton colliders are CLIC~\cite{Roloff_2020}, the LCF~\cite{LinearCollider:2025lya}, the CEPC~\cite{thecepcstudygroup2023cepc} 
or the FCC-ee~\cite{Benedikt:2651299}. 
However, only linear $e^+e^-$ colliders can reach energies sufficient for di-Higgs production.

In more detail, the ILC proposal consists of multiple different stages starting at $\sqrt{s}=\SI{250}{GeV}$ to study the process of Higgs-strahlung. 
A short running period is reserved for a dedicated study of di-top processes at $\sqrt{s}=\SI{350}{GeV}$ . Increasing the centre of mass energy further to 
$\sqrt{s}=\SI{500}{GeV}$ is well suited for BSM searches as well as di-Higgs production. The ILC leaves room for a possible extension to 
$\sqrt{s}=\SI{1}{TeV}$ in the future. A summary of the planned stages of the ILC can be found in \refta{tab:ILC_stages} with the respective integrated luminosities.
Additionally, the ILC concept offers highly polarised electron and positron beams \cite{Moortgat_Pick_2008} with a longitudinal polarisation of 
$\pm 80\%$ and   $\pm 30\%$, respectively. This offers a vast amount of physics applications like rate enhancement, background control, 
access to the chiral structure and many more. For this study, we take advantage of the enhancement of the Higgs production rate at $\sqrt{s}=\SI{500}{GeV}$. 

\begin{table}[htb!]
    \centering
    \renewcommand*{\arraystretch}{1.4}
    \begin{tabular}{c|c|c|c|c}
       $\sqrt{s}$  [GeV]& 250 & 350 & 500 & 1000\\ \hline
       duration [yr] & 11 & 0.75 & 9 & 10 \\
       $\ld_{\rm int}$ [$\rm ab^{-1}$] & 2 & 0.2 & 4 & 8
    \end{tabular}
    \caption{Summary of the planned ILC stages \cite{aryshev2023international}.} \label{tab:ILC_stages}
\end{table}

\smallskip
In this section we analyze whether the N2HDM and the 2HDMS, describing the $95 \gev$ excesses, 
can be distinguished via di-Higgs production in future
$e^+e^-$ colliders. Concretely, we concentrate on the proposed ILC setup with $\sqrt{s}=\SI{500}{GeV}$ (ILC500) and polarized beams with $P_{e^-/e^+} = -80\%/+30\%$.  
The N2HDM/2HDMS scenarios investigated here are the same ones as for the HL-LHC in the previous section.


\subsection{Di-Higgs Production at the ILC}

\begin{figure}[htb!]
    \begin{center}
        \includegraphics[width=0.6\textwidth]{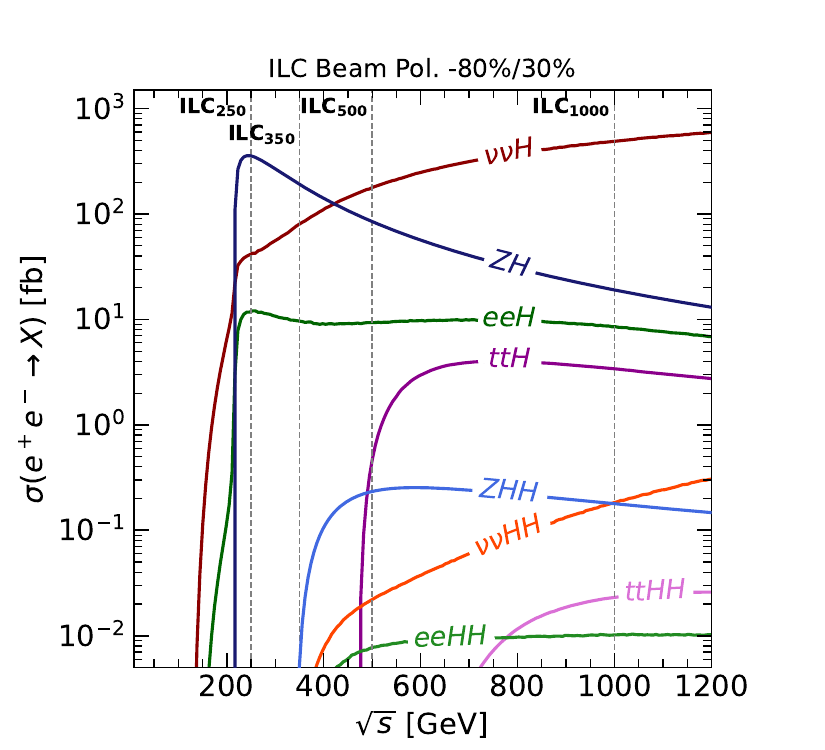}
        \caption{SM single and di-Higgs production cross sections at $e^+e^-$ colliders as a function of $\sqrt{s} = 500 \gev$
        and $P_{e^-/e^+} = -80\%/+30\%$.} 
        \label{fig:DiHiggsILC}
    \end{center}
\end{figure}

At lepton colliders like the ILC, the only available Higgs production channels are Higgs strahlung, vector boson fusion or top pair  production. Consequently, the possible di-Higgs production channels in the SM are:
\begin{align}
    e^+e^-&\to ZHH \qquad & \hspace{-10cm}\text{di-Higgs strahlung},\\
    e^+e^-&\to \bar\nu_e\nu_eHH\qquad &\text{vector boson fusion (via $WW$)},\\
    e^+e^-&\to e^+e^-HH \qquad &\text{vector boson fusion (via $ZZ$)},\\
    e^+e^-&\to \bar ttHH \qquad &\text{top pair associated production}.
\end{align}
In \reffi{fig:DiHiggsILC} we show the various SM single and di-Higgs production cross sections at $e^+e^-$ colliders 
as a function of $\sqrt{s}$ for a beam polarization of $P_{e^-/e^+} = -80\%/+30\%$. 
All $e^+e^-$ production cross sections in this section have been evaluated with {\tt MadGraph5} \cite{Frederix_2018,Alwall_2014,frixione2021lepton,Franzosi_2020}
(more details are described in the following subsections). Indicated by vertical dotted lines
are the possible energy stages of the ILC, see \refse{sec:intro}. As can be seen, at the ILC500
the dominant channel for Higgs pair production is di-Higgs strahlung, where we find
$\sigma_{\mathrm{SM}}(e^+e^- \to ZHH) = 0.2338\,\mathrm{fb}$.


\subsection{Di-Higgs final states at  \boldmath{$\sqrt{s}=500$}\,GeV}
\label{sec:finalstates}

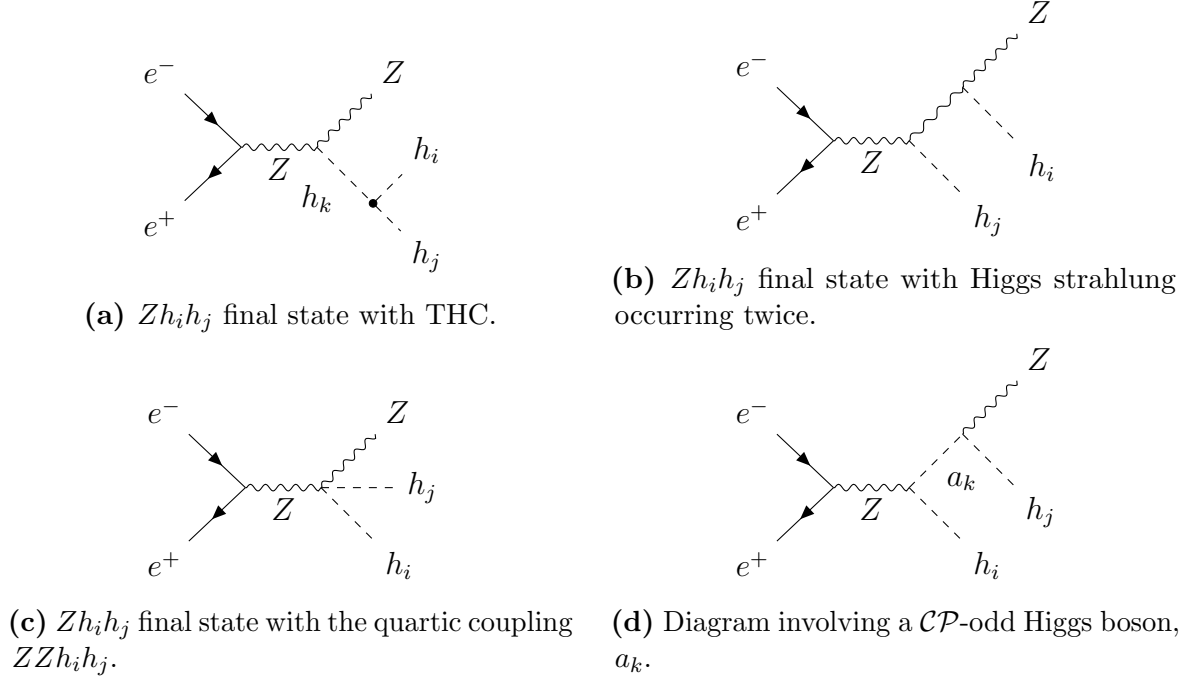
\begin{figure}[htb!]
    \centering
    \begin{subfigure}[b]{0.45\textwidth}
        \centering
        \begin{tikzpicture}
            \begin{feynman}[small]
                \vertex (l1);
                \vertex[below left=of l1] (i1) {\(e^+\)};
                \vertex[above left=of l1] (i2) {\(e^-\)};
                \vertex[right =of l1] (i3);
                \vertex[above right =of i3] (i4) {\(Z\)};
                \vertex[below right =of i3,dot,black] (i5) {};
                \vertex[above right =of i5] (i6) {\(h_i\)};
                \vertex[below right =of i5] (i7) {\(h_j\)};
        
                \diagram* {(i2) --[fermion] (l1) --[fermion] (i1),
                          (l1)   --[boson, edge label'=\(Z\)] (i3)
                          --[boson] (i4),
                          (i3) --[scalar, edge label'=\(h_k\)] (i5), 
                          (i5) --[scalar] (i6),
                          (i5) --[scalar] (i7)
                
                };
            \end{feynman}
            \end{tikzpicture}
        \caption{$Zh_ih_j$ final state with THC.}  \label{fig:model_thc}
    \end{subfigure}
    \quad
    \begin{subfigure}[b]{0.45\textwidth}
        \centering
        \begin{tikzpicture}
            \begin{feynman}[small]
                \vertex (l1);
                \vertex[below left=of l1] (i1) {\(e^+\)};
                \vertex[above left=of l1] (i2) {\(e^-\)};
                \vertex[right =of l1] (i3);
                \vertex[above right =of i3] (i4);
                \vertex[above right =of i4] (i8) {\(Z\)};
                \vertex[below right =of i3] (i5) {\(h_j\)};
                \vertex[above right =of i5] (i6) {\(h_i\)};

                \diagram* {(i2) --[fermion] (l1) --[fermion] (i1),
                          (l1)   --[boson, edge label'=\(Z\)] (i3)
                          --[boson] (i4) --[boson] (i8),
                          (i3) --[scalar] (i5), 
                          (i4) --[scalar] (i6)
                };
            \end{feynman}
            \end{tikzpicture}
            \caption{$Zh_ih_j$ final state with Higgs strahlung occurring twice.} \label{fig:DHS_THC}
    \end{subfigure}
    \begin{subfigure}[b]{0.45\textwidth}
        \centering
        \begin{tikzpicture}
            \begin{feynman}[small]
                \vertex (l1);
                \vertex[below left=of l1] (i1) {\(e^+\)};
                \vertex[above left=of l1] (i2) {\(e^-\)};
                \vertex[right =of l1] (i3);
                \vertex[above right =of i3] (i4){\(Z\)};
                \vertex[below right =of i3] (i5) {\(h_i\)};
                \vertex[right =of i3] (i6) {\(h_j\)};

                \diagram* {(i2) --[fermion] (l1) --[fermion] (i1),
                          (l1)   --[boson, edge label'=\(Z\)] (i3)
                          --[boson] (i4),
                          (i3) --[scalar] (i5), 
                          (i3) --[scalar] (i6)
                
                };
            \end{feynman}
            \end{tikzpicture}
            \caption{$Zh_ih_j$ final state with the quartic coupling $ZZh_ih_j$.}
    \end{subfigure}
    \quad
    \begin{subfigure}[b]{0.45\textwidth}
        \centering
        \begin{tikzpicture}
            \begin{feynman}[small]
                \vertex (l1);
                \vertex[below left=of l1] (i1) {\(e^+\)};
                \vertex[above left=of l1] (i2) {\(e^-\)};
                \vertex[right =of l1] (i3);
                \vertex[above right =of i3] (i4);
                \vertex[above right =of i4] (i8) {\(Z\)};
                \vertex[below right =of i3] (i5) {\(h_i\)};
                \vertex[above right =of i5] (i6) {\(h_j\)};

                \diagram* {(i2) --[fermion] (l1) --[fermion] (i1),
                          (l1)   --[boson, edge label'=\(Z\)] (i3)
                          --[scalar, edge label'=\(a_k\)] (i4) --[boson] (i8),
                          (i3) --[scalar] (i5), 
                          (i4) --[scalar] (i6)
                };
            \end{feynman}
            \end{tikzpicture}
            \caption{Diagram involving a $\cp$-odd Higgs boson, $a_k$.}\label{fig:model_bg}
    \end{subfigure}
    \caption{Diagrams of the process $e^+e^-\to Zh_ih_j$ in the N2HDM and 2HDMS.}
    \label{fig:doubleHiggsModels}
\end{figure}

In this section we discuss the various di-Higgs production channels relevant in the scenarios with a \hnf, i.e.\ describing
the excesses at 95~GeV, see \refse{sec:intro}. 
In the N2HDM and 2HDMS models, we have three $\cp$-even Higgs bosons, $h_{1,2,3}$. 
Consequently, we have the option to investigate different final states ($Zh_ih_j$). 
(Here and in the following we use the notation $h_1 = \hnf$, $h_2 = \hotf$.)
The contributing diagrams are shown in \reffi{fig:doubleHiggsModels}.
The diagrams in \reffi{fig:doubleHiggsModels}(a) contain the THCs and called signal diagrams, whereas the other three types in 
\reffis{fig:doubleHiggsModels}(b)-(d) are the background diagrams. They involve the radiation of two individual Higgs bosons (b), 
a quartic $ZZ$-di-Higgs vertex (c), or the intermediate state of a $\cp$-odd Higgs (d).

The calculation of the cross-sections $\sigma(e^-e^+\to Zh_ih_j)$ is done employing 
\verb|MadGraph5_aMC@NLO_v3.5.0| (MG5)~\cite{Frederix_2018,Alwall_2014,frixione2021lepton,Franzosi_2020}. 
The necessary parameter dependent inputs (e.g., masses and mixing matrices) are provided by \verb|SPheno-4.0.5|~\cite{Porod_2003,Porod_2012}.
The respective model files were generated using \verb|SARAH-4.14.3|~\cite{Staub_2014,staub2012sarah,Staub_2011}. 

To examine the impact of the $\cp$-odd mixing angle $\alpha_4$ on the distinction of the two models (the ``$\al_4$ limits''),
we study the benchmark point given in \refta{tab:BP3}. 
As in our benchmark scenario the $h_3$ is too heavy to be produced at the ILC500, we have access to the final states 
involving the lighter Higgs bosons thus $Zh_{125}h_{125}$, $Zh_{95}h_{125}$ and $Zh_{95}h_{95}$. 
The $\al_4$~limit, see the discussion in \refse{sec:benchmarkscenario}, is found where the 2HDMS prediction for di-Higgs production is excluded
by $2\,\sig$ experimental precision as found in the corresponding N2HDM limit.
To estimate the uncertainty of a potential observation of the various Di-Higgs production channels, we use the results from 
the SM analysis of the International Large Detector (ILD) simulation~\cite{Duerig:310520}. These are rescaled with the number
of events according to the modified cross-section, with the assumption that the background rate remains constant in the BSM scenario. 
Details on the computation of the uncertainties are given in App.~\ref{sec:unc_estimate}. 
In order to demonstrate our method, we we start
from the benchmark point (see \refta{tab:BP3}) and vary $\alpha_4$ in the 2HDMS and compute the di-Higgs cross-sections 
for different $\cp$-odd singlet masses $m_{a_1}={575,550,525} \gev$ and $m_{a_2}=m_A= 600 \gev$.
We first investigate the different di-Higgs production cross sections individually and then perform a statistical combination.


\subsection*{\boldmath{$Zh_{125}h_{125}$}}

We start with the SM-like final state involving two Higgs bosons with a mass of \SI{125}{GeV}, thus 
\begin{equation}
    e^+e^-\to Zh_2h_2.
\end{equation}
The N2HDM cross section is evaluated as $\sig_{h_2h_2}^\NTHDM = 0.162\,\fb$ with a $1\,\sig$ uncertainty of $\pm 0.028\,\fb$, as obtained
from the rescaling. 
The diagrams involving a THC vertex for this process can be mediated via all possible Higgses that couple to the $Z$ boson:
\begin{align}
    \lambda_{h_2h_2h_2}:&\qquad e^+e^- \to Zh_2 \to Zh_2h_2 \qquad (\text{reduced by } c_{h_2VV}),\\
    \lambda_{h_1h_2h_2}:&\qquad e^+e^- \to Zh_1 \to Zh_2h_2 \qquad (\text{suppressed by } c_{h_1VV}),\\
    \lambda_{h_3h_2h_2}:&\qquad e^+e^- \to Zh_3 \to Zh_2h_2 \qquad (\text{negligible}).
\end{align}
In our the benchmark scenario we find $c_{h_3VV}=0.05$, the impact of $\lambda_{h_3h_2h_2}$ is negligible for this study, 
and hence not further discussed even though the cross-section computation within \Code{MadGraph5} considers all possible diagrams. 
\begin{figure}[h!]
    \centering
         \includegraphics[width=0.49\textwidth]{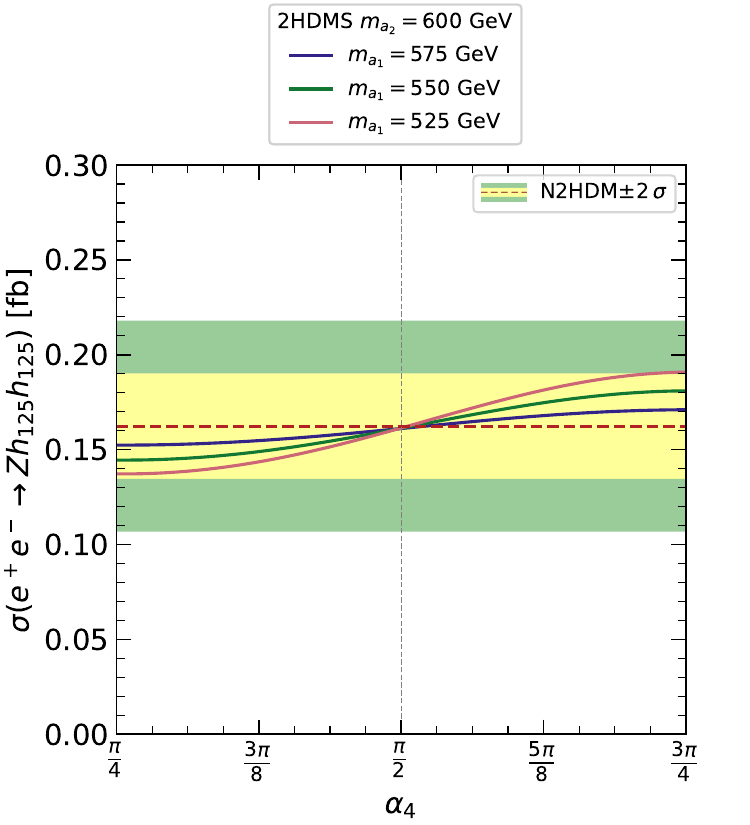}
        \caption{Cross-section for $e^+e^-\to Zh_2h_2$ in the N2HDM and 2HDMS. The 1 and $2\,\sigma$ regions of the N2HDM are colored yellow and green respectively.}
        \label{fig:Zh2h2}
\end{figure}
In \reffi{fig:Zh2h2} we show the cross-sections for the di-Higgs production $Zh_2h_2$ in the 2HDMS and N2HDM including the uncertainty bands of the N2HDM prediction according to our uncertainty estimate. 
For this studied parameter point, the models differ by less than $1\,\sigma$, which is in total agreement with the previous analytical discussion in section \ref{sec:diff}. As the leading contribution comes from the trilinear coupling $\lambda_{h_2h_2h_2}$, it has the smallest differences from all possible THCs, reflected in a weak discrimination power of this channel. 
\FloatBarrier
\subsection*{\boldmath{$Zh_{95}h_{125}$}}

\begin{figure}[htb!]
    \centering
        \includegraphics[width=0.49\textwidth]{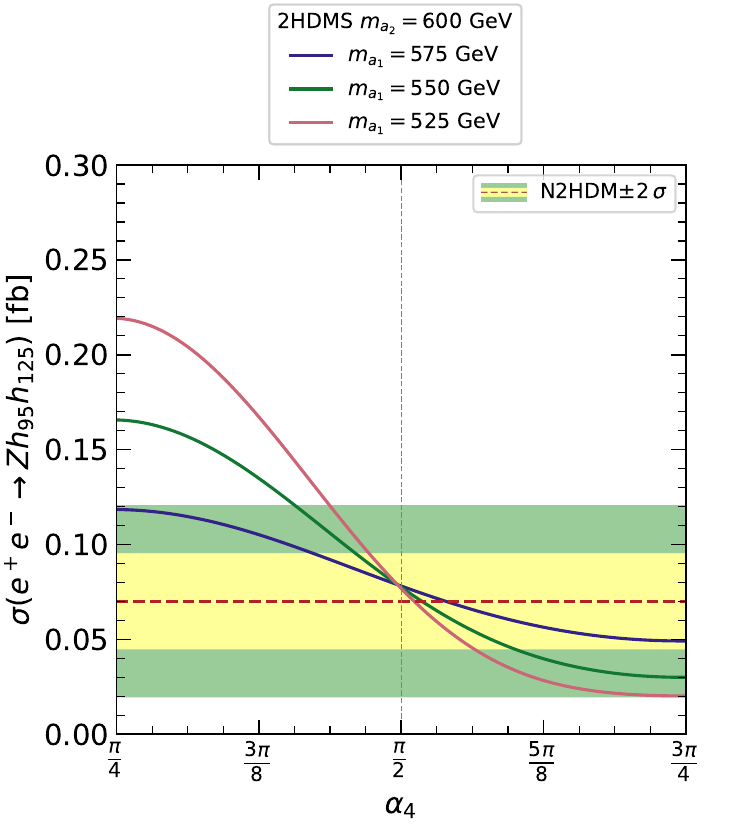}
        \caption{Cross-section for $e^+e^-\to Zh_1h_2$ in the N2HDM and 2HDMS. The 1 and $2\,\sigma$ regions of the N2HDM are colored yellow and green respectively.}
        \label{fig:Zh1h2}
\end{figure}

The next possible option is di-Higgs production of one \SI{125}{GeV} Higgs with one \SI{95}{GeV} Higgs, $Zh_1h_2$. 
The N2HDM cross section is evaluated as $\sig_{h_1h_2}^\NTHDM = 0.07\,\fb$ with a $1\,\sig$ uncertainty of $\pm 0.025\,\fb$, as obtained
from the rescaling. 
For the uncertainty estimation, we take the simplifying assumption, that the \SI{95}{GeV} Higgs boson can be reconstructed in the 
detector with a similar precision as the SM Higgs boson.
The contributing trilinear couplings are then 
\begin{align}
    \lambda_{h_2h_1h_2}:&\qquad e^+e^- \to Zh_2 \to Zh_1h_2 \qquad (\text{reduced by } c_{h_2VV}),\\
    \lambda_{h_1h_1h_2}:&\qquad e^+e^- \to Zh_1 \to Zh_1h_2 \qquad (\text{suppressed by } c_{h_1VV}),\\
    \lambda_{h_3h_1h_2}:&\qquad e^+e^- \to Zh_3 \to Zh_1h_2 \qquad (\text{negligible}).
\end{align}
The computed cross-section ${\sigma(e^+e^-\to Zh_1h_2)}$ is shown in \reffi{fig:Zh1h2}.
First, we observe that in the N2HDM limit of the 2HDMS at $\alpha_4=\frac{\pi}{2}$ the predictions for N2HDM and 2HDMS do not 
match exactly, which can be understood as follows. The discussion in \refse{sec:diff} is only valid in an approximation for degenerate masses of $h_1$ and $h_2$, where the propagators are equal. Our scenarios consider a lighter $h_1$, hence the approximation criteria are not fully satisfied. This yields a non vanishing contribution from the last term of eq. (\ref{eq:thc_diff})  $\propto \frac{1}{v_S}(m_{a_1}^2\sin^2\alpha_4+m_{a_2}^2\cos^2\alpha_4)$. Hence in the N2HDM limit at $\alpha_4\rightarrow\frac{\pi}{2}$ the cross-sections for the models can deviate by a small margin, as manifests itself 
in \reffi{fig:Zh1h2}. 
In the area $\frac{\pi}{4}<\alpha_4<\frac{\pi}{2}$ pronounced difference between the models occur for lower $m_{a_1}$ masses.
There it is possible to determine a limit on $\alpha_4$ where one can distinguish both models on a $2\,\sigma$ level. 
For the opposite side, $\frac{\pi}{2}<\alpha_4<\frac{3\pi}{2}$, we observe only weak deviations. Here the THCs are small, 
hence the total cross-sections is dominated by background processes (see \reffi{fig:doubleHiggsModels} b,c,d), and thus the
differences are less pronounced.
In particular, for $m_{a_1}=\SI{575}{GeV}$ the mass difference to $m_{a_2}$ is not large enough to produce significant differences.


\subsection*{\boldmath{$Zh_{95}h_{95}$}}

\begin{figure}[htb!]
    \centering
        \includegraphics[width=0.49\textwidth]{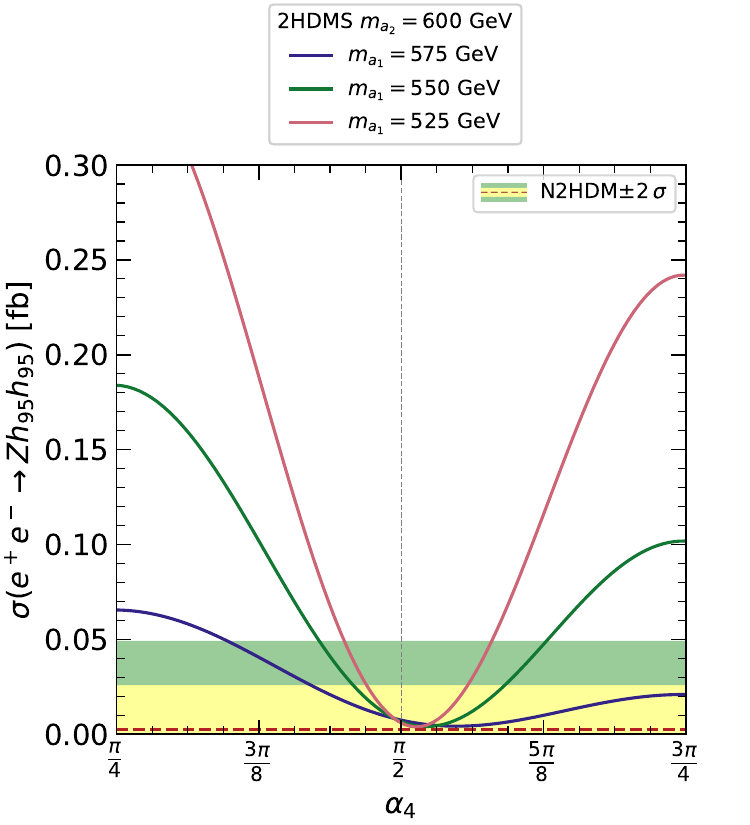}
        \caption{Cross-section for $e^+e^-\to Zh_1h_1$ in the N2HDM and 2HDMS. The 1 and $2\,\sigma$ regions of the N2HDM are colored yellow and green respectively.}
        \label{fig:Zh1h1}
\end{figure}

The last possible option at $\sqrt{s}=\SI{500}{GeV}$ is di-Higgs Production of two light Higgs bosons, $Zh_1h_1$.
The N2HDM cross section is evaluated as $\sig_{h_1h_1}^\NTHDM = 0.0025\,\fb$ with a $1\,\sig$ uncertainty of $\pm 0.0232\,\fb$, as obtained
from the rescaling. 
The involved trilinear couplings are then as follows 
\begin{align}
    \lambda_{h_2h_1h_1}:&\qquad e^+e^- \to Zh_2 \to Zh_1h_1 \qquad (\text{reduced by } c_{h_2VV}),\\
    \lambda_{h_1h_1h_1}:&\qquad e^+e^- \to Zh_1 \to Zh_1h_1 \qquad (\text{suppressed by } c_{h_1VV}),\\
    \lambda_{h_3h_1h_1}:&\qquad e^+e^- \to Zh_3 \to Zh_1h_1 \qquad (\text{negligible}).
\end{align}
The results for the N2HDM and 2HDMS cross section predictions are shown in \reffi{fig:Zh1h1}.
We observe that in the N2HDM limit the three 2HDMS cross section do not agree with each other, nor the N2HDM cross section.
This follows the same argument as discussed for $Zh_{95}h_{125}$.
In the N2HDM limit for $\alpha_4=\frac{\pi}{2}$ the trilinear higgs couplings follow
\begin{equation}
    \lambda^{\rm 2HDMS}_{h_ih_jh_k}-\lambda^{\rm N2HDM}_{h_ih_jh_k}\propto \frac{m_{a_1}^2}{v_S}.
\end{equation}
Consequently, the trilinear Higgs couplings themselves depend on the mass of the singlet-like $\mathcal{CP}$-odd Higgs $a_1$, which leads to slight variations in the cross sections.
Additionally, we observe an almost vanishing cross-section for the N2HDM. 
On the other hand, in the 2HDMS the additional terms (as detailed in \refeq{eq:thc_diff}) can lead to large differences 
in the di-Higgs cross-sections 
thus $2\,\sigma$ limits on $\alpha_4$ close to the N2HDM limit of the 2HDMS are found in all $\alpha_4$ regions 
for $m_{a_1}=\SI{550}{GeV}$ and $m_{a_1}=\SI{525}{GeV}$. However, as expected, in the exact N2HDM limit, $Zh_1h_1$ production 
is not capable of distinguishing the models. For $m_{a_1}=\SI{575}{GeV}$ this process could provide $\alpha_4$ limits 
in the area  $\frac{\pi}{4}<\alpha_4<\frac{\pi}{2}$, however the opposite $\al_4$ range remains below the $1\,\sigma$ level.
Additionally, this particular process has further points of interest. Since the final state consists of two $h_1$, the background diagrams 
are suppressed by $\abs{c_{h_1VV}}^4$. Therefore, the cross-section will be dominated by the diagrams including the Higgs self couplings.
However, in the N2HDM, using equations (\ref{eq:thc_n2hdm}) and (\ref{eq:R_simpls}) with $m_{h_1}\approx m_{h_2}$ one finds, 
\begin{align}
    c_{h_1VV}\lambda_{h_1h_1h_1}&\propto \frac{3c_{h_1VV}c_{h_2VV}^3v+3c_{h_1VV}^4v_S}{vv_S}\approx\frac{3c_{h_1VV}c_{h_2VV}^3}{v_S} \\
    c_{h_2VV}\lambda_{h_2h_1h_1}&\propto \frac{-3c_{h_1VV}c_{h_2VV}^3v+3c_{h_1VV}^2c_{h_2VV}^2v_S}{vv_S}\approx-\frac{3c_{h_1VV}c_{h_2VV}^3}{v_S}
\end{align}
Consequently, the contributions from $c_{h_1VV}\lambda_{h_1h_1h_1}$ and $c_{h_2VV}\lambda_{h_2h_1h_1}$ 
have similar magnitudes but opposite signs, and then almost cancel each other
and we find $\sigma_{Zh_1h_1}^\NTHDM \sim \SI{1}{ab}$. 
On the other hand, the cross-section in the 2HDMS is primarily driven by the extra cubic terms $\propto \mu_{S_1}$ and 
$\propto \mu_{12}$ in the 2HDMS potential. As discussed in \refeq{eq:thc_diff}
for the Higgs configuration involving many singlet Higgses, these contributions are enhanced relative to other configurations.

\subsection*{Combination}
\begin{figure}[h]
    \begin{center}
        \includegraphics[width=0.49\textwidth]{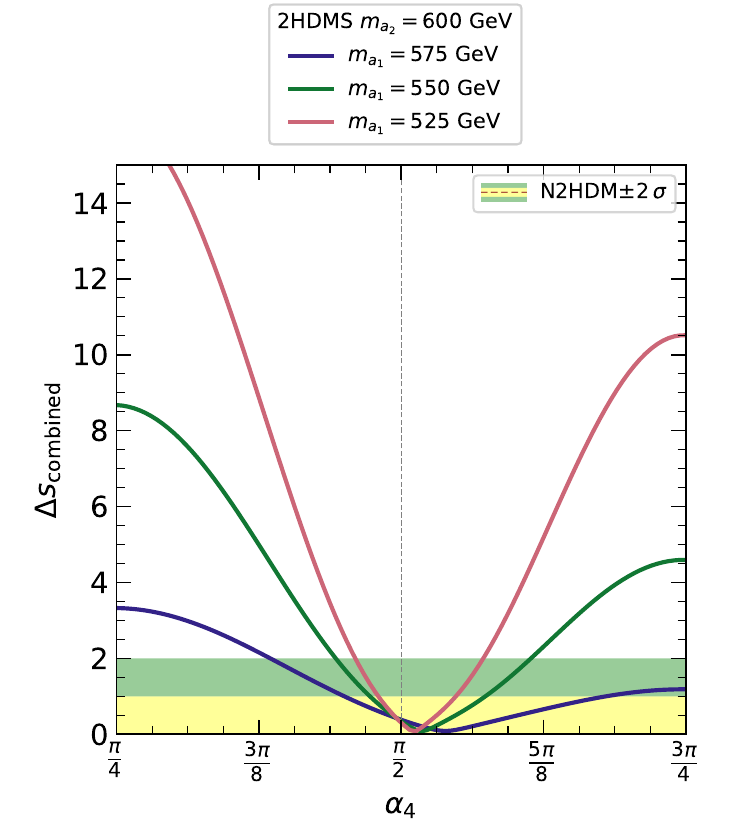}
        \caption{$\Delta s_{\rm combined}$ for different $m_{a_1}$ . The $1\,\sigma$ and $2\,\sigma$ ranges are colored in yellow and green respectively.}
        \label{fig:chi2combination}
    \end{center}
\end{figure}

To achieve maximal possible model distinction, we are interested in combining all available information, hence combining all kinematically accessible di-Higgs final states.
Following \refeq{Deltas} we compute the distinction significance between the two models 
\begin{equation}
    \Delta s^{\rm diHiggs}_{\rm comb.}(\alpha_4{, m_{a_1}})
    = \sqrt{\sum_{i,j\in\{1,2\}}\rbr{\ \frac{\sigma_{Zh_ih_j}^\THDMS(\al_4{, m_{a_1}})- \sigma_{Zh_ih_j}^\NTHDM}
    {\Delta \sigma_{Zh_ih_j}^\NTHDM}^2}}\,.
    \label{eq:chi2comb}
\end{equation}
This approach does not incorporate correlations between the final states, and considers them as independent. 
While such an assumption may be not fully justified, there are no studies of such type of correlations, 
hence we treat the channels as independent to approximate the model differences. 
For the previously determined cross-sections, the combined significances are shown in \reffi{fig:chi2combination}.
First one observes that the distinction power of the model scales strongly with the mass difference of the $\cp$-odd Higgs bosons.
Further, we observe a slight improvement of the combination compared to the $Zh_1h_1$ limits from the other two channels. However,
due to the steep slope of $\Delta s_{Zh_1h_1}$ for this parameter point this improvement is rather small. 
Only where the $Zh_1h_1$ channel does not provide a strong distinction power the addition of the other channels yield a relatively 
stronger improvement on the $\alpha_4$ limits. As an example, for $m_{a_1}=\SI{575}{GeV}$ in the area 
$\frac{\pi}{2}<\alpha_4<\frac{3\pi}{2}$ stand alone the $Zh_1h_1$ stays below the $1\,\sigma$ line. Including the other 
channels pushes the combined significance beyond the  $1\,\sigma$ level. 

\subsection{Scans for \boldmath{$\alpha_4$} limits}
\label{sec:limits}

In this section we apply the derivation of $\al_4$ limits via di-Higgs production at the ILC1000, as described above, 
to the benchmark scenarios defined in \refse{sec:benchmarkscenario}, analogously to the study in the four top case at the HL-LHC.
We study the different 2HDMS scenarios with $\Delta m_{a}=\{50,100,150,200\} \gev$ and determine the $\al_4$ limits 
(corresponding to a model distrinction at the level of $2\,\sig$) by computing the cross-section for the processes 
\begin{align}
    e^+e^-\to Zh_2h_2\,,\qquad
    e^+e^-\to Zh_1h_2\,,\qquad
    e^+e^-\to Zh_1h_1\,.
\end{align}
The uncertainty estimates are derived as outlined in App.~\ref{sec:unc_estimate}, and the significances are 
evaluated following \refeq{eq:chi2comb}. It must be kept in mind, that we are discussing an ``optimistic'' case in which the 
relevant parameters are assumed to be known with sufficiently high precision, such their uncertainties can be neglected. The reason for this
ansatz is that no evaluations of future parametric uncertainties at $e^+e^-$ colliders for the scenarios under investigation 
are available. It is to be expected that including these uncertainties will lead to some degradation of our limits and results.

\subsubsection{Benchmark Scenario 1}

\begin{figure}[htb!]
\begin{subfigure}{0.49\textwidth}
        \centering
        \includegraphics[width=\textwidth]{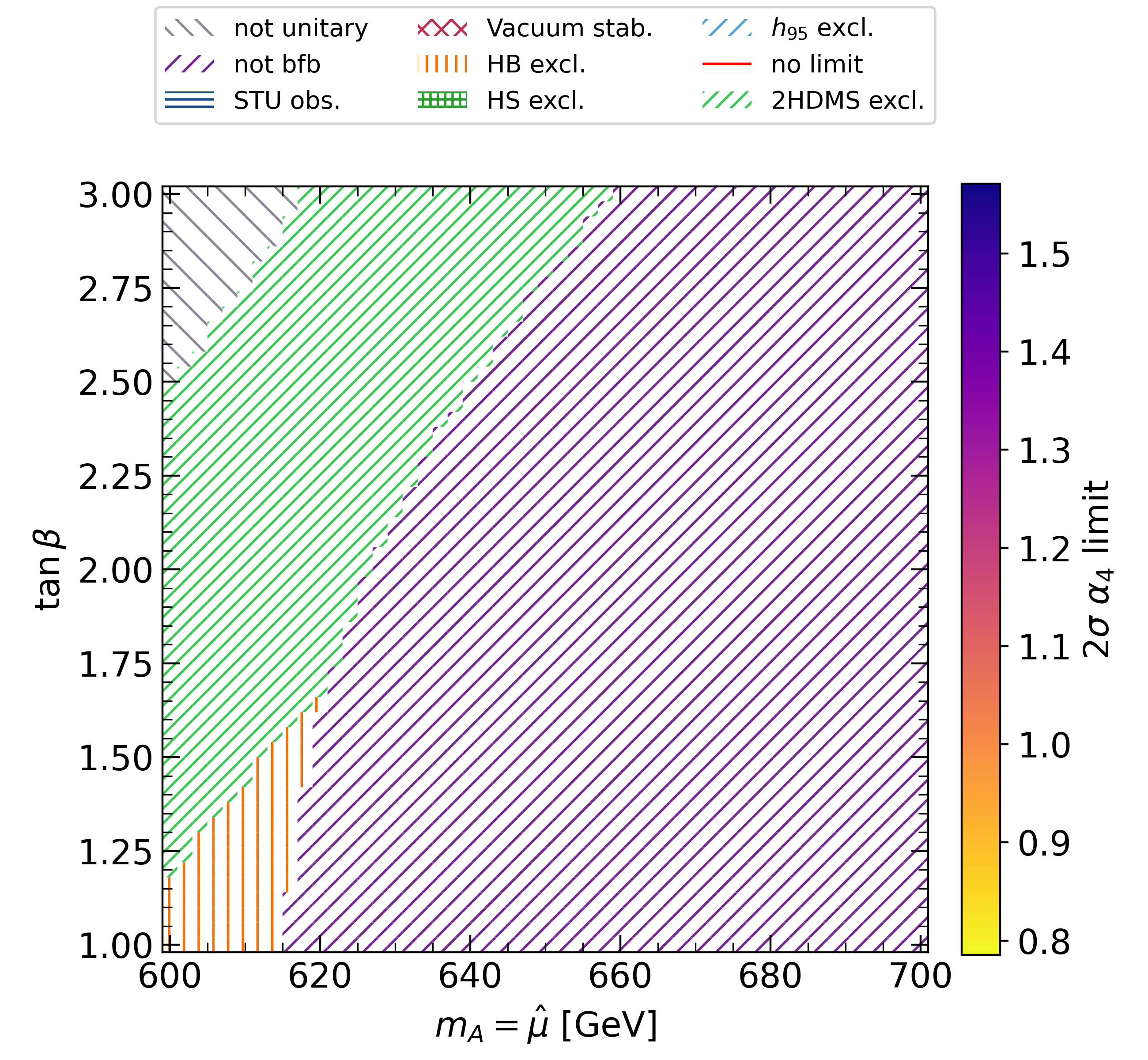}
    \caption{{$\alpha_4$ limits for $\al_4 < \pi/2$, $\De m_a = 50 \gev$.}}
\end{subfigure}
\begin{subfigure}{0.49\textwidth}
        \centering
        \includegraphics[width=\textwidth]{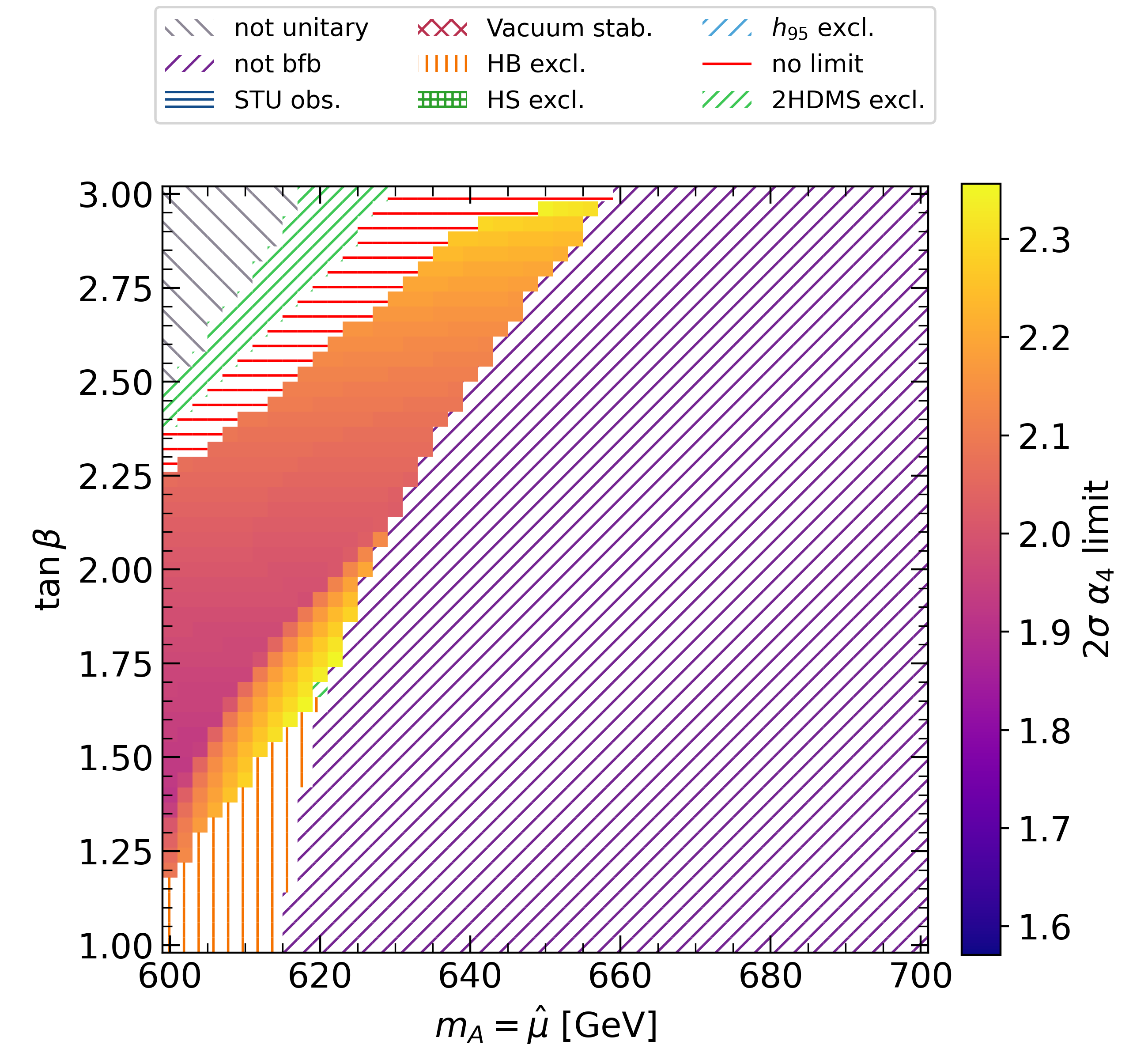}
    \caption{{$\alpha_4$ limits for $\al_4 > \pi/2$, $\De m_a = 50 \gev$.}}
\end{subfigure}
\caption{{$m_A$-$\tb$ plane for $\al_4 < (>) \pi/2$ in the left (right) plot. The hatched areas are excluded, the color code indicates
the value of the $\al_4$ limits (at $2\,\sig$), where in the horizontal red lined areas no $\al_4$ limit was found. 
The parameters are chosen according to our benchmark scenario, see \refta{tab:BP3}.
$m_{a_D}$ is varied, with $\De m_a = 50 \gev$.}}
    \label{fig:limits_C_50}
    \vspace{1em}
\end{figure}

\begin{figure}[htb!]
\begin{subfigure}{0.49\textwidth}
        \centering
        \includegraphics[width=\textwidth]{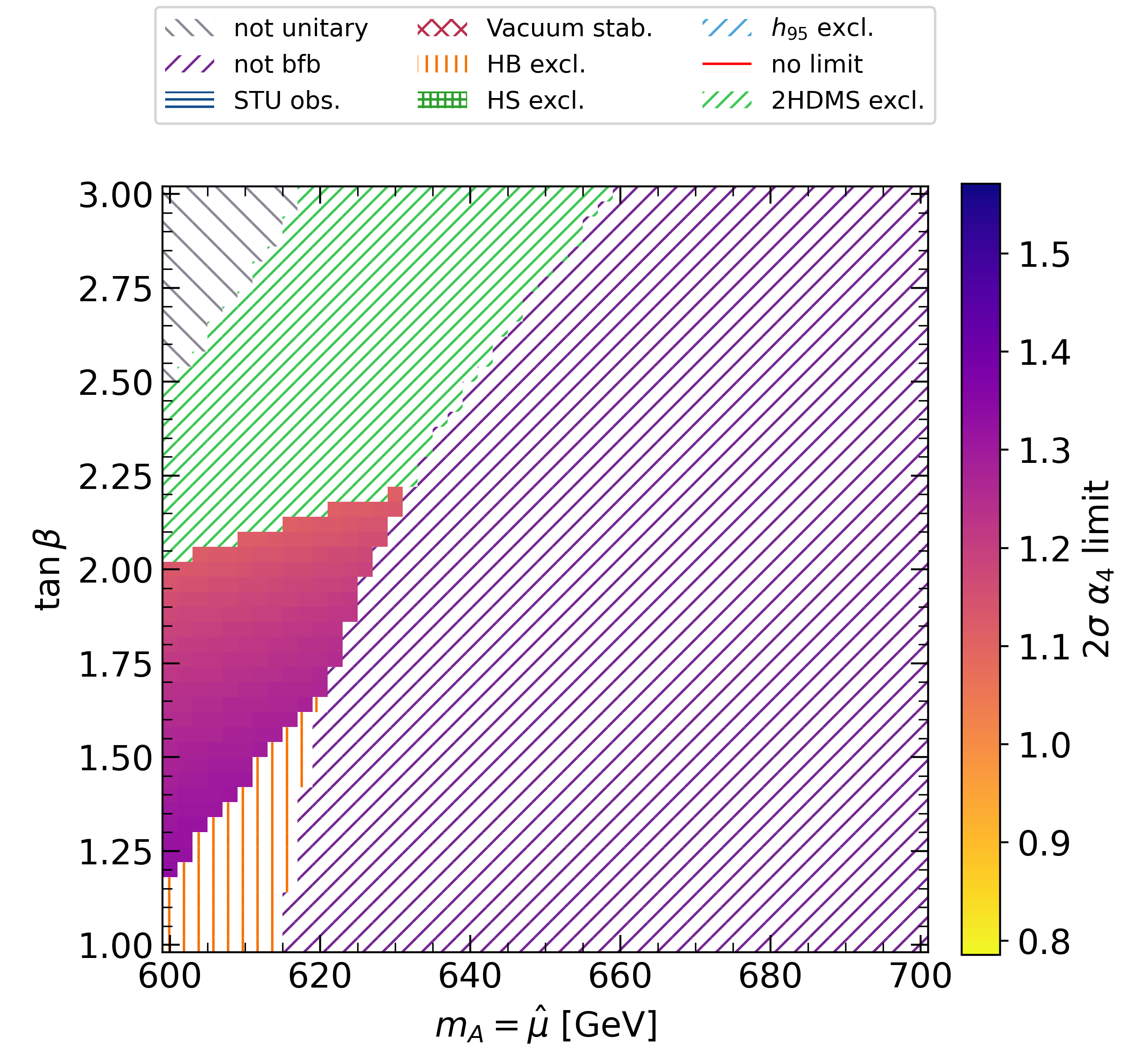}
    \caption{{$\alpha_4$ limits for $\al_4 < \pi/2$, $\De m_a = 100 \gev$.}}
\end{subfigure}
\begin{subfigure}{0.49\textwidth}
        \centering
        \includegraphics[width=\textwidth]{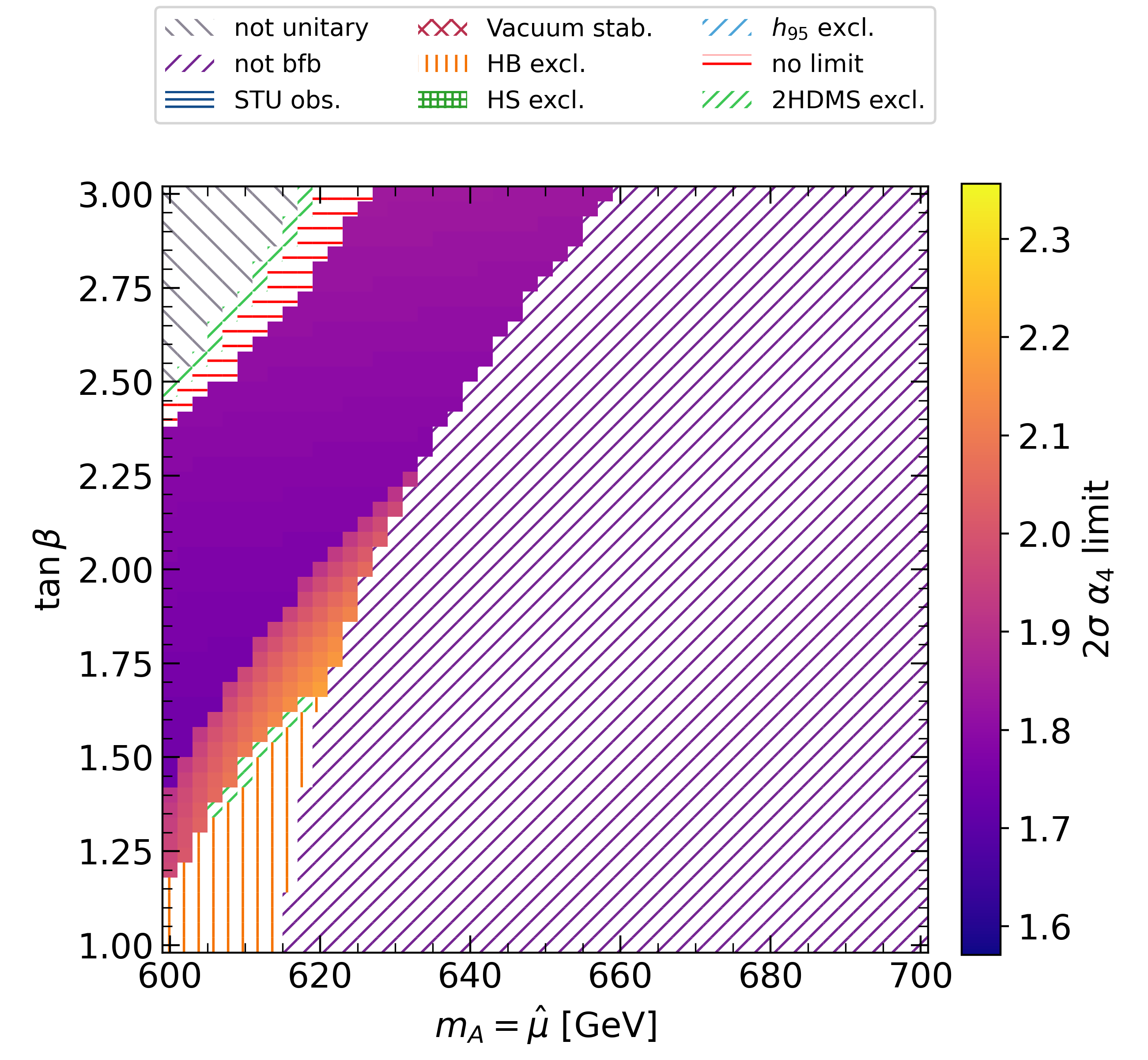}
    \caption{{$\alpha_4$ limits for $\al_4 > \pi/2$, $\De m_a = 100 \gev$.}}
\end{subfigure}
\caption{{$m_A$-$\tb$ plane for $\al_4 < (>) \pi/2$ in the left (right) plot. The hatched areas are excluded, the color code indicates
the value of the $\al_4$ limits (at $2\,\sig$), where in the horizontal red lined areas no $\al_4$ limit was found. 
The parameters are chosen according to our benchmark scenario, see \refta{tab:BP3}.
$m_{a_D}$ is varied, with $\De m_a = 100 \gev$.}}
    \label{fig:limits_C_100}
\end{figure}

\begin{figure}[htb!]
\begin{subfigure}{0.49\textwidth}
        \centering
        \includegraphics[width=\textwidth]{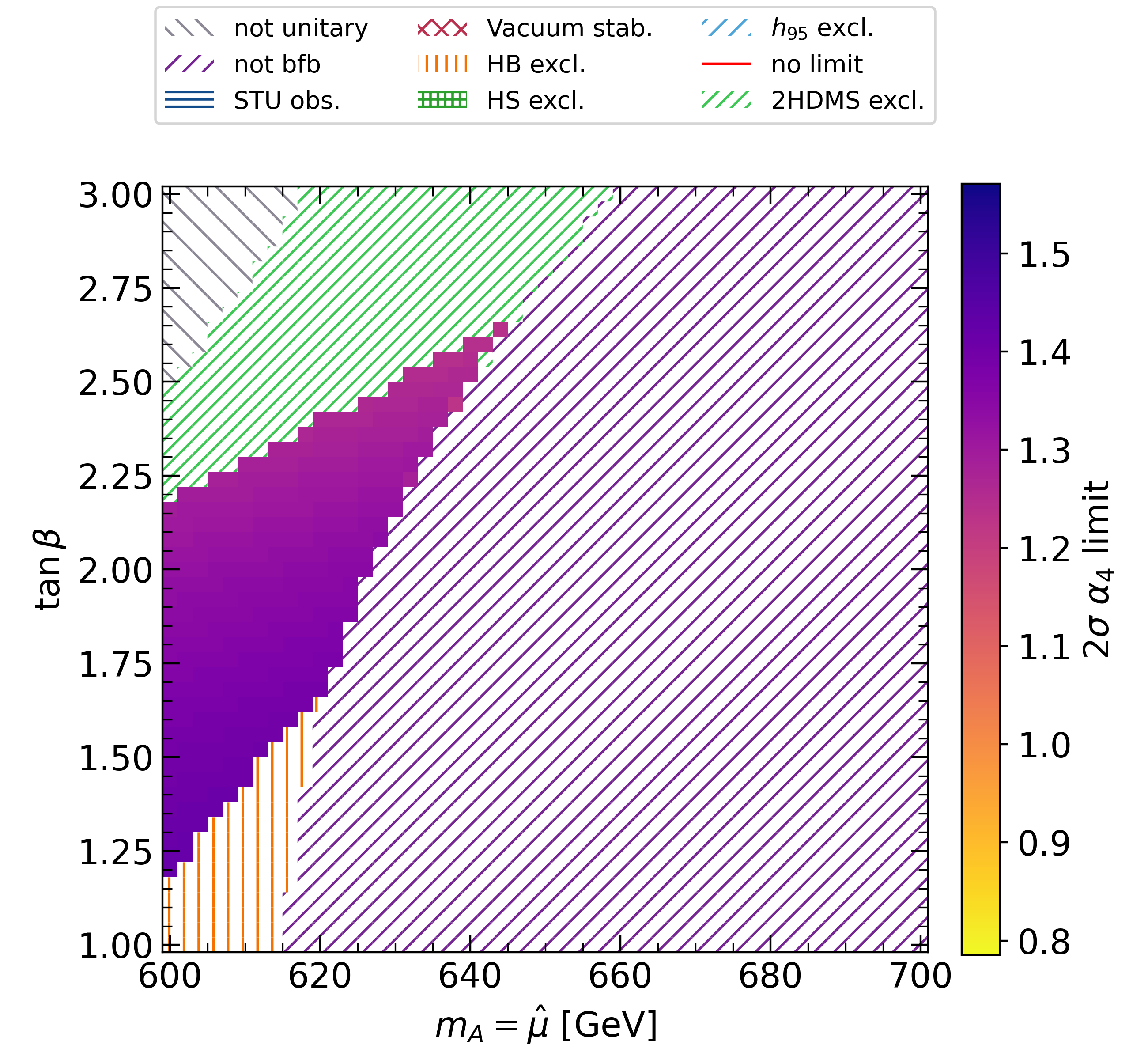}
    \caption{{$\alpha_4$ limits for $\al_4 < \pi/2$, $\De m_a = 150 \gev$.}}
\end{subfigure}
\begin{subfigure}{0.49\textwidth}
        \centering
        \includegraphics[width=\textwidth]{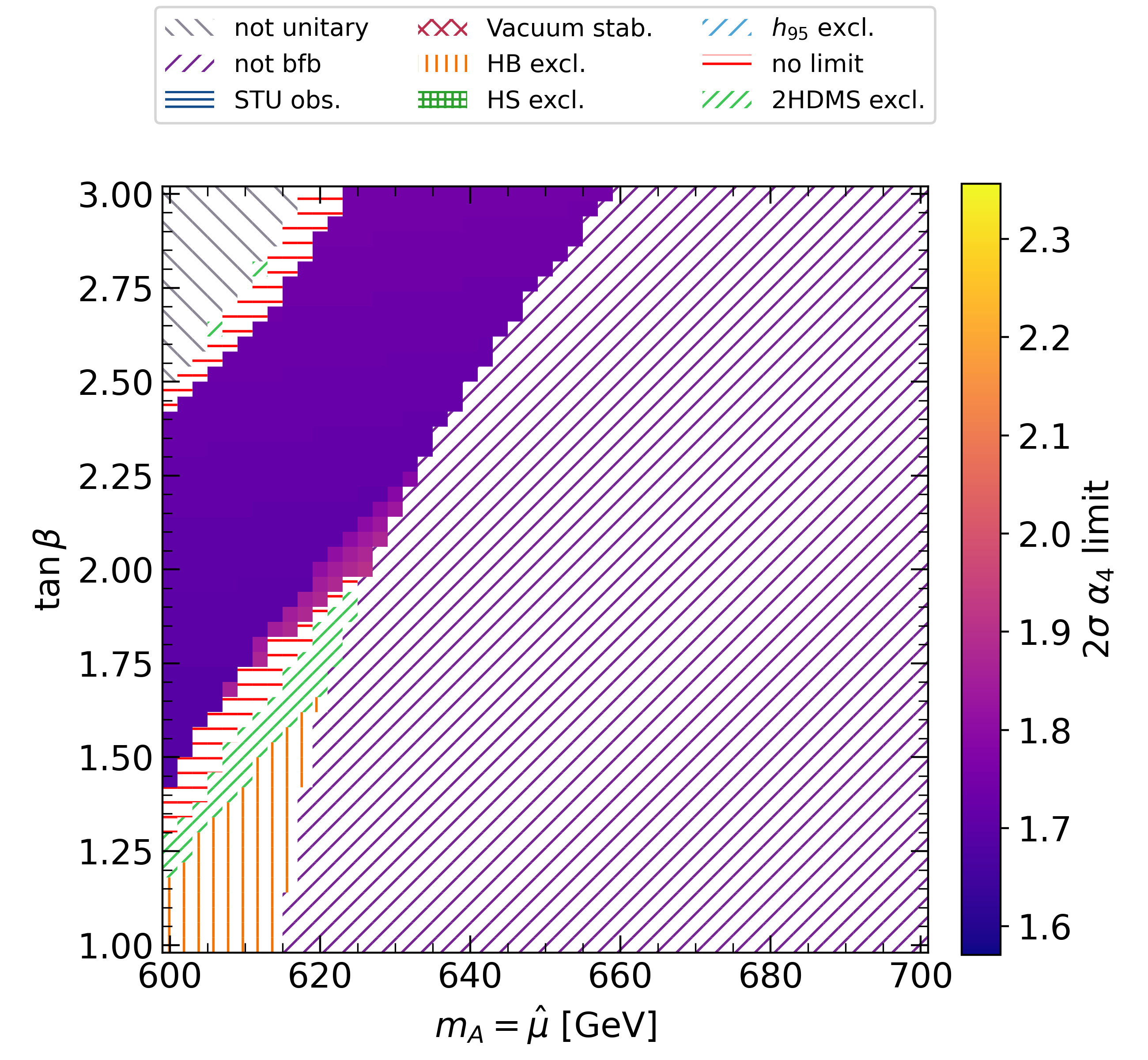}
    \caption{{$\alpha_4$ limits for $\al_4 > \pi/2$, $\De m_a = 150 \gev$.}}
\end{subfigure}
\caption{{$m_A$-$\tb$ plane for $\al_4 < (>) \pi/2$ in the left (right) plot. The hatched areas are excluded, the color code indicates
the value of the $\al_4$ limits (at $2\,\sig$), where in the horizontal red lined areas no $\al_4$ limit was found. 
The parameters are chosen according to our benchmark scenario, see \refta{tab:BP3}.
$m_{a_D}$ is varied, with $\De m_a = 150 \gev$.}}
    \label{fig:limits_C_150}
    \vspace{1em}
\end{figure}

\begin{figure}[htb!]
\begin{subfigure}{0.49\textwidth}
        \centering
        \includegraphics[width=\textwidth]{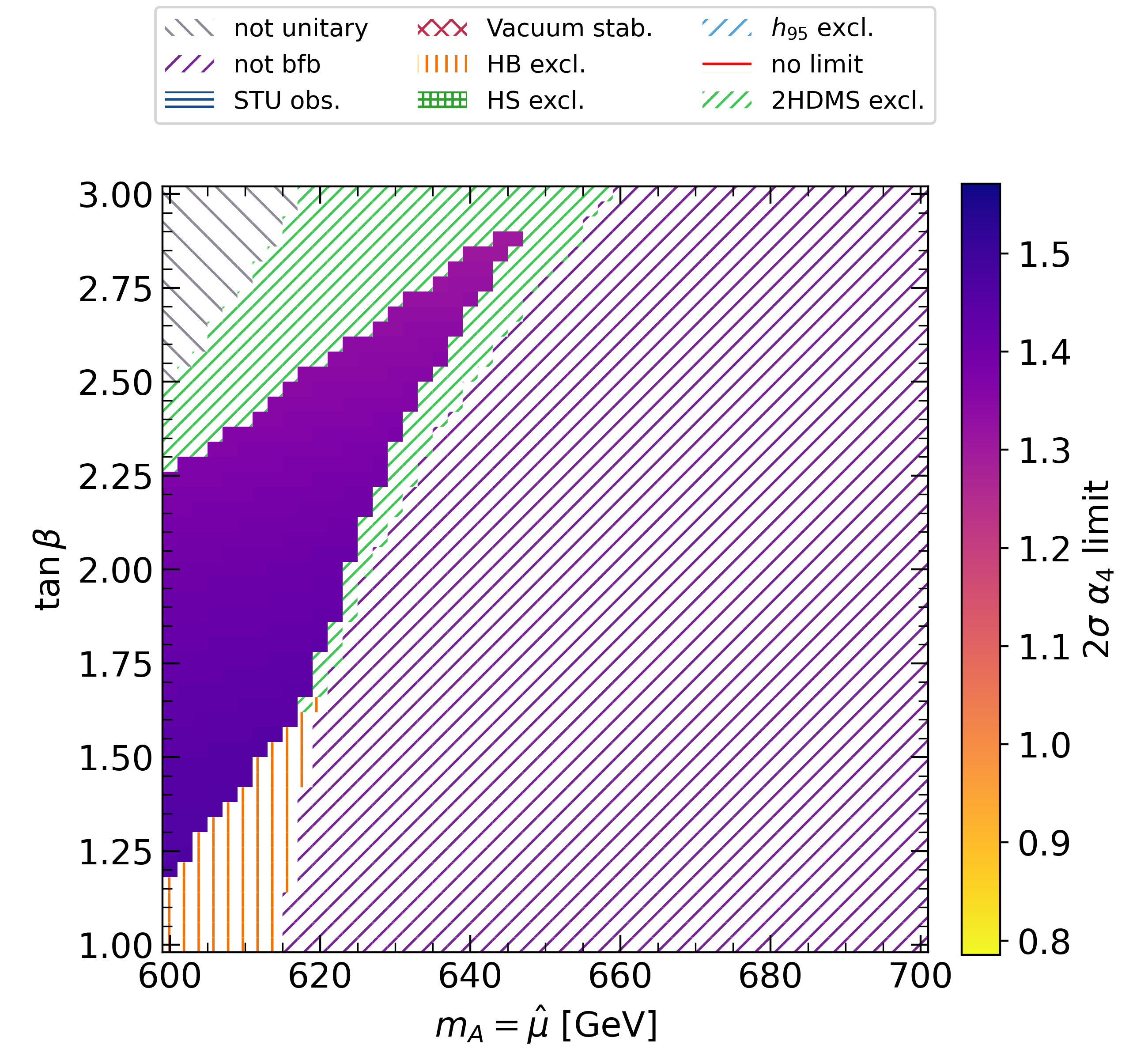}
    \caption{{$\alpha_4$ limits for $\al_4 < \pi/2$, $\De m_a = 200 \gev$.}}
\end{subfigure}
\begin{subfigure}{0.49\textwidth}
        \centering
        \includegraphics[width=\textwidth]{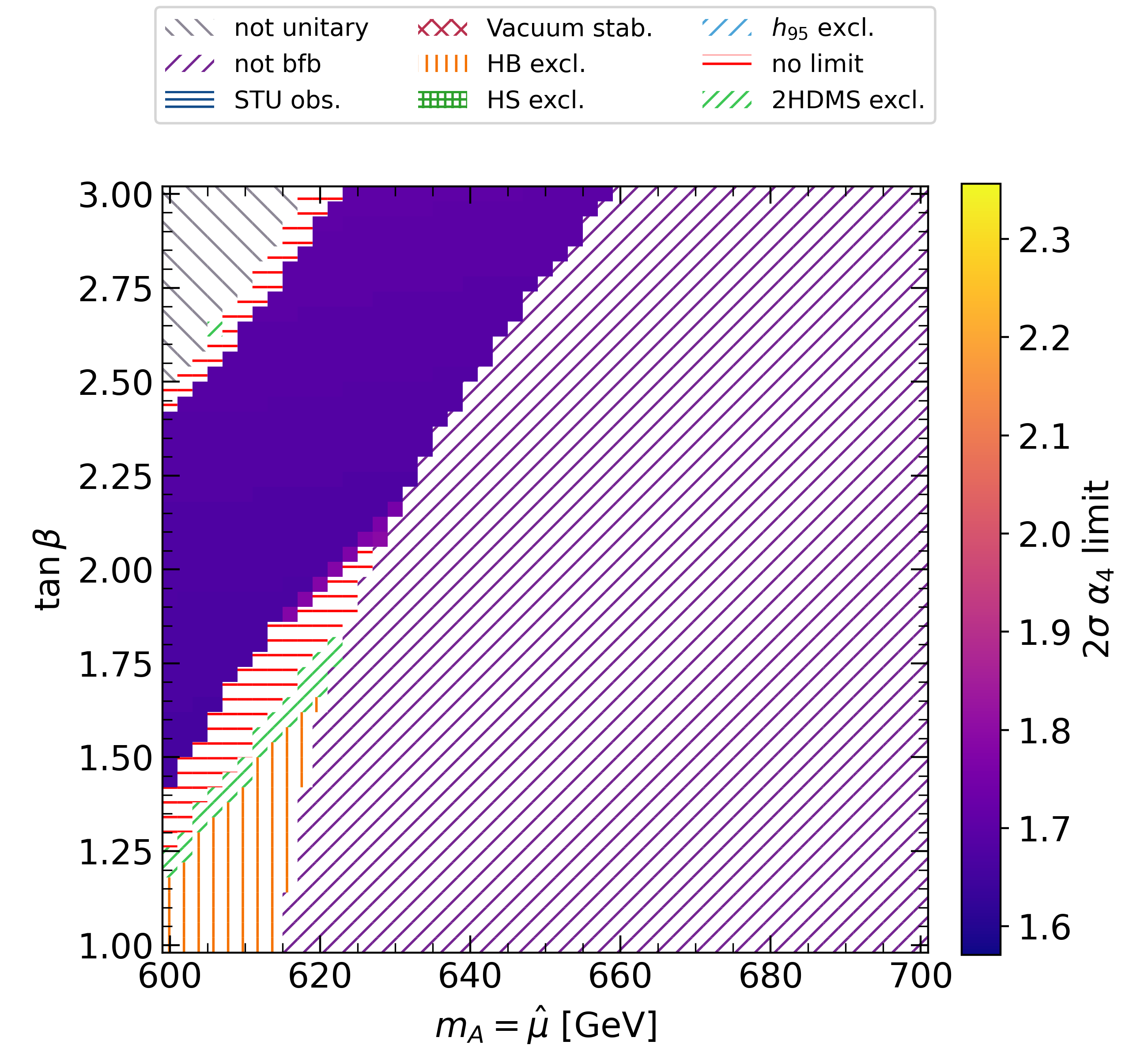}
    \caption{{$\alpha_4$ limits for $\al_4 > \pi/2$, $\De m_a = 200 \gev$.}}
\end{subfigure}
\caption{{$m_A$-$\tb$ plane for $\al_4 < (>) \pi/2$ in the left (right) plot. The hatched areas are excluded, the color code indicates
the value of the $\al_4$ limits (at $2\,\sig$), where in the horizontal red lined areas no $\al_4$ limit was found.
The parameters are chosen according to our benchmark scenario, see \refta{tab:BP3}.
$m_{a_D}$ is varied, with $\De m_a = 200 \gev$.}}
    \label{fig:limits_C_200}
\end{figure}

\reffis{fig:limits_C_50} - \ref{fig:limits_C_200} show the $\alpha_4$ limits for fixed $\cp$-odd mass differences, 
$\Delta m_a= 50 \gev$, $100 \gev$, $150 \gev$ and $200 \gev$, respectively, where the N2HDM can be distinguished from the 2HDMS at the level of $2\,\sigma$. 
The left (right) plots show the regions with $\al_4 < (>) \pi/2$ (i.e.\ "region~3 (4)"). Overall, 
we observe a similar picture as for the 4 top limits. The $\cp$ Higgs masses are the key feature to distinguish the models, 
as larger mass differences lead to limits close to the N2HDM limit at $\al_4 = \pi/2$. 

For $\Delta m_a=\SI{50}{GeV}$, \reffi{fig:limits_C_50}, one observes that the 2HDMS is fully excluded by the constraints for $\alpha_4 \leq \pi/2$.
$\al_4$ limits are found for large parts of the allowed parameter space, ranging from $\sim 1.8$ to $\sim 2.3$. However, for the larger $\tb$ values
(allowed depending on $m_A$) the 2HDMS cannot be distinguished from the N2HDM via di-Higgs production (as indicated by the horizontally red striped region).
For $m_A \lsim 630 \gev$ we observed a sudden worsening of the $\al_4$ limits for the smallest allowed $\tb$ values. The reason for this is found in
the constraints in the allowed parameter space (see \reffi{fig:2hdms_region}). In this part of the parameter space the vacuum stability constraint
in the 2HDMS excludes all values close to $\alpha_4=\pi/2$,
and only larger values are found as $\al_4$ limit, corresponding to the situation represented by the green curve in \reffi{fig:limitscenarios}. 
For $\Delta m_a=\SI{100}{GeV}$, \reffi{fig:limits_C_100}, nearly all allowed parameter regions yield an $\al_4$ limit, allowing the distinction of the
2HDMS from the N2HDM (under the given assumptions). Where limits are obtained, they are found relatively close to $\al_4 = \pi/2$, except for very low $M_A$ and
the lowest $\tb$ values.
Continuing the trend, for $\Delta m_a=\SI{150}{GeV}$, \reffi{fig:limits_C_150}, and $\Delta m_a=\SI{200}{GeV}$, \reffi{fig:limits_C_200},
exhibit an even strong possibility for model distinction, with $\al_4$ limits very close to $\pi/2$ over nearly the total allowed parameter space,
both for $\al_4 < \pi/2$ and for $\al_4 > \pi/2$. Only slim slivers at the border of the allowed parameter space do not yield any $\al_4$ limit.


\FloatBarrier
\subsubsection{Limits for larger singlet VEVs \boldmath{$v_S$}}
\label{sec:vs_limits}

\begin{figure}[htb!]
    \centering
    \begin{subfigure}{0.49\textwidth}
        \includegraphics[width=\textwidth]{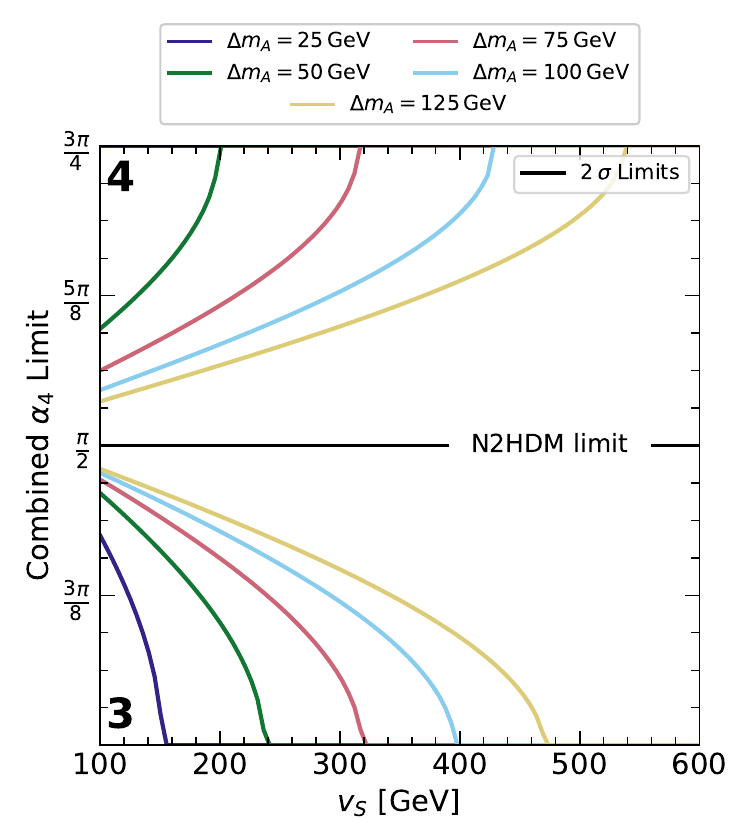}
    \end{subfigure}
    \caption{The $1\,\sigma$, $2\,\sigma$ $\alpha_4$ limits for different $v_S$, $m_{a_1}$ {(see text). The labels ``3'' and ``4'' correspond
    to the two parameter regions labelled as such in \protect\reffi{fig:2hdms_region}.}} 
    \label{fig:vs_limits_34}
\end{figure}

To complete our analysis, we include a brief discussion of the dependence of $\al_4$ limits on the VEV of the singlet, $v_S$, 
as the scenarios analyzed above the value was fixed to $v_S=\SI{120}{GeV}$. 
From the analytical expressions (see \refse{sec:diff}), one observes a preference for low $v_S$ to find significant differences between the 
N2HDM and the 2HDMS. In this subsection we will expand on this and compute the $\alpha_4$ limits for increasing $v_S$ and different 2HDMS scenarios 
defined by $m_{a_1}=\Delta m_a=\{575,550,525,500,475\} \gev$.
The results for the combined limits are shown in \reffi{fig:vs_limits_34}.
The $\De m_a$ values are indicated by the color coding, and the labels ``3'' and ``4'' correspond
    to the two parameter regions labelled as such in \reffi{fig:2hdms_region}.
For low mass differences corresponding to $m_{a_1}\sim\SI{575}{GeV}$, the limits are declining quickly $\al_4 = \pi/2$ as a function of $v_S$,
and differences between the models are barely observable.
For larger mass differences $m_{a_1}\sim\SI{500}{GeV}$, one can extract $2\,\sigma$ limits up to $v_S=\SI{400}{GeV}$. 
This trend increases for increasing $\De m_a$. Furthermore, one can observe that
region~3 shows the possibility of the limits to get much closer to $\al_4 = \pi/2$, as compared to region~4.
This trend is more pronounced for smaller $\De m_a$, whereas for
larger differences both regions show a similar behavior.

%% file: section5.tex
\section{Conclusions and Outlook}
\label{sec:conclusions}

We have investigated the phenomenological differences between the N2HDM and the
2HDMS in the context of the Yukawa type~II realization, focusing on scenarios that
can accommodate the Higgs-boson excesses around $95 \gev$ in the $\ga\ga$ and $b\bar b$ final states. 
Although both models contain a similar $\mathcal{CP}$-even Higgs sector and can provide an good description
of the excesses while reproducing the observed properties of the $125 \gev$ Higgs
boson, they differ in the structure of their scalar potential and in the $\mathcal{CP}$-odd sector.
Our aim has been to determine whether these differences can be seen
experimentally to distinguish the two models.

We derived compact analytical expressions for the trilinear Higgs couplings in both
models in terms of physical input parameters, allowing us to identify the origin
of the differences between the two models. We have shown that the additional cubic
interactions present in the 2HDMS yield contributions to the trilinear Higgs 
couplings that depend on the singlet--doublet mixing angle $\alpha_4$ (the 2HDMS $\mathcal{CP}$-odd mixing angle,
where $\al_4 \to \pi/2$ decouples the second $\mathcal{CP}$-odd Higgs boson), 
the singlet vacuum expectation value $v_S$, and the mass splitting in the $\mathcal{CP}$-odd sector.
These differences are even present in the N2HDM limit, in which the particle content of the two models
is indistinguishable. It was also found that sizeable increased deviations can
appear already for relatively small departures from this limit, particularly for small
$v_S$ and large pseudoscalar mass splittings. 

Based on these analytical results we introduced the concept of the
``$\alpha_4$ limit'', defined as the smallest deviation from the N2HDM limit (corresponding to 
$\al_4 = \pi/2$) that allows the two models to be distinguished experimentally at the $2\sigma$ level.
We evaluated this limit in two complementary collider environments for two sets of benchmark planes.

At the HL-LHC we studied the four-top final state, where heavy neutral Higgs bosons
are produced in association with top quarks and subsequently decay into
$t\bar t$. We demonstrated that the modified $\mathcal{CP}$-odd Higgs couplings and branching
ratios in the 2HDMS can lead to measurable deviations of the four-top production rate
with respect to the N2HDM prediction over large parts of the parameter space. The sensitivity increases for larger mass
splittings between the doublet- and singlet-like pseudoscalars, where additional Higgs
decay channels become accessible. 

At a future high-energy $e^+e^-$ collider we investigated the complete set of
di-Higgs production channels in association with a $Z$ boson involving the $125 \gev$ and the $95 \gev$ Higgs bosons.
Owing to the clean experimental environment, these processes provide an important probe of the scalar
potential. We found that the different Higgs pair-production channels probe
complementary combinations of trilinear couplings, leading to a significantly improved
discrimination power when combined. In particular, the sensitivity remains strong even
in parameter regions where the four-top analysis alone becomes less effective.
However, also here some parts of the parameter space do not yield an experimental discrimination beyond the $2\,\sig$ level.

Overall, our results demonstrate that the combination of HL-LHC measurements and
precision di-Higgs studies at future high-energy $e^+e^-$ colliders provide a powerful strategy to
discriminate between extended Higgs sectors that otherwise exhibit very similar
low-energy phenomenology. While the present analysis has been motivated by the
$95 \gev$ Higgs excesses, our analysis should be regarded as a show case: 
the methodology developed here is considerably more
general and can be applied to distinguish a broad class of extended Higgs-sector 
models through searches at the HL-LHC, as well as in precision measurements of di-Higgs 
cross sections at future $e^+e^-$ colliders.

The introduction of the ``$\alpha_4$ limit'' provides a simple and quantitative measure
of the experimental distinguishability between the N2HDM and the 2HDMS. It may also
serve as a useful benchmark for future studies of other extended Higgs sectors with
similar scalar spectra but different Higgs potentials.

Future work could include the incorporation of higher-order corrections to the
trilinear Higgs couplings and production cross sections, a dedicated detector-level
analysis of the proposed observables, and the extension of the present framework to
other Yukawa types and/or collider facilities. Such studies will further clarify
the potential of future experiments to unravel the underlying structure of an
extended Higgs sector.

%% file: appendixA.tex
\section{Tree level Higgs masses}
\label{appendix:a}
The mass matrices can be obtained by taking the second derivative of the potential
\begin{equation}
    M^2_\rho=\frac{\partial^2 V}{\partial \rho_i \partial \rho_j },\qquad     M^2_\eta=\frac{\partial^2 V}{\partial \eta_i \partial \eta_i },\qquad M^2_\chi=\frac{\partial^2 V}{\partial \chi^+_i \partial \chi^-_j }.
\end{equation}
The basis change can be derived using
\begin{equation}
   M^2_{ij}=\sum_{n=1}^3 m^2_{h_n}R_{ni}R_{nj}. 
\end{equation}
\subsection{N2HDM}
\begin{subequations}
\begin{align}
M_{\rho11}^2&=2\lambda_1v^2\cos^2\beta+m_{12}^2\tan\beta,\\
M_{\rho22}^2&=2\lambda_2v^2\sin^2\beta+m_{12}^2\cot\beta,\\
M_{\rho12}^2&=(\lambda_{3}+\lambda_4+\lambda_5)v^2\sin\beta\cos\beta-m_{12}^2,\\
M_{\rho1S}^2&=2\lambda_7v_S\cos\beta v,\\
M_{\rho2S}^2&=2\lambda_8v_S\sin\beta v,\\
M_{\rho SS}^2&=\lambda_6 v_S.
	\end{align}
\end{subequations}
\begin{subequations}
\begin{align}
M_{\eta11}^2&=(m_{12}^2-\lambda_5v_1v_2)\tan\beta\\
M_{\eta12}^2&=-(m_{12}^2-\lambda_5v_1v_2)\\
M_{\eta22}^2&=(m_{12}^2-\lambda_5v_1v_2)\cot\beta\\
	\end{align}
\end{subequations}
\begin{subequations}
\begin{align}
M_{\chi11}^2&=(m_{12}^2-(\lambda_4+\lambda_5)v_1v_2)\tan\beta\\
M_{\chi12}^2&=-(m_{12}^2-(\lambda_4+\lambda_5)_5v_1v_2)\\
M_{\chi22}^2&=(m_{12}^2-(\lambda_4+\lambda_5)v_1v_2)\cot\beta\\
	\end{align}
\end{subequations}
The basis change in the N2HDM then reads
\begin{align}
&\lambda_1=\frac{1} {v^2\cos^2\beta}\left(\sum_im^2_{h_i}R^2_{i1}-\hat{\mu}^2\sin^2\beta\right), \label{eq:lambda1_n2hdm}\\
&\lambda_2=\frac{1}{v^2\sin^2\beta}\left(\sum_im^2_{h_i}R^2_{i2}-\hat{\mu}^2\cos^2\beta\right),\\
&\lambda_3=\frac{1}{v^2}\left(\frac{1}{\sin\beta\cos\beta}\sum_im^2_{h_i}R_{i1}R_{i2}+2m^2_{h^\pm}-\hat{\mu}^2\right),\\
&\lambda_4=\frac{1}{v^2}\left(\hat{\mu}^2+m_A^2-2m^2_{h^\pm}\right),\\
&\lambda_5=\frac{1}{v^2}\left(\hat{\mu}^2-m_A^2\right),\\
&\lambda_6=\frac{1}{v_S^2}\sum_i m^2_{h_i}R_{i3}^2,\\
&\lambda_7=\frac{1}{v v_S \cos\beta}\left(\sum_i m^2_{h_i}R_{i1}R_{i3}\right),\\
&\lambda_8=\frac{1}{v v_S \sin\beta}\left(\sum_i m^2_{h_i}R_{i2}R_{i3}\right),  \label{eq:lambda8_n2hdm}
\end{align}
with the abbreviation 
\begin{equation}
    \hat\mu^2=\frac{m_{12}^2}{\sin\beta\cos\beta}.
\end{equation}


\subsection{2HDMS}
\label{sec:2hdms-app}

\begin{subequations}
\begin{align}
M_{\rho11}^2&=\lambda_1v^2\cos^2\beta+\rbr{m_{12}^2-\frac{\mu_{12}v_S}{\sqrt{2}}}\tan\beta,\\
M_{\rho22}^2&=\lambda_2v^2\sin^2\beta+\rbr{m_{12}^2-\frac{\mu_{12}v_S}{\sqrt{2}}}\cot\beta,\\
M_{\rho12}^2&=(\lambda_{3}+\lambda_4)v^2\cos\beta\sin\beta-\rbr{m_{12}^2-\frac{\mu_{12}v_S}{\sqrt{2}}},\\
M_{\rho1S}^2&=(\lambda'_1v_S\cos\beta+\frac{\mu_{12}}{\sqrt{2}}\sin\beta)v,\\
M_{\rho2S}^2&=(\lambda'_2v_S\sin\beta+\frac{\mu_{12}}{\sqrt{2}}\cos\beta)v,\\
M_{\rho SS}^2&=\frac{\mu_{s1}}{2\sqrt{2}}v_S+\frac{\lambda''_3v_S^2}{2}-\mu_{12}\frac{v^2}{\sqrt{2}v_S}\sin\beta\cos\beta.
\label{eq:masscpeven}
\end{align}
\end{subequations}
\begin{subequations}
	\begin{align}
	M_{\eta11}^2&=\rbr{m_{12}^2-\frac{\mu_{12}v_S}{\sqrt{2}}}\tan\beta,\\
	M_{\eta22}^2&=\rbr{m_{12}^2-\frac{\mu_{12}v_S}{\sqrt{2}}}\cot\beta,\\
	M_{\eta12}^2&=-\rbr{m_{12}^2-\frac{\mu_{12}v_S}{\sqrt{2}}},\\
	M_{\eta1S}^2&=\frac{\mu_{12}v}{\sqrt{2}}\sin\beta,\\
	M_{\eta2S}^2&=-\frac{\mu_{12}v}{\sqrt{2}}\cos\beta,\\
	M_{\eta SS}^2&=-\frac{3}{2\sqrt{2}}\mu_{S1}v_S-\mu_{12}\frac{v^2}{\sqrt{2}v_S}\sin\beta\cos\beta. 
	\label{eq:masscpodd}
	\end{align}
\end{subequations}
\begin{subequations}
\begin{align}
M_{\chi11}^2&=\rbr{m_{12}^2-\frac{\lambda_4v_1v_2}{2}-\frac{\mu_{12}v_S}{\sqrt{2}}}\tan\beta\\
M_{\chi12}^2&=-\rbr{m_{12}^2-\frac{\lambda_4v_1v_2}{2}-\frac{\mu_{12}v_S}{\sqrt{2}}}\\
M_{\chi22}^2&=\rbr{m_{12}^2-\frac{\lambda_4v_1v_2}{2}-\frac{\mu_{12}v_S}{\sqrt{2}}}\cot\beta\\
\end{align}
\end{subequations}
The basis change in the 2HDMS then reads

\begin{align}
&\mu_{12}=\frac{\sqrt{2}(m_{a_2}^2-m_{a_1}^2)}{v}\sin\alpha_4\cos\alpha_4\label{eq:mu12-a4}~,\\
&m^2_{12}=\frac{v_S\mu_{12}}{\sqrt{2}}+\tilde{\mu}^2\sin\beta\cos\beta~,\\
&\mu_{S1}=-\frac{2\sqrt{2}}{3v_S}\left(\sin^2\alpha_4m_{a_1}^2+\cos^2\alpha_4m_{a_2}^2+\frac{v^2}{\sqrt{2}v_S}\sin\beta\cos\beta\mu_{12} \right),\label{eq:mus1-ma1}\\
&\lambda_1=\frac{1}{v^2\cos^2\beta}\left(\sum_im^2_{h_i}R^2_{i1}-\tilde{\mu}^2\sin^2\beta\right),\\
&\lambda_2=\frac{1}{v^2\sin^2\beta}\left(\sum_im^2_{h_i}R^2_{i2}-\tilde{\mu}^2\cos^2\beta\right),\\
&\lambda_3=\frac{1}{v^2}\left(\frac{1}{\sin\beta\cos\beta}\sum_im^2_{h_i}R_{i1}R_{i2}+2m^2_{h^\pm}-\tilde{\mu}^2\right),\\
&\lambda_4=\frac{2(\tilde{\mu}^2-m^2_{h^\pm})}{v^2},\\
&\lambda'_1=\frac{1}{2v_Sv\cos\beta}\left(\sum_i m^2_{h_i}R_{i1}R_{i3}-\mu_{12}v\sin\beta \right),\label{eq:lam1p}\\
&\lambda'_2=\frac{1}{2v_Sv\sin\beta}\left(\sum_i m^2_{h_i}R_{i2}R_{i3}-\mu_{12}v\cos\beta \right),\\
&\lambda''_3=\frac{1}{v_S^2}\left(\sum_im^2_{h_i}R^2_{i3}+\mu_{12}\frac{v^2}{2v_S}\sin2\beta-\frac{\mu_{S1}}{2}v_S\right),\label{eq:lambdapp3_2hdms}
\end{align}
with the abbreviation $\Tilde\mu$
\begin{equation}
\tilde{\mu}^2=\frac{m^2_{12}-v_S\mu_{12}}{\sin\beta\cos\beta}
\equiv \cos^2\alpha_4m_{a_1}^2+\sin^2\alpha_4m_{a_2}^2. \label{eq:mutilde2HDMS}
\end{equation}

%% file: appendixB.tex
\section{Uncertainty Estimate for Di-Higgs Production at the ILC500}
\label{sec:unc_estimate}

In order to study the significance of the differences in the di-Higgs production, we need information about the experimental uncertainties of a potential 
observation at the ILC. On the other hand, we will assume that the theory uncertainties will be at a negligible level at the time of the ILC500 runs. In 
\citere{aryshev2023international} the ILC collaboration reports a predicted significance of $8\,\sigma$ for the observation of di-Higgs strahlung after 
the full running scenario of $\mathcal{L}_{\rm int}=\SI{4}{\atto\per\barn}$ at  $\sqrt{s}=\SI{500}{GeV}$. This value results from the observation of the final states 
$Zbbbb$ and $ZbbWW$, taking into account simulations of the signal and background regions at the International Large Detector \cite{theildcollaboration2020international} 
(ILD). The $4b$ final state study, performed in \citere{Durig:2016jrs}, derives the uncertainty using a log-likelihood fit for $\ld^{\rm ILD}_{\rm int}=\SI{2}
{\atto\per\barn}$ and $\sigma^{\rm SM}_{ZHH}=\SI{0.198}{fb}$. 
This is slightly different from our value of $\sigma_{\rm SM}=\SI{0.2338}{fb}$, due to the employed value of $\al_{\mathrm{em}}$.
For this work we use
\begin{equation}
\alpha_{\rm em}^{-1}\, = \, {\al_{\mathrm{em}}^{-1}(G_F) \, = \, }\frac{\pi}{\sqrt{2}G_Fm_W^2\sin^2\theta_W} = 132.5.    
\end{equation}
The likelihood function is given by 
\begin{equation}
    L(N_{ZHH})=\prod_{i\in\{ee,\mu\mu,\nu\nu,bb,qq\}} \left({\frac{N_{ ZHH} {\mathcal{BR}}_i \varepsilon_i+b_i}{b_i}}\right)^{s_i+b_i}\exp(-N_{ZHH} {\mathcal{BR}}_i \varepsilon_i),
\end{equation}
where $N_{ZHH}=\sigma \mathcal{L}$ is the expected number of $ZHH$ events before applying any cuts or efficiencies,
and $s_i$ and $b_i$ are the signal and background rates predictions for the different $Z$ decay channels at the ILD including tagging efficiencies, that can be taken from \cite{Durig:2016jrs}. ${\mathcal{BR}}_i$ is the respective branching ratio, given by
\begin{equation}
    {\mathcal{BR}}_i={\mathcal{BR}}(Z\to ff){\mathcal{BR}}(H\to bb)^2.
\end{equation}
The $\varepsilon_i$ denotes the efficiencies of observing the respective channels with the ILD. They can be derived using the given $s_i$ with 
\begin{equation}
    \varepsilon_i\approx\frac{s_i}{\sigma^{\rm SM}_{ ZHH}\cdot \ld_{\rm int}^{\rm ILD}\cdot{\mathcal{BR}}_i }.
\end{equation}
All used input values for the uncertainty estimation are summarized in \refta{tab:unc_sum} where we use $\mathcal{BR}(H\to bb)=0.584$ \cite{CERN_YELLOW_2017}.

\begin{table}[htb!]
  \begin{center}
    \begin{tabular}{c|c|c|c|c}
        & $s_i$ &$b_i$& ${\mathcal{BR}}(Z\to ff)$ & $\varepsilon_i$\\\hline
        $ee$ & 3.9 & 7.0 & 0.034 & 0.857 \\
        $\mu\mu$  & 5.1 & 8.9 & 0.034 & 1.00\footnotemark[1] \\
        $\nu\nu$  & 5.6 & 6.9 & 0.2 & 0.2\\
        $bb$  & 8.5 & 21.9 & 0.15 & 0.42\\
        $qq$  & 12.6 & 55.0 & 0.7 & 0.13\\
    \end{tabular}
    \caption{Signal and background estimates for the ILD from \citere{Durig:2016jrs} with the branching ratios \cite{ParticleDataGroup:2022pth} and efficiencies of the respective channel for ${\ld_{\rm int}=\SI{2}{\atto\per\barn}}$ and $\sigma^{\rm SM}_{ZHH}=\SI{0.198}{fb}$.}
    \label{tab:unc_sum}
\end{center}
\end{table}
\footnotetext[1]{The efficiency $\varepsilon_{\mu\mu}$ is approximate and rounded to 1.00.} 

\begin{figure}[htb!]
    \centering
    \includegraphics[width=0.8\textwidth]{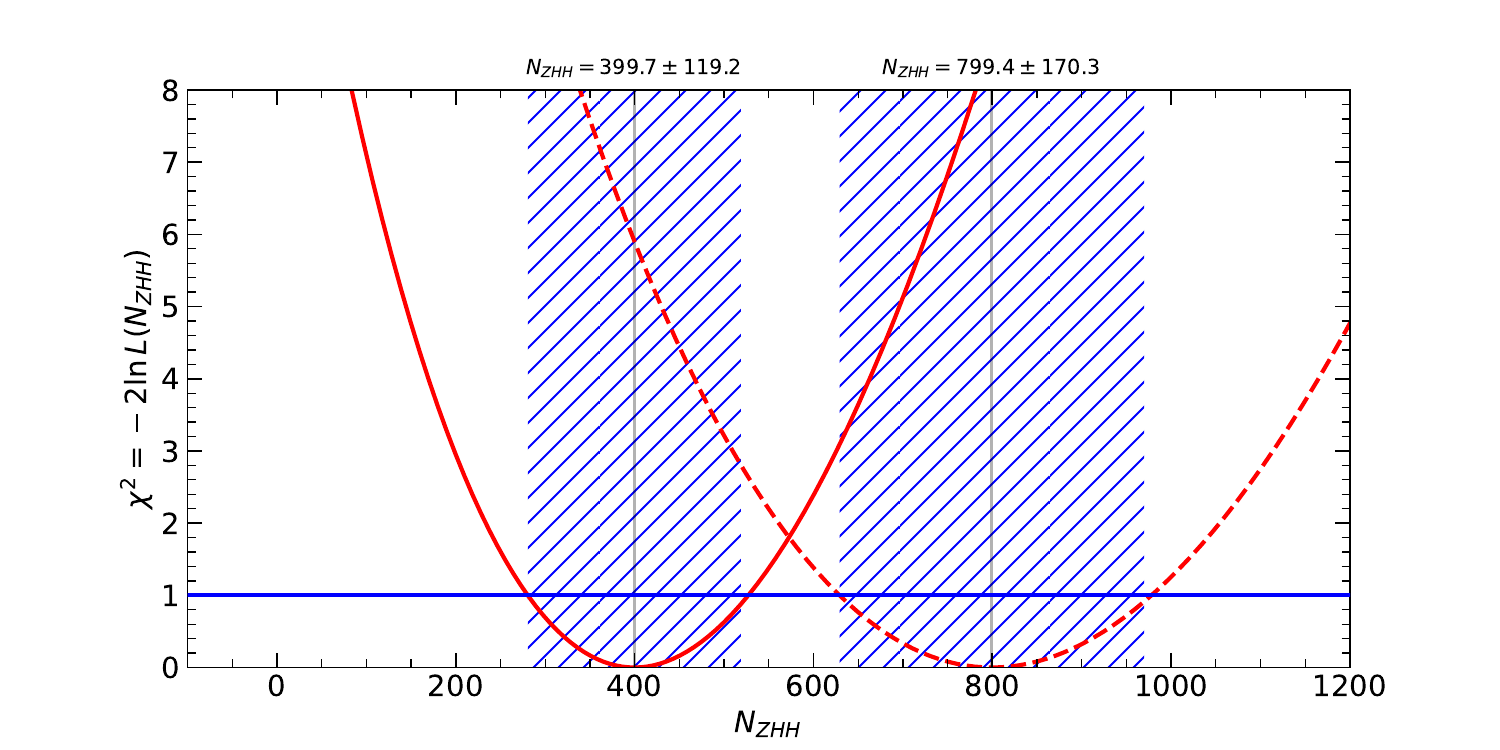}
    \caption{Log likelihood plot for $\ld_{\rm int}=\SI{2}{\atto\per\barn}$ (solid red, as evaluated in \citere{Durig:2016jrs}) and ${\ld_{\rm int}=\SI{4}{\atto\per\barn}}$ (dashed red).
    The uncertainty is given by $\chi^2_{\rm min}+1$. For simplicity, we are considering only symmetrical uncertainties.}
    \label{fig:ILC_Err_1}
\end{figure}

Since our study considers an integrated luminosity of $\ld_{\rm int}=\SI{4}{\atto\per\barn}$ after the full running scenario, the signal and background regions need to be adjusted accordingly, thus
\begin{equation}
    s_i\to s_i\frac{\ld_{\rm int}}{\ld_{\rm int}^{\rm ILD}}, \qquad \qquad   b_i\to b_i\frac{\ld_{\rm int}}{\ld_{\rm int}^{\rm ILD}}.
\end{equation} 
The Log-likelihood function 
\begin{equation}
    \chi^2=-2\ln L(N_{ZHH})
\end{equation}
for $\ld_{\rm int}=\SI{2}{\atto\per\barn}$ and ${\ld_{\rm int}=\SI{4}{\atto\per\barn}}$ is shown in \reffi{fig:ILC_Err_1}. The uncertainties $\Delta N_{ ZHH}$ are given 
by the interval $\chi^2_{\rm min}+1$. For $\ld_{\rm int}=\SI{4}{\atto\per\barn}$ following \citere{Durig:2016jrs}
we get (with the error obtained as in \reffi{fig:ILC_Err_1}),

\begin{equation}
    N_{ZHH}^{\SI{4}{\atto\per\barn}}=799.4\pm170.3\,.
\end{equation}
This corresponds to a significance of $4.7\;\sigma$. The reported total significance in \citere{aryshev2023international} of $8\;\sigma$ takes also the decay
$ZHH\to ZbbWW$ \cite{kurata_2013} into account. In order to technically yield $8\,\sig$ with our slightly higher SM di-Higgs cross section, we rescale 
our uncertainties: using $\sigma^{\rm SM}_{ZHH}=\SI{0.2338}{fb}$ and adding the $ZbbWW$ final state corresponds to
an improvement of \SI{32.6}{\%} in the total accuracy, yielding an improvement factor for the uncertainty of $(1 - 32.6\%)$ that we assume
unchanged throughout the analysis (see below).

\begin{figure}[h]
    \centering
    \includegraphics[width=0.8\textwidth]{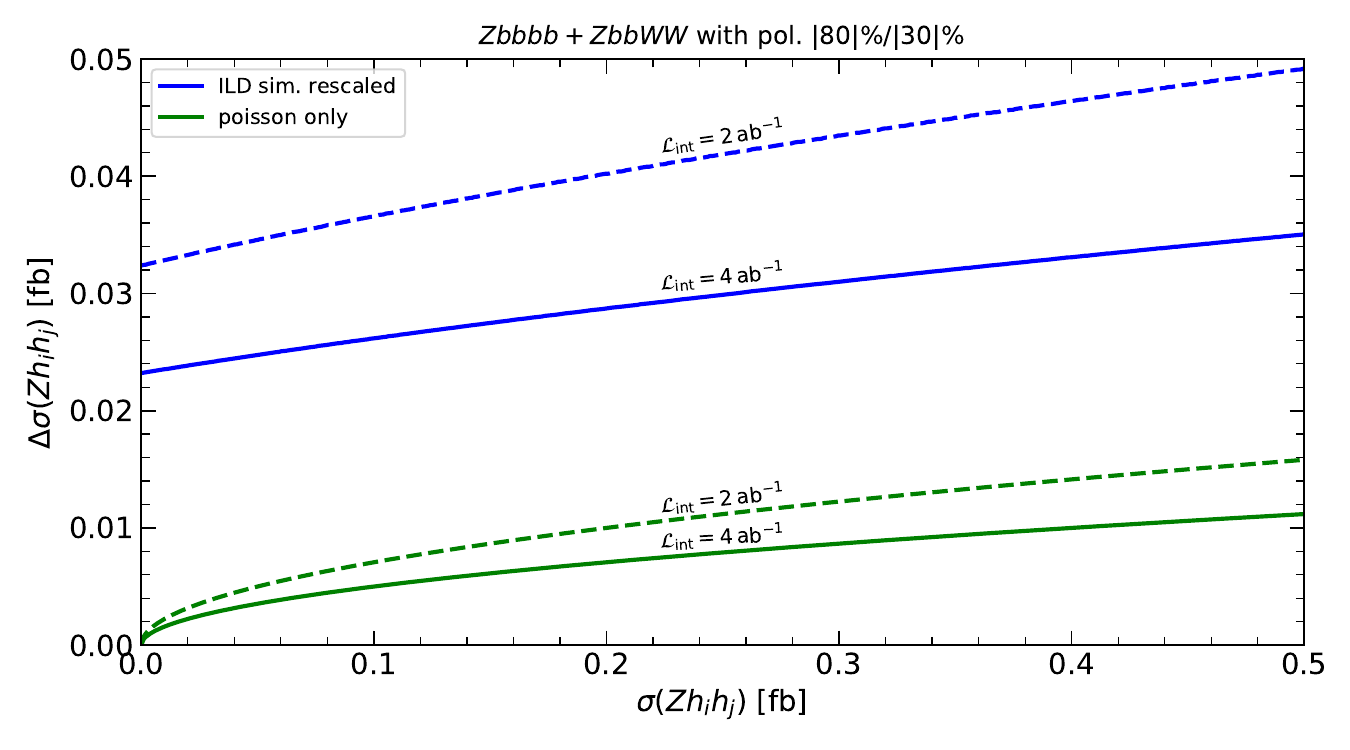}
    \caption{Rescaled uncertainties for the cross-sections $\sigma(Zh_ih_j)$ including the signal and background rates from the ILD for $\ld_{\rm int}=\SI{2}{\atto\per\barn}$ and ${\ld_{\rm int}=\SI{4}{\atto\per\barn}}$ in blue and the uncertainties given by the Poisson statistic in green.}
    \label{fig:ILC_Err_2}
\end{figure}

In the N2HDM and 2HDMS, the cross-section of double Higgs strahlung $\sigma^{\mathrm{model}}_{Zh_ih_j}$ varies over the parameter space.
We assume, that the BSM Higgs sectors have no impact on the background rates and only alter the signal rates according to the different cross-section, thus
\begin{equation}
    s_i\to s_i\frac{\sigma^{{\mathrm{model}}}_{Zh_ih_j}}{\sigma^{\rm SM}_{ZHH}},\qquad \qquad b_i\to b_i.
\end{equation}
However, this has to be considered an approximation, since the background region also contains, e.g., events from the type $ZZH\to Zbbbb$, 
which would be reduced by $\abs{c_{h_{125}VV}}^2$. For this study, we neglect such effects, and leave the $b_i$ unchanged. 
Furthermore, we do not take into account possible changes of the branching ratios ${{\mathcal{BR}}(h_i\to bb)}$, as these change only little
over our studied parameter space, where we remain close to the SM alignment limit.
The uncertainties are then given by,
\begin{equation}
    \Delta \sigma(Zh_ih_j)=\frac{\Delta N_{Zh_ih_j}}{\ld_{\rm int}}\cdot \underbrace{0.674}_{1-32.6\,\%},
\end{equation}
where the factor $0.674$ accounts for the improvement, when combining the $Zbbbb$ and $ZbbWW$ channels and taking into account our $\sig_{ZHH}^{\mathrm{SM}}$. 
We assume that this improvement factor remains unchanged in the
two BSM models. 
The results, as well as the uncertainties when only considering Poisson statistics, are shown in \reffi{fig:ILC_Err_2}.
The blue rescaled uncertainty from the ILD simulations appears to approximately follow the green Poisson curve, which would be the uncertainty in a perfect detector without any background, with an offset that accounts for such imperfections.


%% file: appendixC.tex
\section{Calculation of the oblique parameters \boldmath{$S$, $T$ and $U$}}
\label{sec:stu-calc}
\label{sec:A_STU}

We derive the oblique parameters $S,T,U$ for the N2HDM and 2HDMS following the results for multi-Higgs-doublet model and singlets (mHDSM)~\cite{Grimus_2008}. The mHDSM contains $n_d$ SU(2) doublets, $n_c$ SU(2) singlets with $Y=1$ and $n_n$ SU(2) real singlets with $Y=0$.
We compute the $S,T$ and $U$ parameters according to equations (30), (28) and (31) of \citere{Grimus_2008}, respectively.
We define the mixing matrices $\mathcal{U}$ and $\mathcal{V}$, that rotate the mass eigen fields to the interaction states.  
\begin{equation}
    \mathcal{U}: n_d \times n,\quad \mathcal{V}: n_d \times m,
\end{equation}
with $n=n_d+n_c$ and $m=2n_d+n_n$. The N2HDM contains two complex doublets and one singlet, thus $n_d=2$, $n_c=0$ and $n_n=1$. Therefore, $\mathcal{U}$ is a $2 \times 2$ matrix where 
\begin{equation}
    \zv{\chi_1^\pm}{\chi_2^\pm}= \mathcal{U}\zv{\xi_{c}^0}{H^\pm}\quad \to \quad \mathcal{U}= \begin{pmatrix} \cos\beta &-\sin\beta \\ \sin\beta& \cos\beta\,.
    \end{pmatrix} \label{eq:mathcalU}
\end{equation}
$\mathcal{V}$ is then a $2\times5$ matrix, that fulfills
\begin{equation}
    \zv{\rho_1+i\eta_1}{\rho_2+i\eta_2}= \mathcal{V}\begin{pmatrix}
        \xi^0_a \\ A \\ h_1\\h_2\\h_3
    \end{pmatrix}\,.
\end{equation}
In other words, $\mathcal{V}$ describes the doublet mixing content of our model. In the N2HDM it is given by 
\begin{equation}
    \mathcal{V}=\begin{pmatrix}
        iR^A_{11} & iR^A_{21} & R_{11}& R_{21} & R_{31} \\
        iR^A_{12} & iR^A_{22} & R_{12}& R_{22} & R_{32} \\
    \end{pmatrix}.
\end{equation}
where $R^A$ is given by eq. \ref{eq:RA} and $R$ by eq. \ref{eq:rot}. Further, we define the charged and neutral mass vectors
\begin{equation}
    m^2_a=\zv{0}{m_{H^\pm}^2}, \qquad  \mu_b^2=\begin{pmatrix}
        0 \\ m_A^2 \\ m_{h_1}^2 \\ m_{h_2}^2 \\ m_{h_3}^2
    \end{pmatrix}.
\end{equation}
The 2HDMS is a model with two singlets $\rho_S$ and $\eta_S$, thus $n_d=2,\:n_c=0,\:n_n=2 \to n=2,\:m=6$. The charged sector is equal in both model, hence $\mathcal{U}$ remains a $2\times 2$ matrix and is given by eq. \ref{eq:mathcalU}. The now $2\times 6$ matrix $\mathcal{V}$ is given by 
\begin{equation}
    \mathcal{V}=\begin{pmatrix}
        iR^A_{11} & iR^A_{21} & iR^A_{31} & R_{11}& R_{21} & R_{31} \\
        iR^A_{12} & iR^A_{22} & iR^A_{32} & R_{12}& R_{22} & R_{32} \\
    \end{pmatrix},
\end{equation}
where $R$ is given by eq. \ref{eq:rot} and $R^A$ is given by eq. \ref{eq:RA}
so that 
\begin{equation}
    \zv{\rho_1+i\eta_1}{\rho_2+i\eta_2}= \mathcal{V}\begin{pmatrix}
        \xi^0_a \\ a_1 \\ a_2 \\ h_1\\h_2\\h_3
    \end{pmatrix}.
\end{equation}
The 2HDMS mass vectors are then defined as follows:
\begin{equation}
    m^2_a=\zv{0}{m_{H^\pm}^2}, \qquad  \mu_b^2=\begin{pmatrix}
        0 \\ m_{a_1}^2\\ m_{a_2}^2 \\ m_{h_1}^2 \\ m_{h_2}^2 \\ m_{h_3}^2
    \end{pmatrix}.
\end{equation}
We compute the agreement of the $S,T,U$ parameters with 
\begin{equation}
    \chi^2=\begin{pmatrix} S^\prime-S & T^\prime-T & U^\prime-U \end{pmatrix}   \cdot  \textbf{cov}^{-1} \cdot \begin{pmatrix} S^\prime-S \\ T^\prime-T \\ U^\prime-U \end{pmatrix}  
\end{equation}
The inverse covariance matrix is given by
{\small
\begin{equation}
    \textbf{cov}=\begin{pmatrix}
        0.012& -0.014& -0.008\\
         -0.014& 0.020&  -0.013 \\
         -0.008& -0.013 & 0.012
    \end{pmatrix},\:
    \textbf{cov}^{-1}=\begin{pmatrix}
        879.5& -933.5& -435.6\\
         -933.5& 1200.7&  694.7 \\
        -435.6&  694.7 & 555.7
    \end{pmatrix}.
\end{equation}}

\label{sec:coup-eval}

%% file: reference.bib
@article{ATLAS:2012tfa,
    author = "Aad, Georges and others",
    collaboration = "ATLAS",
    title = "{Observation of a new particle in the search for the Standard Model Higgs boson with the ATLAS detector at the LHC}",
    eprint = "1207.7214",
    archivePrefix = "arXiv",
    primaryClass = "hep-ex",
    reportNumber = "CERN-PH-EP-2012-218",
    doi = "10.1016/j.physletb.2012.08.020",
    journal = "Phys. Lett. B",
    volume = "716",
    pages = "1--29",
    year = "2012"
}

@article{Chatrchyan:2012xdj,
    author = "Chatrchyan, Serguei and others",
    collaboration = "CMS",
    title = "{Observation of a New Boson at a Mass of 125 GeV with the CMS Experiment at the LHC}",
    eprint = "1207.7235",
    archivePrefix = "arXiv",
    primaryClass = "hep-ex",
    reportNumber = "CMS-HIG-12-028, CERN-PH-EP-2012-220",
    doi = "10.1016/j.physletb.2012.08.021",
    journal = "Phys. Lett. B",
    volume = "716",
    pages = "30--61",
    year = "2012"
}

@article{Khachatryan:2016vau,
    author = "Aad, Georges and others",
    collaboration = "ATLAS, CMS",
    title = "{Measurements of the Higgs boson production and decay rates and constraints on its couplings from a combined ATLAS and CMS analysis of the LHC pp collision data at $ \sqrt{s}=7 $ and 8 TeV}",
    eprint = "1606.02266",
    archivePrefix = "arXiv",
    primaryClass = "hep-ex",
    reportNumber = "CERN-EP-2016-100, ATLAS-HIGG-2015-07, CMS-HIG-15-002",
    doi = "10.1007/JHEP08(2016)045",
    journal = "JHEP",
    volume = "08",
    pages = "045",
    year = "2016"
}

@article{Barate:2003sz,
    author = "Barate, R. and others",
    collaboration = "LEP Working Group for Higgs boson searches, ALEPH, DELPHI, L3, OPAL",
    title = "{Search for the standard model Higgs boson at LEP}",
    eprint = "hep-ex/0306033",
    archivePrefix = "arXiv",
    reportNumber = "CERN-EP-2003-011",
    doi = "10.1016/S0370-2693(03)00614-2",
    journal = "Phys. Lett. B",
    volume = "565",
    pages = "61--75",
    year = "2003"
}

@article{Sirunyan:2018aui,
    author = "Sirunyan, Albert M and others",
    collaboration = "CMS",
    title = "{Search for a standard model-like Higgs boson in the mass range between 70 and 110 GeV in the diphoton final state in proton-proton collisions at $\sqrt{s}=$ 8 and 13 TeV}",
    eprint = "1811.08459",
    archivePrefix = "arXiv",
    primaryClass = "hep-ex",
    reportNumber = "CMS-HIG-17-013, CERN-EP-2018-207",
    doi = "10.1016/j.physletb.2019.03.064",
    journal = "Phys. Lett. B",
    volume = "793",
    pages = "320--347",
    year = "2019"
}

@article{Biekotter:2019kde,
    author = {Biek\"otter, T. and Chakraborti, M. and Heinemeyer, S.},
    title = "{A 96 GeV Higgs boson in the N2HDM}",
    eprint = "1903.11661",
    archivePrefix = "arXiv",
    primaryClass = "hep-ph",
    reportNumber = "IFT-UAM/CSIC-19-034",
    doi = "10.1140/epjc/s10052-019-7561-2",
    journal = "Eur. Phys. J. C",
    volume = "80",
    number = "1",
    pages = "2",
    year = "2020"
}

@inproceedings{Biekotter:2020ahz,
    author = {Biek\"otter, T. and Chakraborti, M. and Heinemeyer, S.},
    title = "{The ''96 GeV excess'' at the ILC}",
    booktitle = "{International Workshop on Future Linear Colliders}",
    eprint = "2002.06904",
    archivePrefix = "arXiv",
    primaryClass = "hep-ph",
    reportNumber = "IFT-UAM/CSIC-20-021, DESY-20-019",
    month = "2",
    year = "2020"
}

@article{Biekotter:2021qbc,
    author = {Biek\"otter, Thomas and Grohsjean, Alexander and Heinemeyer, Sven and Schwanenberger, Christian and Weiglein, Georg},
    title = "{Possible indications for new Higgs bosons in the reach of the LHC: N2HDM and NMSSM interpretations}",
    eprint = "2109.01128",
    archivePrefix = "arXiv",
    primaryClass = "hep-ph",
    reportNumber = "IFT--UAM/CSIC--21-041, IFT-UAM/CSIC-21-041, DESY 21-132",
    doi = "10.1140/epjc/s10052-022-10099-1",
    journal = "Eur. Phys. J. C",
    volume = "82",
    number = "2",
    pages = "178",
    year = "2022"
}

@article{Domingo:2018uim,
    author = "Domingo, Florian and Heinemeyer, Sven and Pa\ss{}ehr, Sebastian and Weiglein, Georg",
    title = "{Decays of the neutral Higgs bosons into SM fermions and gauge bosons in the $\mathcal{CP}$-violating NMSSM}",
    eprint = "1807.06322",
    archivePrefix = "arXiv",
    primaryClass = "hep-ph",
    reportNumber = "DESY-18-084, IFT-UAM/CSIC-17-125, DESY--18--084, IFT--UAM/CSIC--17--125",
    doi = "10.1140/epjc/s10052-018-6400-1",
    journal = "Eur. Phys. J. C",
    volume = "78",
    number = "11",
    pages = "942",
    year = "2018"
}

@article{Biekotter:2017xmf,
    author = {Biek\"otter, T. and Heinemeyer, S. and Mu\~noz, C.},
    title = "{Precise prediction for the Higgs-boson masses in the $\mu \nu $ SSM}",
    eprint = "1712.07475",
    archivePrefix = "arXiv",
    primaryClass = "hep-ph",
    reportNumber = "IFT--UAM-CSIC--17-118",
    doi = "10.1140/epjc/s10052-018-5978-7",
    journal = "Eur. Phys. J. C",
    volume = "78",
    number = "6",
    pages = "504",
    year = "2018"
}

@article{Biekotter:2019gtq,
    author = {Biek\"otter, T. and Heinemeyer, S. and Mu\~noz, C.},
    title = "{Precise prediction for the Higgs-Boson masses in the ${{\mu }}{\nu }$ SSM with three right-handed neutrino superfields}",
    eprint = "1906.06173",
    archivePrefix = "arXiv",
    primaryClass = "hep-ph",
    reportNumber = "IFT--UAM/CSIC--19-030",
    doi = "10.1140/epjc/s10052-019-7175-8",
    journal = "Eur. Phys. J. C",
    volume = "79",
    number = "8",
    pages = "667",
    year = "2019"
}

@article{Muhlleitner:2016mzt,
    author = "Muhlleitner, Margarete and Sampaio, Marco O. P. and Santos, Rui and Wittbrodt, Jonas",
    title = "{The N2HDM under Theoretical and Experimental Scrutiny}",
    eprint = "1612.01309",
    archivePrefix = "arXiv",
    primaryClass = "hep-ph",
    doi = "10.1007/JHEP03(2017)094",
    journal = "JHEP",
    volume = "03",
    pages = "094",
    year = "2017"
}

@article{Ferreira:2019,
   title={Vacuum instabilities in the N2HDM},
   volume={2019},
   ISSN={1029-8479},
   url={https://doi.org/10.1007/JHEP09(2019)006},
   DOI={10.1007/jhep09(2019)006},
   number={9},
   journal={Journal of High Energy Physics},
   publisher={Springer Science and Business Media LLC},
   author={Ferreira, P. M. and Santos, Rui and Mühlleitner, Margarete and Weiglein, Georg and Wittbrodt, Jonas},
   year={2019}
}

@article{Hollik:2019,
   title={Impact of vacuum stability constraints on the phenomenology of supersymmetric models},
   volume={2019},
   ISSN={1029-8479},
   url={https://doi.org/10.1007/JHEP03(2019)109},
   DOI={10.1007/JHEP03(2019)109},
   number={3},
   journal={Journal of High Energy Physics},
   publisher={Springer Science and Business Media LLC},
   author={Hollik, Wolfgang G. and Weiglein, Georg and Wittbrodt, Jonas},
   year={2019}
}

@misc{evade:online,
  author = {J. Wittbrodt},
  howpublished = "\url{https://gitlab.com/jonaswittbrodt/EVADE}"
}

@article{Bechtle:2013xfa,
    author = "Bechtle, Philip and Heinemeyer, Sven and St\r{a}l, Oscar and Stefaniak, Tim and Weiglein, Georg",
    title = "{$HiggsSignals$: Confronting arbitrary Higgs sectors with measurements at the Tevatron and the LHC}",
    eprint = "1305.1933",
    archivePrefix = "arXiv",
    primaryClass = "hep-ph",
    reportNumber = "BONN-TH-2013-07, DESY-13-078",
    doi = "10.1140/epjc/s10052-013-2711-4",
    journal = "Eur. Phys. J. C",
    volume = "74",
    number = "2",
    pages = "2711",
    year = "2014"
}

@article{Bechtle:2014ewa,
    author = "Bechtle, Philip and Heinemeyer, Sven and St\r{a}l, Oscar and Stefaniak, Tim and Weiglein, Georg",
    title = "{Probing the Standard Model with Higgs signal rates from the Tevatron, the LHC and a future ILC}",
    eprint = "1403.1582",
    archivePrefix = "arXiv",
    primaryClass = "hep-ph",
    reportNumber = "DESY-14-026, BONN-TH-2014-05",
    doi = "10.1007/JHEP11(2014)039",
    journal = "JHEP",
    volume = "11",
    pages = "039",
    year = "2014"
}

@article{Baum:2018zhf,
    author = "Baum, Sebastian and Shah, Nausheen R.",
    title = "{Two Higgs Doublets and a Complex Singlet: Disentangling the Decay Topologies and Associated Phenomenology}",
    eprint = "1808.02667",
    archivePrefix = "arXiv",
    primaryClass = "hep-ph",
    reportNumber = "NORDITA-2018-067, WSU-HEP-1803, WSU-HEP-1804",
    doi = "10.1007/JHEP12(2018)044",
    journal = "JHEP",
    volume = "12",
    pages = "044",
    year = "2018"
}

@inproceedings{peskin:1995QFT,
	author = {Peskin, Michael Edward},
	month = {10},
	publisher = {Westview Press},
	title = {{An introduction to quantum field theory}},
	year = {1995},
}

@article{stal:2013HS,
    author = {Stål,Oscar and Stefaniak,Tim},
    title = "{Constraining extended Higgs sectors with HiggsSignals}",
    eprint = "1310.4039",
    archivePrefix = "arXiv",
    primaryClass = "hep-ph",
    doi = "10.48550/arXiv.1310.4039",
    year = "2013"
}

@article{bechtle:2021HS,
    author = {Bechtle, Philip and others},
    title = "{HiggsSignals-2: probing new physics with precision Higgs measurements in the {LHC} 13 TeV era}",
    eprint = "1434-6052",
    archivePrefix = "arXiv",
    primaryClass = "hep-ph",
    journal={The European Physical Journal C},
    doi = "10.1140/epjc/s10052-021-08942-y",
    year = "2021"
}

@article{Bahl_2023,
   title={{HiggsTools: BSM scalar phenomenology with new versions of HiggsBounds and HiggsSignals}},
   volume={291},
   ISSN={0010-4655},
   DOI={10.1016/j.cpc.2023.108803},
   journal={Computer Physics Communications},
   publisher={Elsevier BV},
   author={Bahl, Henning and Biekötter, Thomas and Heinemeyer, Sven and Li, Cheng and Paasch, Steven and Weiglein, Georg and Wittbrodt, Jonas},
   year={2023},
   pages={108803} 
}

@misc{HT_2019,
  author = {Henning Bahl et al.},
title = {{HiggsTools, A toolbox for BSM scalar phenomenogy}},
  howpublished = "\url{https://gitlab.com/higgsbounds/higgstools}"
}

@article{PhysRevLett.114.191803,
  title = {{Combined Measurement of the Higgs Boson Mass in $pp$ Collisions at $\sqrt{s}=7$ and 8 TeV with the ATLAS and CMS Experiments}},
  author = {ATLAS Collaboration and CMS Collaboration},
  journal = {Phys. Rev. Lett.},
  volume = {114},
  issue = {19},
  pages = {191803},
  numpages = {33},
  year = {2015},
  publisher = {American Physical Society},
  doi = {10.1103/PhysRevLett.114.191803}
}

@misc{CERN_YELLOW_2017,
  doi = {10.23731/CYRM-2017-002},
  author = {{CERN}},
  language = {en},
  title = {{CERN Yellow Reports: Monographs, Vol 2 (2017): Handbook of LHC Higgs cross sections: 4. Deciphering the nature of the Higgs sector}},
  publisher = {CERN},
  year = {2017},
  copyright = {This work is licensed under a Creative Commons Attribution 4.0 International License.}
}

@article{Bechtle_2010,
   title={{HiggsBounds: Confronting arbitrary Higgs sectors with exclusion bounds from LEP and the Tevatron}},
   volume={181},
   ISSN={0010-4655},
   DOI={10.1016/j.cpc.2009.09.003},
   number={1},
   journal={Computer Physics Communications},
   publisher={Elsevier BV},
   author={Bechtle, P. and Brein, O. and Heinemeyer, S. and Weiglein, G. and Williams, K.E.},
   year={2010},
   pages={138–167} 
}

@article{Bechtle_2011,
   title={{HiggsBounds 2.0.0: Confronting neutral and charged Higgs sector predictions with exclusion bounds from LEP and the Tevatron}},
   volume={182},
   ISSN={0010-4655},
   DOI={10.1016/j.cpc.2011.07.015},
   number={12},
   journal={Computer Physics Communications},
   publisher={Elsevier BV},
   author={Bechtle, P. and Brein, O. and Heinemeyer, S. and Weiglein, G. and Williams, K.E.},
   year={2011},
   pages={2605–2631} 
}

@misc{bechtle2013recent,
      title={{Recent Developments in HiggsBounds and a Preview of HiggsSignals}}, 
      author={Philip Bechtle and Oliver Brein and Sven Heinemeyer and Oscar Stål and Tim Stefaniak and Georg Weiglein and Karina Williams},
      year={2013},
      eprint={1301.2345},
      archivePrefix={arXiv},
      primaryClass={hep-ph}
}

@article{Bechtle_2014_3,
   title={{HiggsBounds-4: improved tests of extended Higgs sectors against exclusion bounds from LEP, the Tevatron and the LHC}},
   volume={74},
   ISSN={1434-6052},
   DOI={10.1140/epjc/s10052-013-2693-2},
   number={3},
   journal={The European Physical Journal C},
   publisher={Springer Science and Business Media LLC},
   author={Bechtle, Philip and Brein, Oliver and Heinemeyer, Sven and Stål, Oscar and Stefaniak, Tim and Weiglein, Georg and Williams, Karina E.},
   year={2014}
}

@article{Bechtle_2015,
   title={{Applying exclusion likelihoods from LHC searches to extended Higgs sectors}},
   volume={75},
   ISSN={1434-6052},
   DOI={10.1140/epjc/s10052-015-3650-z},
   number={9},
   journal={The European Physical Journal C},
   publisher={Springer Science and Business Media LLC},
   author={Bechtle, Philip and Heinemeyer, Sven and Stål, Oscar and Stefaniak, Tim and Weiglein, Georg},
   year={2015}
}

@article{Bechtle_2020,
   title={{HiggsBounds-5: testing Higgs sectors in the LHC 13 TeV Era}},
   volume={80},
   ISSN={1434-6052},
   DOI={10.1140/epjc/s10052-020-08557-9},
   number={12},
   journal={The European Physical Journal C},
   publisher={Springer Science and Business Media LLC},
   author={Bechtle, Philip and Dercks, Daniel and Heinemeyer, Sven and Klingl, Tobias and Stefaniak, Tim and Weiglein, Georg and Wittbrodt, Jonas},
   year={2020}
}

@article{Bahl_2022,
   title={{Testing exotic scalars with HiggsBounds}},
   volume={82},
   ISSN={1434-6052},
   DOI={10.1140/epjc/s10052-022-10446-2},
   number={7},
   journal={The European Physical Journal C},
   publisher={Springer Science and Business Media LLC},
   author={Bahl, Henning and Lozano, Victor Martin and Stefaniak, Tim and Wittbrodt, Jonas},
   year={2022}
}

@article{Haller_2018,
   title={{Update of the global electroweak fit and constraints on two-Higgs-doublet models}},
   volume={78},
   ISSN={1434-6052},
   DOI={10.1140/epjc/s10052-018-6131-3},
   number={8},
   journal={The European Physical Journal C},
   publisher={Springer Science and Business Media LLC},
   author={Haller, J. and Hoecker, A. and Kogler, R. and Mönig, K. and Peiffer, T. and Stelzer, J.},
   year={2018}
}

@article{PhysRevD.46.381,
  title = {{Estimation of oblique electroweak corrections}},
  author = {Peskin, Michael E. and Takeuchi, Tatsu},
  journal = {Phys. Rev. D},
  volume = {46},
  issue = {1},
  pages = {381--409},
  numpages = {0},
  year = {1992},
  publisher = {American Physical Society},
  doi = {10.1103/PhysRevD.46.381}
}

@article{Grimus_2008,
   title={{The oblique parameters in multi-Higgs-doublet models}},
   volume={801},
   ISSN={0550-3213},
   DOI={10.1016/j.nuclphysb.2008.04.019},
   number={1–2},
   journal={Nuclear Physics B},
   publisher={Elsevier BV},
   author={Grimus, W. and Lavoura, L. and Ogreid, O.M. and Osland, P.},
   year={2008},
   pages={81–96} 
}

@article{Porod_2012,
   title={{SPheno 3.1: extensions including flavour, CP-phases and models beyond the MSSM}},
   volume={183},
   ISSN={0010-4655},
   DOI={10.1016/j.cpc.2012.05.021},
   number={11},
   journal={Computer Physics Communications},
   publisher={Elsevier BV},
   author={Porod, W. and Staub, F.},
   year={2012},
   pages={2458–2469} 
}

@article{Porod_2003,
   title={{SPheno, a program for calculating supersymmetric spectra, SUSY particle decays and SUSY particle production at electron  positron colliders}},
   volume={153},
   ISSN={0010-4655},
   DOI={10.1016/s0010-4655(03)00222-4},
   number={2},
   journal={Computer Physics Communications},
   publisher={Elsevier BV},
   author={Porod, W.},
   year={2003},
   pages={275–315} 
}

@misc{aryshev2023international,
      title={{The International Linear Collider: Report to Snowmass 2021}}, 
      author={ILC-Collaboration},
      year={2023},
      eprint={2203.07622},
      archivePrefix={arXiv},
      primaryClass={physics.acc-ph}
}

@article{Roloff_2020,
   title={{Double Higgs boson production and Higgs self-coupling extraction at CLIC}},
   volume={80},
   ISSN={1434-6052},
   DOI={10.1140/epjc/s10052-020-08567-7},
   number={11},
   journal={The European Physical Journal C},
   publisher={Springer Science and Business Media LLC},
   author={Roloff, Philipp and Schnoor, Ulrike and Simoniello, Rosa and Xu, Boruo},
   year={2020}
}

@misc{thecepcstudygroup2023cepc,
      title={{CEPC Technical Design Report -- Accelerator}}, 
      author={The CEPC Study Group},
      year={2023},
      eprint={2312.14363},
      archivePrefix={arXiv},
      primaryClass={physics.acc-ph}
}

@techreport{Benedikt:2651299,
      author        = "Benedikt Michael et al.",
      title         = "{{FCC-ee: The Lepton Collider: Future Circular Collider
                       Conceptual Design Report Volume 2. Future Circular
                       Collider}}",
      institution   = "CERN",
      reportNumber  = "CERN-ACC-2018-0057",
      address       = "Geneva",
      number        = "2",
      year          = "2019",
      doi           = "10.1140/epjst/e2019-900045-4",
}

@article{Moortgat_Pick_2008,
   title={{Polarized positrons and electrons at the linear collider}},
   volume={460},
   ISSN={0370-1573},
   DOI={10.1016/j.physrep.2007.12.003},
   number={4–5},
   journal={Physics Reports},
   publisher={Elsevier BV},
   author={G. Moortgat-Pick et al.},
   year={2008},
   pages={131–243} 
}

@article{Frederix_2018,
   title={{The automation of next-to-leading order electroweak calculations}},
   volume={2018},
   ISSN={1029-8479},
   DOI={10.1007/jhep07(2018)185},
   number={7},
   journal={Journal of High Energy Physics},
   publisher={Springer Science and Business Media LLC},
   author={Frederix, R. and Frixione, S. and Hirschi, V. and Pagani, D. and Shao, H.-S. and Zaro, M.},
   year={2018}
}

@article{Alwall_2014,
   title={{The automated computation of tree-level and next-to-leading order differential cross sections, and their matching to parton shower simulations}},
   volume={2014},
   ISSN={1029-8479},
   DOI={10.1007/jhep07(2014)079},
   number={7},
   journal={Journal of High Energy Physics},
   publisher={Springer Science and Business Media LLC},
   author={Alwall, J. and Frederix, R. and Frixione, S. and Hirschi, V. and Maltoni, F. and Mattelaer, O. and Shao, H.-S. and Stelzer, T. and Torrielli, P. and Zaro, M.},
   year={2014}
}

@misc{frixione2021lepton,
      title={{Lepton collisions in {MadGraph5\_aMC@NLO}}}, 
      author={Stefano Frixione and Olivier Mattelaer and Marco Zaro and Xiaoran Zhao},
      year={2021},
      eprint={2108.10261},
      archivePrefix={arXiv},
      primaryClass={hep-ph}
}

@article{Franzosi_2020,
   title={{Automated predictions from polarized matrix elements}},
   volume={2020},
   ISSN={1029-8479},
   DOI={10.1007/jhep04(2020)082},
   number={4},
   journal={Journal of High Energy Physics},
   publisher={Springer Science and Business Media LLC},
   author={Franzosi, Diogo Buarque and Mattelaer, Olivier and Ruiz, Richard and Shil, Sujay},
   year={2020}
}

@article{Staub_2014,
   title={{SARAH   4: A tool for (not only SUSY) model builders}},
   volume={185},
   ISSN={0010-4655},
   DOI={10.1016/j.cpc.2014.02.018},
   number={6},
   journal={Computer Physics Communications},
   publisher={Elsevier BV},
   author={Staub, Florian},
   year={2014},
   pages={1773–1790} }

@misc{staub2012sarah,
      title={{Sarah}}, 
      author={F. Staub},
      year={2012},
      eprint={0806.0538},
      archivePrefix={arXiv},
      primaryClass={hep-ph}
}

@article{Staub_2011,
   title={{Automatic calculation of supersymmetric renormalization group equations and loop corrections}},
   volume={182},
   ISSN={0010-4655},
   DOI={10.1016/j.cpc.2010.11.030},
   number={3},
   journal={Computer Physics Communications},
   publisher={Elsevier BV},
   author={Staub, Florian},
   year={2011},
   pages={808–833} 
}

@phdthesis{Durig:2016jrs,
    author = {D\"urig, Claude Fabienne},
    title = "{{Measuring the Higgs Self-coupling at the International Linear Collider}}",
    reportNumber = "DESY-THESIS-2016-027",
    doi = "10.3204/PUBDB-2016-04283",
    school = "Hamburg U.",
    address = "Hamburg",
    year = "2016"
}

@misc{theildcollaboration2020international,
      title={{International Large Detector: Interim Design Report}}, 
      author={The ILD Collaboration},
      year={2020},
      eprint={2003.01116},
      archivePrefix={arXiv},
      primaryClass={physics.ins-det}
}

@misc{kurata_2013,
      title={{The Higgs Self Coupling Analysis Using the Events containing $H\rightarrow WW^*$ Decay}}, 
      author={Kurata, M. and Tanabe, T. and Tian, J. and Fujii, K.},
      year={2013}
}

@article{Craig:2016ygr,
     author = "Craig, Nathaniel and Hajer, Jan and Li, Ying-Ying and Liu, Tao and Zhang, Hao",
     title = "{Heavy Higgs bosons at low $\tan \beta$: from the LHC to 100 TeV}",
     eprint = "1605.08744",
     archivePrefix = "arXiv",
     primaryClass = "hep-ph",
     doi = "10.1007/JHEP01(2017)018",
     journal = "JHEP",
     volume = "01",
     pages = "018",
     year = "2017"
 }

@PHDTHESIS{Duerig:310520,
      author       = {Duerig, Claude Fabienne},
      othercontributors = {List, Jenny and Garutti, Erika},
      title        = "Measuring the Higgs Self-coupling at the International Linear Collider",
      issn         = {1435-8085},
      school       = {Universität Hamburg},
      type         = {Dissertation},
      address      = {Hamburg},
      publisher    = {Verlag Deutsches Elektronen-Synchrotron},
      reportid     = {PUBDB-2016-04283, DESY-THESIS-2016-027},
      series       = {DESY-THESIS},
      pages        = {246},
      year         = {2016},
      note         = {Dissertation, Universität Hamburg, 2016},
      cin          = {FLC},
      cid          = {I:(DE-H253)FLC-20120731},
      pnm          = {611 - Fundamental Particles and Forces (POF3-611) / SFB 676
                      B01 - Optimierung des ILC setups: Physikprogramm,
                      Betriebsszenarien und Designentscheidungen (B01) (28895157)
                      / PHGS, VH-GS-500 - PIER Helmholtz Graduate School
                      $(2015_IFV-VH-GS-500)$},
      pid          = {G:(DE-HGF)POF3-611 / G:(GEPRIS)28895157 /
                      $G:(DE-HGF)2015_IFV-VH-GS-500$},
      experiment   = {EXP:(DE-MLZ)NOSPEC-20140101},
      typ          = {PUB:(DE-HGF)3 / PUB:(DE-HGF)11},
      doi          = {10.3204/PUBDB-2016-04283},
      url          = {https://bib-pubdb1.desy.de/record/310520},
}

@phdthesis{Paasch:2023gdz,
    author = "Paasch, Steven",
    title = "Phenomenology and Constraints in Singlet Extensions of Two Higgs Doublet Models",
    school = "Hamburg U.",
    year = "2023"
}

@article{Bahl:2022igd,
    author = {Bahl, Henning and Biek{\"o}tter, Thomas and Heinemeyer, Sven and Li, Cheng and Paasch, Steven and Weiglein, Georg and Wittbrodt, Jonas},
    title = "{HiggsTools: BSM scalar phenomenology with new versions of HiggsBounds and HiggsSignals}",
    eprint = "2210.09332",
    archivePrefix = "arXiv",
    primaryClass = "hep-ph",
    doi = "10.1016/j.cpc.2023.108803",
    journal = "Comput. Phys. Commun.",
    volume = "291",
    pages = "108803",
    year = "2023"
}

@article{Heinemeyer:2021msz,
    author = "Heinemeyer, S. and Li, C. and Lika, F. and Moortgat-Pick, G. and Paasch, S.",
    title = "{Phenomenology of a 96~GeV Higgs boson in the 2HDM with an additional singlet}",
    eprint = "2112.11958",
    archivePrefix = "arXiv",
    primaryClass = "hep-ph",
    reportNumber = "DESY 21-230, IFT-UAM/CSIC-21-158",
    doi = "10.1103/PhysRevD.106.075003",
    journal = "Phys. Rev. D",
    volume = "106",
    number = "7",
    pages = "075003",
    year = "2022"
}

@article{Biekotter:2023oen,
    author = {Biek{\"o}tter, Thomas and Heinemeyer, Sven and Weiglein, Georg},
    title = "{95.4~GeV diphoton excess at ATLAS and CMS}",
    eprint = "2306.03889",
    archivePrefix = "arXiv",
    primaryClass = "hep-ph",
    reportNumber = "KA-TP-11-2023, DESY-23-071, IFT--UAM/CSIC-23-062",
    doi = "10.1103/PhysRevD.109.035005",
    journal = "Phys. Rev. D",
    volume = "109",
    number = "3",
    pages = "035005",
    year = "2024"
}

@article{Azevedo:2023zkg,
    author = {Azevedo, Duarte and Biek{\"o}tter, Thomas and Ferreira, P. M.},
    title = "{2HDM interpretations of the CMS diphoton excess at 95 GeV}",
    eprint = "2305.19716",
    archivePrefix = "arXiv",
    primaryClass = "hep-ph",
    reportNumber = "KA-TP-10-2023",
    doi = "10.1007/JHEP11(2023)017",
    journal = "JHEP",
    volume = "11",
    pages = "017",
    year = "2023"
}

@article{Gunion_2003,
   title={CP-conserving two-Higgs-doublet model: The approach to the decoupling limit},
   volume={67},
   ISSN={1089-4918},
   url={http://dx.doi.org/10.1103/PhysRevD.67.075019},
   DOI={10.1103/physrevd.67.075019},
   number={7},
   journal={Physical Review D},
   publisher={American Physical Society (APS)},
   author={Gunion, John F. and Haber, Howard E.},
   pages={075019},
   year={2003}
}

@article{LinearCollider:2025lya,
    author = "Abramowicz, H. and others",
    collaboration = "Linear Collider",
    title = "{The Linear Collider Facility (LCF) at CERN}",
    eprint = "2503.24049",
    archivePrefix = "arXiv",
    primaryClass = "hep-ex",
    reportNumber = "DESY-25-054, FERMILAB-PUB-25-0239-CSAID",
    month = "3",
    year = "2025"
}

@phdthesis{Li:2023hsr,
    author = "Li, Cheng",
    title = "{Phenomenology of extended Two-Higgs-Doublets models}",
    school = "U. Hamburg (main), Hamburg U.",
    year = "2023"
}

@article{Biekotter:2023jld,
    author = {Biek{\"o}tter, Thomas and Heinemeyer, Sven and Weiglein, Georg},
    title = "{The CMS di-photon excess at 95 GeV in view of the LHC Run 2 results}",
    eprint = "2303.12018",
    archivePrefix = "arXiv",
    primaryClass = "hep-ph",
    reportNumber = "KA-TP-03-2023, DESY-23-033, IFT-UAM/CSIC-23-028",
    doi = "10.1016/j.physletb.2023.138217",
    journal = "Phys. Lett. B",
    volume = "846",
    pages = "138217",
    year = "2023"
}

@article{Ellwanger:2023zjc,
    author = "Ellwanger, Ulrich and Hugonie, Cyril",
    title = "{Additional Higgs Bosons near 95 and 650 GeV in the NMSSM}",
    eprint = "2309.07838",
    archivePrefix = "arXiv",
    primaryClass = "hep-ph",
    doi = "10.1140/epjc/s10052-023-12315-y",
    journal = "Eur. Phys. J. C",
    volume = "83",
    number = "12",
    pages = "1138",
    year = "2023"
}

@article{CMS:2022arx,
    collaboration = "CMS",
    title = "{Search for dilepton resonances from decays of (pseudo)scalar bosons produced in association with a massive vector boson or top quark anti-top quark pair at $\sqrt{s}=13~\mathrm{TeV}$}",
    reportNumber = "CMS-PAS-EXO-21-018",
    year = "2022"
}

@article{CMS:2022goy,
    author = "Tumasyan, Armen and others",
    collaboration = "CMS",
    title = "{Searches for additional Higgs bosons and for vector leptoquarks in $\tau\tau$ final states in proton-proton collisions at $\sqrt{s}$ = 13 TeV}",
    eprint = "2208.02717",
    archivePrefix = "arXiv",
    primaryClass = "hep-ex",
    reportNumber = "CMS-HIG-21-001, CERN-EP-2022-137",
    doi = "10.1007/JHEP07(2023)073",
    journal = "JHEP",
    volume = "07",
    pages = "073",
    year = "2023"
}

@article{ATLAS:2023jzc,
    collaboration = "ATLAS",
    title = "{Search for diphoton resonances in the 66 to 110 GeV mass range using 140 fb$^{-1}$ of 13 TeV $pp$ collisions collected with the ATLAS detector}",
    reportNumber = "ATLAS-CONF-2023-035",
    year = "2023"
}

@article{CMS:2023yay,
    collaboration = "CMS",
    title = "{Search for a standard model-like Higgs boson in the mass range between 70 and 110$~\mathrm{GeV}$  in the diphoton final state in proton-proton collisions at $\sqrt{s}=13~\mathrm{TeV}$}",
    reportNumber = "CMS-PAS-HIG-20-002",
    year = "2023"
}

@article{ParticleDataGroup:2022pth,
    author = "Workman, R. L. and others",
    collaboration = "Particle Data Group",
    title = "{Review of Particle Physics}",
    doi = "10.1093/ptep/ptac097",
    journal = "PTEP",
    volume = "2022",
    pages = "083C01",
    year = "2022"
}
